\documentclass[12pt,useAMS]{article}
\usepackage{natbib}
\usepackage{graphicx}
\usepackage{booktabs}
\usepackage{multirow}
\usepackage{enumitem}
\usepackage{varwidth}
\usepackage{amsmath, amsthm, amssymb}
\usepackage{float,rotating,subfigure}
\def\aln{\begin{align*}}
\def\eln{\end{align*}}
\def\be{\begin{equation}}
\def\ee{\end{equation}}
\def\bse{\begin{eqnarray*}}
\def\ese{\end{eqnarray*}}
\def\bea{\begin{eqnarray}}
\def\eea{\end{eqnarray}}
\def\Cor{\hbox{Corr}}

\def\bet{\boldsymbol\beta}

\def\bel{\boldsymbol\epsilon}
\def\bga{\boldsymbol\gamma}
\def\bmu{\boldsymbol\mu}
\def\bea{\boldsymbol\eta}
\def\bta{\boldsymbol\tau}

\def\mby{\mathbf y}
\def\mbx{\mathbf x}
\def\mbr{\mathbf r}
\def\mbz{\mathbf z}

\def\ANNALS{{\it The Annals of Statistics}}
\def\AAS{{\it The Annals of Applied Statistics}}

\def\BIOK{{\it Biometrika}}
\def\BIOI{{\it Bioinformatics}}

\def\JASA{{\it Journal of the American Statistical Association}}

\def\JCGS{{\it Journal of Computational and Graphical Statistics }}
\def\JE{{\it Journal of Econometrics}}

\def\JSS{{\it Journal of Statistical Software}}
\def\JRSSB{{\it Journal of the Royal Statistical Society, Series B}}
\def\JMLR{{\it Journal of Machine Learning Research}}

\usepackage{hyperref}
\hypersetup{
    bookmarks=true,							
    unicode=true,         					
    pdftoolbar=true,							
    pdfmenubar=false,						
    pdffitwindow=false,						
    pdftitle={Some Advances in Model Selection of Nonparametric and Semiparametric Modeling},					
    pdfauthor={Zaili Fang},									
    pdfsubject={Ph.D. Dissertation Proposal of Statistics},	
    pdfcreator={Zaili Fang},				
    pdfnewwindow=true,						
    colorlinks=false,						
}
\newcommand{\Appendix}{\appendix\def\thesection{Appendix~\Alph{section}}\def\thesubsection{\Alph{section}.\arabic{subsection}}}

\begin{document}

\font\fmain=cmbx10 scaled\magstep4
\title{A Graphical View of Bayesian Variable Selection}
\author{Zaili Fang and Inyoung Kim$^{*}$}
\date{\today}
\maketitle
\thispagestyle{empty}
\baselineskip=26pt
\hskip 5mm \\
Department of Statistics, Virginia Polytechnic
  Institute and State University, Blacksburg, Virginia, U.S.A.\\
\hskip 5mm \\
\hskip 5mm \\
\noindent
*To whom correspondence should be addressed:\\
Inyoung Kim, Ph.D.\\
Department of Statistics, Virginia Polytechnic
  Institute and State University, 410A Hutcheson Hall, Blacksburg, VA 24061-0439, U.S.A.\\
Tel: (540) 231-5366\\
Fax: (540) 231-3863\\
Email: inyoungk$@$vt.edu\\
\hskip 5mm \\

\newpage
\begin{abstract}
In recent years, Ising prior with the network information for the ``in'' or ``out'' binary random variable in Bayesian variable selections has received more and more attentions. In this paper, we discover that even without the informative prior a Bayesian variable selection problem itself can be considered as a complete graph and described by a Ising model with random interactions. There are many advantages of treating variable selection as a graphical model, such as it is easy to employ the single site updating as well as the cluster updating algorithm, suitable for problems with small sample size and larger variable number, easy to extend to nonparametric regression models and incorporate graphical prior information and so on. In a Bayesian variable selection Ising model the interactions are determined by the linear model coefficients, so we systematically study the performance of different scale normal mixture priors for the model coefficients by adopting the global-local shrinkage strategy. Our results prove that the best prior of the model coefficients in terms of variable selection should maintain substantial weight on small shrinkage instead of large shrinkage. We also discuss the connection between the tempering algorithms for Ising models and the global-local shrinkage approach, showing that the shrinkage parameter plays a tempering role. The methods are illustrated with simulated and real data.
\vskip 5mm
\noindent
\underline{\hbox{\bf Keywords}}: Cluster Algorithm; Global-Local Shrinkage; Graphical Model; Ising Model; KM Model; Long Tail Prior; Mixture Normals; Tempering Algorithm; Variable Selection.
\vskip 5mm
\noindent
\underline{\hbox{\bf Running Title} }:
\thispagestyle{empty}
\end{abstract}
\newpage

\section{Introduction}\label{b.sec1}
In this paper, we consider the standard multiple linear regression model $[\mby|\bet,\phi]\sim N(X\bet, \phi^{-1}I)$, where $\mby$ is $n\times1$ vector of the response variable, $X=(\mbx_1,...,\mbx_p)$ is an $n\times p$ matrix of predictors, $\bet=(\beta_1,...,\beta_p)^T$ is $p\times 1$ model coefficient vector of the full model with $\beta_j, j=1,...,p$ corresponding to the $j$th predictor, and $\phi$ is the precision parameter. The ``in'' or ``out'' of the predictor is represented by a binary random variable $\gamma_j$. The Bayesian spike and slab approaches to sampling $\gamma_j$'s have been introduced by different authors and maintain one of the most active research areas in Bayesian statistics, such as the Stochastic Search Variable Selection (SSVS) \citep{c7} and rescaled spike and slab model \citep{c12}. In recent years, incorporating networked prior information of the predictor into those Bayesian variable selection models has received many attentions \citep{c15,c20,c31,c33}. In all these papers, the network information of the predictors are introduced through an informative prior for $\gamma_j$'s, which is a binary random graph, but none of them treat the variable selection as a graphical model when the prior is noninformative. A binary random graphical model for the random vector $\bga=(\gamma_1,...,\gamma_p)^T$ is represented by an undirected graph $G=(V,E)$, where $V$ represents the set of $p$ vertices or nodes corresponding to $p$ predictors and $E$ is a set of edges connecting neighboring nodes. In this paper, based on a reparameterized Bayesian variable selection model, KM model \citep{c13}, we generalize the Bayesian variable selection problem into a Bayesian graphical model, referred to Bayesian Variable Selection Graphical Model (BVGM), and demonstrate that with the noninformative prior for $\bga$ the model is essentially a complete graphical model.

The Markov chain random process on a random binary graph can be well modeled by a Ising model conditional on $\bet$ and $\phi$. Thus the posterior distribution of $\gamma_j=1$ or the posterior distribution of $j$th predictor ``in'' the model can be achieved by sampling the random binary variable. As one of the most active research areas, abundant theories and sampling procedures for Ising model have been reported. A nice review can be found in \citep{c11,c21}. One difficulty to sample $\bga$ in BVGM is that the interactions are random since they are expressed by the product of $\beta_j$'s and $\phi$. Another difficulty is due to the long-range interaction of the complete graph where each node is coupled or neighboring with all other nodes. In the literature, the well known approaches to handle random and long-range interactions are the cluster algorithm and a family of exchange Monte Carlo, parallel tempering and simulated tempering algorithm \citep{c11}. For the issue of the cluster algorithm, \citet{c22} introduced the Swendsen-Wang algorithm \citep{c32} into Bayesian variable selection and \citet{c20} discussed the Wolff algorithm \citep{c36} in the study of network-structured genomics data. However, both algorithms are constructed based on the graph prior for $\bga$ and consider fixed interaction only. Therefore, both are not applicable to the more general random complete graphical model. In this paper, we generalize the cluster algorithm based on Wolff's approach so that the cluster is formed even with the random interactions among nodes. Furthermore, our generalized Wolff algorithm is introduced for complete graph with noninformative prior for $\bga$, and it is straightforward to combine the graphical prior information.

For the issue of tempering algorithm, so far to our best knowledge there are no work discussing the connection between the tempering algorithm and Bayesian variable selection. In all the Bayesian variable selection models, there is always a critical parameter associated with penalization or shrinkage. By showing the variable selection problem as a Ising model, we address that the well known shrinkage parameter in Bayesian variable selection is equivalent to the temperature parameter in a Ising model. However, in the regular tempering algorithm, there is only one global temperature as a random variable. In BVGM, we adopt the global shrinkage and local acting strategy \citep{c25,c26}, which is employed by assigning the priors of scale normal mixtures for $\beta_j$'s \citep{c0,c35}. Each $\beta_j$ has a local shrinkage parameter, its normal precision parameter, as the local temperature, and there is another global shrinkage parameter to place a constrain on all local parameters. Furthermore, assigning different prior for $\beta_j$'s precision parameter leads to different performance. The widely known priors for $p(\beta_j)$ in this area include Student-t (normal/gamma) prior \citep{c34}, Laplace (normal/inverse gamma) prior \citep{c2,c8,c19}, horseshoe prior \citep{c3}, and Jeffrey's prior \citep{c1}.

Another issue we concern in this paper is the dynamics of the selection probability under different shrinkage. In Bayesian variable selection with large $p$, instead of the appearance frequency of one of the $2^p$ possible models, usually the predictors are selected according to the posterior marginal selection probability, $p(\gamma_j=1|\mby)$, since the frequency of one specific model is extremely small. We define the curves of the selection probabilities of all predictors against the shrinkage parameter as the profile curves of BVGM. These profile curves are important because they provide a direct view about how to select the shrinkage parameter. They also assess the performance of different priors for $\bet$. Unfortunately, we have not seen any work study the overall profile of selection probability under a wide range of shrinkage expect \citet{c17} where they studied the selection probability against the shrinkage with Laplace prior only. Hence one purpose of this paper is to systematically study the dynamics of the selection probabilities, and compare different $\beta_j$ priors with different weight on shrinkage. We address this issue by focusing on the orthogonal design. Interestingly, instead of priors with much weight on large shrinkage, our results indicates that the best performance of a prior is obtained by placing substantial weight on small shrinkage but not zero shrinkage. Among those $\beta_j$ prior candidates, horseshoe prior is the one capable to maintain such shrinkage proportion for the widest range of shrinkage parameter, thus considered as the best.

We also consider one extension of BVGM to Bayesian sparse additive model (BSAM). Unlike the popular topic of Bayesian variable selection, there are only few papers discuss Bayesian variable selection with nonparametric regression \citep{c28,c29,c30}. Based on the KM model, our BVGM is very straightforward to extended to BSAM. We employ the Lancaster and \v{S}alkauskas (LS) spline basis \citep{c4,c14} to express the nonparametric function components. To our best knowledge, our paper is the first one capable to connect the graphical model with the nonparametric regressors such that we can select an appropriate subset of the  function components and estimate the flexible function curves simultaneously.

We first introduce the KM hierarchical model and full conditional distributions for sampling the parameters except $\bga$ in Section \ref{b.sec2}. In Section \ref{b.sec3} we discuss the connection between Bayesian variable selection and binary random graphical model and express our model as the Ising model with noninformative prior for $\bga$. Then in Section \ref{b.sec4}, we first introduce the single site algorithm for sampling $\bga$, then present a generalized Wolff cluster algorithm. In Section \ref{b.sec5}, we focus on understanding the selection probability profile including the dynamics of the selection probability under different shrinkage priors, and we discuss the connection between the simulated tempering algorithm and priors of the scale mixture of normals. In Section \ref{b.sec6} and \ref{b.sec7} we consider two extensions, one is how to incorporate prior network information for $\bga$, another one is how to extend to BSAM with LS basis. In Section \ref{b.sec8} and \ref{b.sec9}, we illustrate our model with simulations and real data analysis. Finally, in the last section, we conclude our work and discuss other potential extensions of our model.

\section{Bayesian Variable Selection with Normal Mixture Priors}\label{b.sec2}
We are interested in selecting a subset of predictors from the $p$ potential candidates. Thus we introduce the binary random vector
\[\label{b.e1}
\bga=(\gamma_1,...,\gamma_p)^T,
\]
where $\gamma_j\in (0,1), j=1,...,p$ is the binary indicator random variable corresponding to the $j$th predictor.  With $\gamma_j=1$ we selecte predictor $\mbx_j$ otherwise exclude it from the model. To implement the stochastic search for $\gamma_j$'s, SSVS considers a multi-mode point mass and Gaussian mixture prior for $\beta_j$'s, $[\beta_j|\gamma_j,\tau_\beta]\sim(1-\gamma_j)\delta(0)+\gamma_jN(0,\tau^{-1}_\beta)$, where $\delta(0)$ represents the point mass density at zero.

In this paper we consider the KM model, which is expressed as
\be\label{b.e2}
\mby=\sum_j^p\gamma_j\mbx_j\beta_j+\bel,
\ee
where $\bel\sim N(\mathbf0,\phi^{-1}I)$ is independent identical noise vector. We standardize the data set $X$ and center the response $\mby$ such that  $\sum_{i=1}^nx_{ij}^2=1, \sum_{i=1}^nx_{ij}=0, j=1,...p$ and $\sum_{i=1}^ny_i=0$. We may also include an intercept term $\mu$ in model (\ref{b.e2}) with a normal prior, which requires only a simple extra step in the sampling procedure. In Section \ref{b.sec7}, this parametric linear regression model is easy to extend to nonparametric additive model by using some basis function to express the $j$th individual function component with $\bet_j$ as a parameter vector for the $j$th predictor.

The reasons we employ KM model in this paper are: first, it is more natural as a variable selection model, where $\gamma_j=0$ indicates that $j$th predictor has no effect in the response. Second, spike and slab models such as SSVS consider a multi-mode prior for $\beta_j$'s which may have a mixing problem for sampling $\beta_j$'s since $\beta_j$'s may get trapped in the point mass mode for a long time. This problem becomes worse when we extend the SSVS to nonparametric additive model \citep{c29}, because the chance of moving between the point mass and the normal model for $\bet_j$ becomes lower in higher dimensional space. The third reason can be demonstrated in next section where we can see that it is very straightforward to express a KM model in a Ising model, while it is difficult for SSVS models.

In usual Bayesian variable selection, the normal prior assigned to $\beta_j$'s has form $[\bet|\tau_\beta]\sim N(0,\tau_\beta^{-1}I_p)$, where $\tau_\beta$ is a common precision parameter for all $\beta_j$'s and usually assigned an gamma prior with scale $a/2$ and rate $b/2$. Similar prior can be $[\bet|\tau_\beta]\sim N\left(0,\tau_\beta^{-1}(X^TX)^{-1}\right)$ assuming $\tau_\beta=g^{-1}\phi$, where $g$ is a positive number called $g$-factor \citep{c16}. Because of this simplicity, $\beta_j$'s and $\phi$ can be integrated out and a closed form of the posterior distribution for $\bga$ can be achieved. However, we realized this simplicity has many disadvantages for variable selection purpose. For example, if we integrate out $\tau_\beta$ and achieve the marginal prior of $\bet$ as $p(\bet|a,b)\propto(\bet^T\bet+b)^{-{{p+a}\over2}}$, we can see the prior of $\beta_j$'s are no longer statistically independent to each other. This is not a good idea to explore the whole joint distribution space of $(\bga, \bet)$ since the main purpose of Bayesian variable selection is to explore the space of $\bga$, while dependent $\bet$ prior will limit the stochastic searching space. Based on this argument, we follow the shrink globally act locally scheme suggested by \citet{c25} to assign independent normal mixture priors for $\beta_j$'s. In next section, we will see that the interactions of the Ising model are determined by $\beta_j$'s. The larger $\beta_j$'s of two predictors, the larger the interaction between them, then the corresponding nodes have high probability to be dependent, meaning they are either ``aligned'' (both $\gamma_j$ equal to 1, or both equal to 0), or ``anti-aligned'' (one $\gamma_j$ equals to 1, and another equals to 0). On the other hand, if $\beta_j$'s between two nodes are very small, then the two predictors are independent to flip their $\gamma_j$ values. Therefore, we want $\beta_j$'s to be as flexible as possible to explore the configuration space, while we do not want to lose the control so we constrain the overall variability of the interaction through a global parameter, which we refer to $b$.

The shrink globally act locally scheme is easy to be implemented by following hierarchical model with scale normal variance mixture priors for $\beta_j$'s.
\be\begin{split}\label{b.e3}
[\beta_j|\tau_j,b]&\sim N(0,b^2\tau^{-1}_j),\\
[\tau_j]&\sim p(\tau_j),\\
[\phi]&\sim p(\phi),
\end{split}\ee
where $\tau_j$ is the precision parameter for the conditional normal prior of $\beta_j$ and plays the role of local tempering. $p(\tau_j)$ and $p(\phi)$ are the priors for $\tau_j$'s  and $\phi$ respectively. Similar hierarchical model in SSVS setting also has been discussed by \citet{c9}.  With these settings, we can easily achieve the full conditional distribution for $\bet_c\subseteq\bet$
\be\label{b.e4}
[\bet_c|\mby,\bga, \bet_{\bar{c}},\phi]\sim\left\{\begin{array}{l l}
  N(\bmu_c, \Sigma_c)   &\hbox{ if }\bga_c=\mathbf1           \\
  N(\mathbf0,D^{-1}_c)  &\hbox{ if }\bga_c=\mathbf0.\end{array}\right.
\ee
Here we use a general subscript``$c$'' to stands for subset of the index $\{1,...,p\}$. We use $\bar{c}$ to present the complementary index set of $c$. In above expression, $D_c$ is a $|c|\times|c|$ diagonal matrix with $\tau_j/b^2, j\in c$ as the diagonal elements, where $|c|$ stands for the cardinality of $c$. $\Sigma_c$ and $\bmu_c$ are expressed as
\be\begin{split}\label{b.e5}
\Sigma_c&=\left(\phi X_c^TX_c+D_c\right)^{-1},\\
\bmu_c&=\phi\Sigma_cX_c^T\left(\mby-X_{\bga_{\bar{c}}}\bet_{\bar{c}}\right).
\end{split}\ee
With some notations abuse here, $X_c$ stands for the sub-matrix of $X$ corresponding the predictors in $c$, and $X_{\bga_{\bar{c}}}$ is the sub-matrix of $X_{\bga}=(\gamma_1\mbx_1,...,\gamma_p\mbx_p)$ corresponding to predictors in $\bar{c}$.

We simply assign a noninformative prior for $\phi: [\phi]\sim\phi^{-1}$ and the posterior distribution of $\phi$ is simply a gamma distribution.
\be\label{b.e6}
[\phi|\mby,\bet,\bga]\sim G\left({n\over2}, {1\over2}\|\mby-X_{\bga}\bet\|^2\right).
\ee

The prior for $\tau_j$'s are critical since it determines how the local action of the sampling process. Many different type of $p(\tau_j)$ can be considered. A very general review of different choice for $p(\tau_j)$ can be found in \citet{c27,c25}. In this paper we only consider three widely known $p(\tau_j)$'s that result in three typical marginal $\beta_j$'s priors with characteristics of heavy tail, heavy mass around zero and both. We refer to these marginal priors of $\beta_j$ as Cauchy, Laplace and horseshoe priors which are achieved by assigning gamma prior $[\tau_j]\sim G(1/2, 1/2)$, inverse gamma prior $[\tau_j]\sim IG(1,1/2)$ and half Cauchy prior $[\tau_j^{1/2}]\sim C^+(0,1)$ to $\tau_j$ respectively. The density forms for these three normal mixture settings are list in Table \ref{t1} respectively. Notice, to avoid the confusion, the terms of ``Cauchy'', ``Laplace'' and ``horseshoe'' not only refer to the marginal priors of $\beta_j$'s but also represent the normal mixture settings. For example, in the context, ``Cauchy prior'' stands for normal/gamma setting such that the marginal prior of $\beta_j$ is Cauchy and the prior for $p(\tau_j)$ is $G(1/2, 1/2)$.
\begin{table}[h]
\centering
\caption{Summary of Cauchy, Laplace and horseshoe priors for the marginal prior of $\beta_j$'s, corresponding priors for $p(\tau_j)$ and the density functions of the shrinkage parameter $\kappa_j$.}
\begin{tabular}{llll}
\hline\hline
Marginal prior &$p(\beta_j|b)$                    & Prior for $\tau_j$                        &Distribution for $\kappa_j$\\
\hline
Cauchy         &$\pi b(\beta_j^2+b^2)^{-1}$    &$\tau_j^{-{1\over2}}\exp(-{{\tau_j}\over2})$ &$\kappa_j^{-{1\over2}}(1-\kappa_j)^{-{3\over2}}\exp(-{{b^2\kappa_j}\over{2(1-\kappa_j)}}) $ \\
Laplace        &$(2b)^{-1}\exp(-|\beta_j|/b)$ &$\tau_j^{-2}\exp(-{2\over{\tau_j}})$ & $\kappa_j^{-2}\exp(-{{(1-\kappa_j)}\over{2b^2\kappa_j}})$  \\
Horseshoe      &-                                 &$\tau_j^{-{1\over2}}(1+\tau_j)^{-1}$               &$\kappa_j^{-{1\over2}}(1-\kappa_j)^{-{1\over2}}\left[1-\kappa_j+b^2\kappa_j\right]^{-1} $              \\
\hline
\hline
\end{tabular}
\label{t1}
\end{table}

By defining a scaleless parameter $b^*=b/\sqrt{\phi}$, The full conditional distribution for $\tau_j$ in Cauchy and Laplace settings are
\begin{align}
[\tau_j|\beta_j,b]&\propto\tau_j^{-3/2}\exp\left[-{{(\tau_j-b^*/|\beta_j|)^2}\over{2\tau_j{b^*}^2/\beta_j^2}}\right],&\hbox{Laplace prior}\label{b.e7}\\
[\tau_j|\beta_j,b]&\propto\exp\left[-{1\over2}(\beta_j^2/{b^*}^2+1)\tau_j\right],&\hbox{Cauchy prior}\label{b.e8}
\end{align}
and the full condition distribution for $\tau_j$ in horseshoe prior setting is obtained by
\begin{align}
[u_j|\beta_j,b,v_j]&\propto\exp\left[-{1\over2}\left({\beta_j^2\over{{b^*}^2v_j}}+1\right)u_j\right],&\nonumber\\
[v_j|\beta_j,b,u_j]&\propto v_j^{-1}\exp\left[-{1\over2}\left({{\beta_j^2u_j}\over{{b^*}^2}}{1\over v_j}+v_j\right)\right],&\hbox{Horseshoe prior}\label{b.e9}\\
\tau_j&=u_j/v_j,&\nonumber
\end{align}
(\ref{b.e7}) is an inverse Gaussian distribution $ING({b^*}/|\beta_j|, 1)$ with mean ${b^*}/|\beta_j|$ and shape parameter 1. (\ref{b.e8}) is an exponential distribution or Gamma distribution $G\left(1,(\beta_j^2/{b^*}^2+1)/2\right)$. The Gibbs sampler for horseshoe prior is implemented by using the redundant multiplicative reparameterization technique similar to \citet{c5}. Reparameterize $\tau_j$ as $\tau_j=u_j/v_j$ where $u_j$ and $v_j$ are independently distributed with prior $G(1/2, 1/2)$ respectively, then the prior for $\tau_j^{1/2}\sim C^+(0,1)$, and the prior for $\beta_j$ is the horseshoe prior. In (\ref{b.e9}), the full conditional distribution for $u_j$ is a Gamma distribution $G(1,(\beta_j^2/({b^*}^2v_j)+1)/2)$ and the full conditional distribution for $v_j$ is generalized inverse Gaussian distribution $GING\left(0,{{\beta_j^2u_j}\over{{b^*}^2}}, 1\right)$.
\begin{figure}[bth]
\begin{center}
\includegraphics[height=70mm]{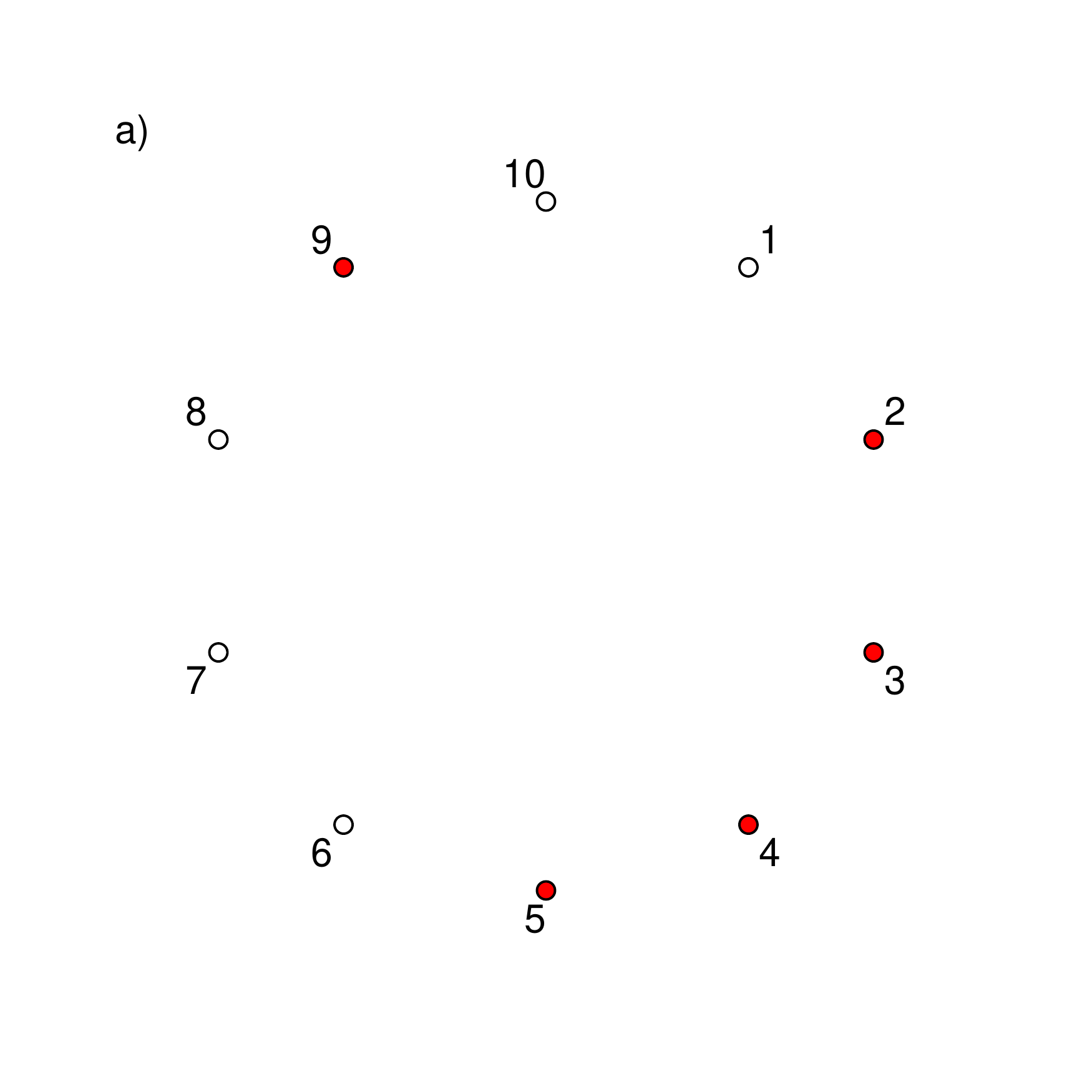}\includegraphics[height=70mm]{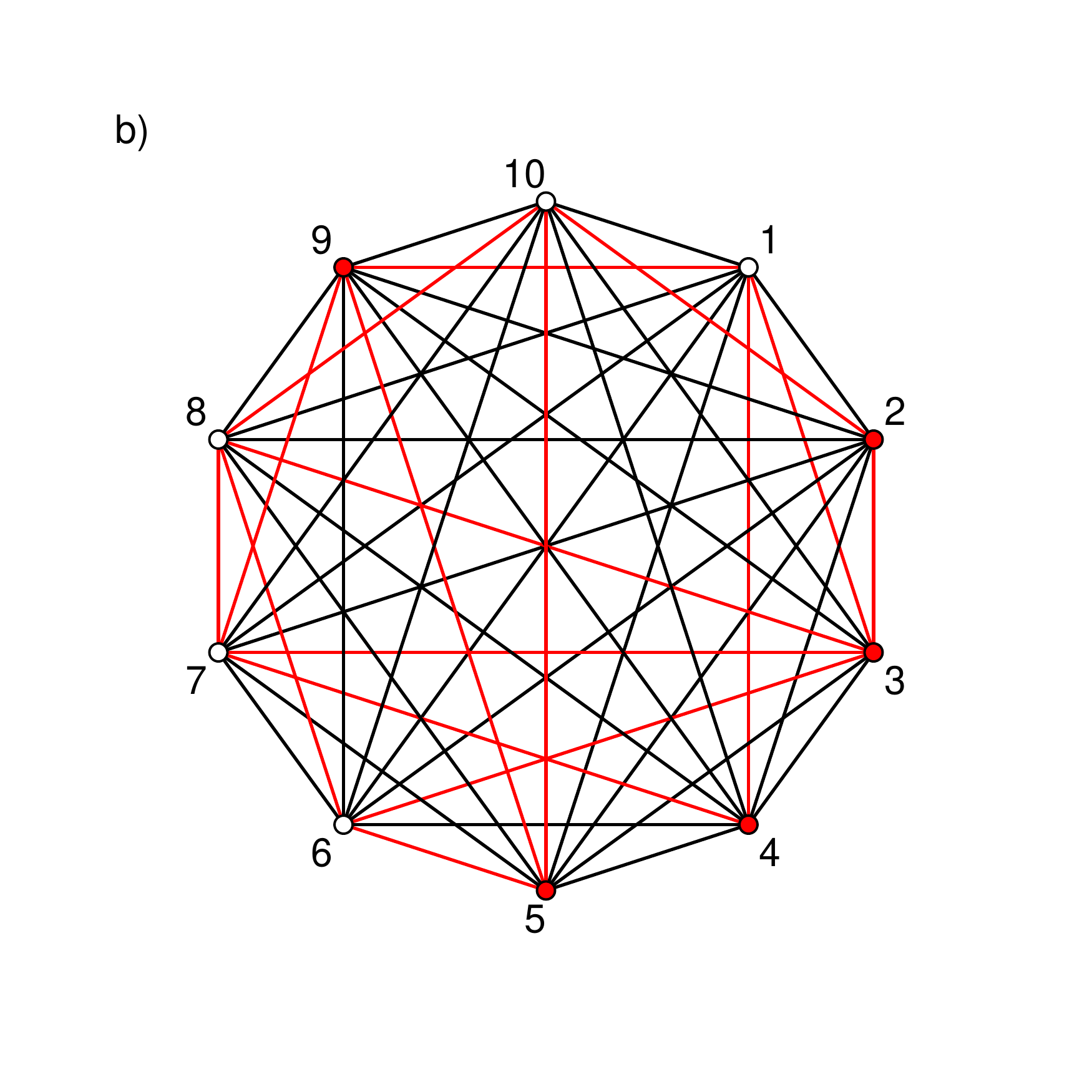}
\caption{Diagram of variable selection as a random graph model with  selected nodes (filled circles), excluded nodes (circles), edges of positive interaction (black lines), and edges of negative interaction (red lines). Independent variable selection: no interactions among nodes (a). General variable selection: a complete graph (b). }\label{b.pic1}
\end{center}
\end{figure}
\section{Bayesian Variable Selection and Binary Random Graphical Model}\label{b.sec3}
The noninformative prior for $\bga$ is $\bga\sim \left(1\over2\right)^p$. Thus the full conditional distribution of $\bga|\bet, \phi$ is directly derived from the likelihood of $\bga$ given $\bet$ and $\phi$. Given $\bet$, consider the matrix of marginal regression functions $R=(\mbr_1,...,\mbr_p)=(\beta_1\mbx_1,...,\beta_p\mbx_p)$, with each column as the marginal regression vector for $j$th predictor vector. In additive nonparametric model (see Section \ref{b.sec7}), $\mbr_j=f_j(\mbx_j)=Z_j\bet_j$ is the nonparametric function component of $\mbx_j$ expanding on the $n\times M_j$ basis matrix $Z_j$ with $1\times M_j$ coefficient vector $\bet_j$ ($M_j$ is the dimension of the basis). Here we consider parametric regression model only, thus the full conditional distribution of $\bga$ is
\be\begin{split}\label{b.e10}
p(\bga|\mby,\bet,\phi)&\propto p(\mby|\bga,\bet,\phi)\\
&\propto\exp\left(-{1\over2}\phi\bga^TR^TR\bga+\phi\mby^TR\bga\right).
\end{split}\ee
This is nothing more than a Boltzman distribution of Ising model, ${1\over Z}\exp(-U(\bga))$, with
\be\begin{split}\label{b.e11}
U(\bga)&=-\bga^T J\bga-\mathbf h^T\bga,\\
J&=-{{\phi R^TR}\over2},\\
\mathbf h&=\phi R^T\mby,
\end{split}\ee
where $Z=\sum_{\bga}\exp(-U(\bga))$ is called the partition (normalized) function and $U(\bga)$ is called the energy of state $\bga$ given $\bet$ and $\phi$, $J$ is the interaction matrix and $\mathbf h$ is called ``external field''. Above expression of Ising model is equivalent to following model:
\be\label{b.e12}
p(\bga|\mby,\bet,\phi)\propto\exp\left(\sum_{i<j}J_{ij}\delta_{ij}+\sum_jh^*_j\gamma_j\right),
\ee
where the first summation is on all $i<j, j=1,...,p$, $\delta_{ij}=1$ if $\gamma_i=\gamma_j$ otherwise $\delta_{ij}=0$, $J_{ij}=\phi\beta_i(\mbx^T_i\mbx_j)\beta_j$ is the non diagonal element of matrix $J$ and $h_j^*$ is the $j$th element of vector $\mathbf h^*=\phi R^T(\mby-R\mathbf 1/2)$. Above expression is achieved by plugging in following transformation into (\ref{b.e10})
\be\label{b.e13}
2\left[{1\over4}+\left(\gamma_i-{1\over2}\right)\left(\gamma_j-{1\over2}\right)\right]=\delta_{ij}=\left\{\begin{array}{l l}
  1& \gamma_i=\gamma_j           \\
  0& \gamma_i\neq\gamma_j.\end{array} \right.
\ee
In the literature, the model with $\bga$ distributed as (\ref{b.e10}) is called spin glass model (consider $\gamma_j$ has two spin states, up and down, corresponding to 1 and 0 respectively) when the coupling parameter $J_{ij}$ follows some random distribution with positive or negative values. In our Bayesian variable selection model, because $J_{ij}$ is the product of random variable $\beta_j$'s and $\phi$ each has a prior, the distribution for $J_{ij}$ is some unknown distribution usually is neither iid nor tractable. Therefore, numerical method, such as MCMC sampling to simulate the distribution of $\bga$ is required. Now we can see that the choice of the prior for $\beta_j$'s is important since it directly effects the interaction among the nodes. The independent scale normal mixture prior for $\beta_j$'s is a nice choice since it is similar to the well known tempering algorithm in Ising model, and we can also derive some cluster algorithms.  Both algorithms are expected to improve the mixing issue of the sampler \citep{c22,c32,c36}.

Based on the Ising model, considering the $p$ predictors as a set of nodes, we assign a binary random variable $\gamma_j$ for each nodes. Those nodes may interact or couple with each other as described by a Ising model, so we have the following proposition:

\noindent{\bf Proposition 1:} {\it The $p$ dimension binary random variable $\bga\in\{0,1\}^p$ of the Bayesian variable selection problem based on KM model (\ref{b.e2}) is a class of stochastic processes on a finite random undirect graph model $G=(V, E)$, where $V=\{1,..., p\}$ is the set of nodes, corresponding to $p$ predictors, and $E\subset V\times V$ is the set of edges. $\bga\in\Gamma=\{(\gamma_1,...,\gamma_p):\gamma_j\in(0,1),j=1,...,p\}$ is indexed by $V$ with probability measure on $\Gamma$ as (\ref{b.e12}), in which $J_{ij}$'s and $h^*_j$'s are all random with some distributions determined by the priori distributions of $\beta_j$'s and $\phi$.
}

This is a complete graph model, since we don't limit the connection between any two nodes of $V$ and the coupling between two nodes are long-range interaction. Figure \ref{b.pic1} is the diagram of the graphical model for Bayesian variable selection. In Figure \ref{b.pic1} (a), the interaction between any nodes $J_{ij}=0$ thus this is a complete independent setting with which the configuration of $\bga$ depends on the ``external field'' $\mathbf h$ only. Figure \ref{b.pic1} (b) is a more general diagram for the complete graphical model. However, since any possible $J$ is allowed, for a given $J$, a specific configuration of the edges will be given. For example, for a one dimension Ising model, the nodes form a one dimension chain, and one node only interacts with its two nearest neighbor nodes. This means the matrix $J$ is a sparse matrix with non zero elements in positions $|i-j|\le1$ only. Furthermore, we can also consider the external field $\mathbf h^*$ as a node indexed by $0$ which represents the response variable $\mby$, except that $\gamma_0=1$ is fixed. Then (\ref{b.e12}) can be expressed by a more compact form:
\[\label{b.e14}
p(\bga|\mby,\bet,\phi)\propto\exp\left(\sum_{i<j}J_{ij}\delta_{ij}\right),
\]
where $i,j=0, 1,...,p$, and $J_{ij}$ is the extended matrix with the first row and column equal to $\mathbf h^*$. However, in this paper we keep focus on expression (\ref{b.e12}) for explicitness.
\section{Updating of $\bga$}\label{b.sec4}
\subsection{Single Site Algorithm}\label{b.sec4.1}
The joint posterior distribution of $\bga, \bet$ and $\phi$ will directly give this posterior distribution for $\bga$. Because (\ref{b.e10}) and (\ref{b.e12}) are the direct form of Ising model, we can direct apply the Gibbs sampler procedure for $\bga$ based on (\ref{b.e10}) and (\ref{b.e12}) after sampling $\bet$ and $\phi$. This means we assign a noninformative prior for $\bga$, $\bga\sim \left({1\over2}\right)^p$, and the full conditional distribution given the data for single site updating is
\be\begin{split}\label{b.e15}
[\gamma_j&|\mby,\bga_{\bar{j}},\bet,\phi]\sim Ber\left({1\over{1+\pi}}\right),\\
\pi&=\exp\left\{-[U(\gamma_j=1|\bga_{\bar{j}})-U(\gamma_j=0|\bga_{\bar{j}})]\right\}\\
&=\exp(-J_{jj}-2J_{j\bar{j}}\bga_{\bar{j}}-h_j),
\end{split}\ee
where $Ber$ stands for the Bernoulli distribution and $J_{j\bar{j}}$ is the $j$th row of $J$ with $j$th column removed. $U(\gamma_j=1|\bga_{\bar{j}})-U(\gamma_j=0|\bga_{\bar{j}})$ is the ``energy'' difference of two state configurations, $\gamma_j=0|\bga_{\bar{j}}$ and $\gamma_j=1|\bga_{\bar{j}}$, where $\bga_{\bar{j}}$ is the vector of $\bga$ with $\gamma_j$ removed.  Therefore the complete full conditional distributions of Gibbs sampler to update $\bga, \bet$ and $\phi$ involves expression (\ref{b.e4}), (\ref{b.e5}), (\ref{b.e6}) and one of (\ref{b.e7}-\ref{b.e9}). This procedure is very simple and works well for most cases with moderate size $p$. The main advantage of our procedure is there is only one tuning parameter $b$, and the ``tuning'' process is extremely simple: just choose a $b$ that separates the signals and noises with the largest gap in the marginal selection probability.
\begin{figure}[bth]
\begin{center}
\includegraphics[height=55mm]{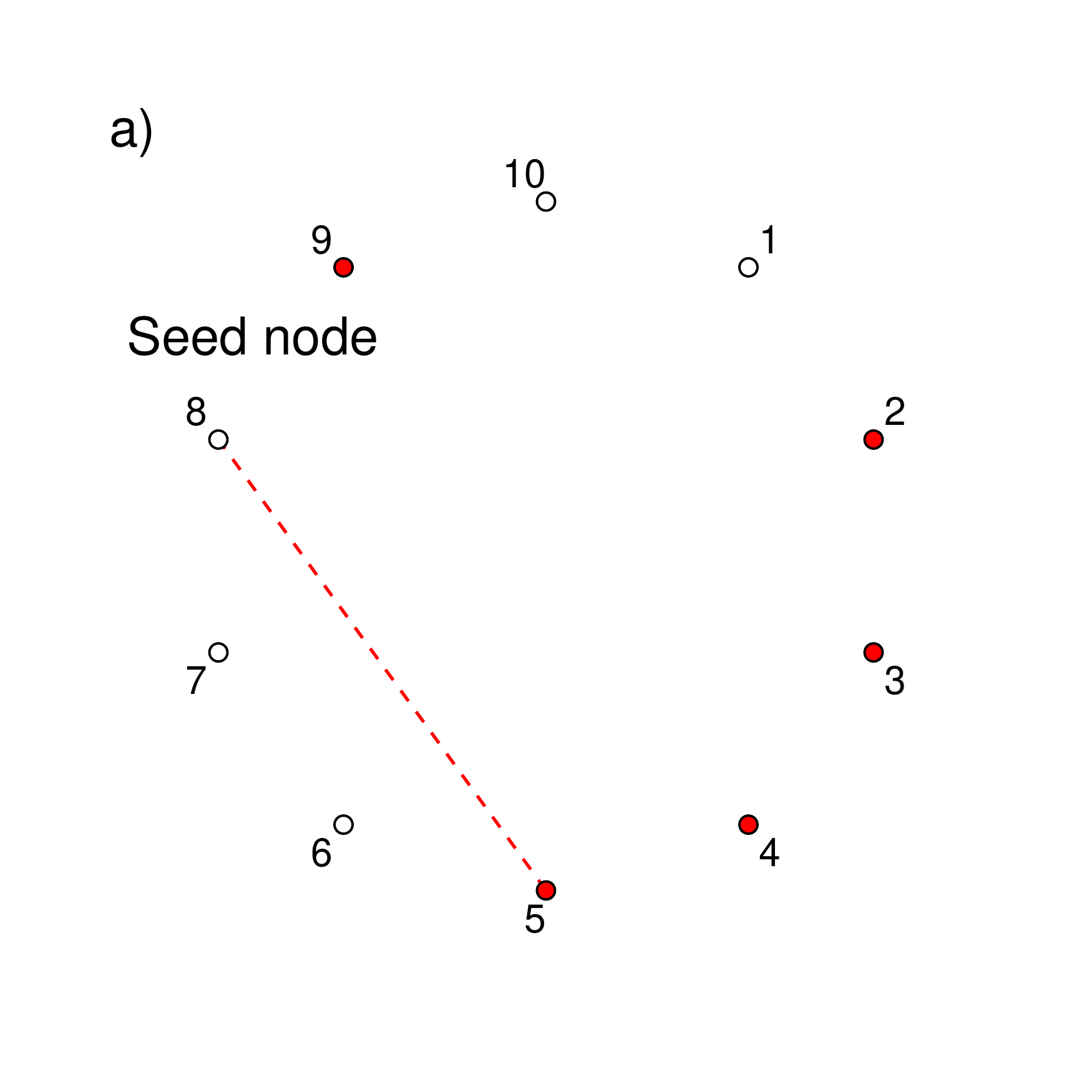}\includegraphics[height=55mm]{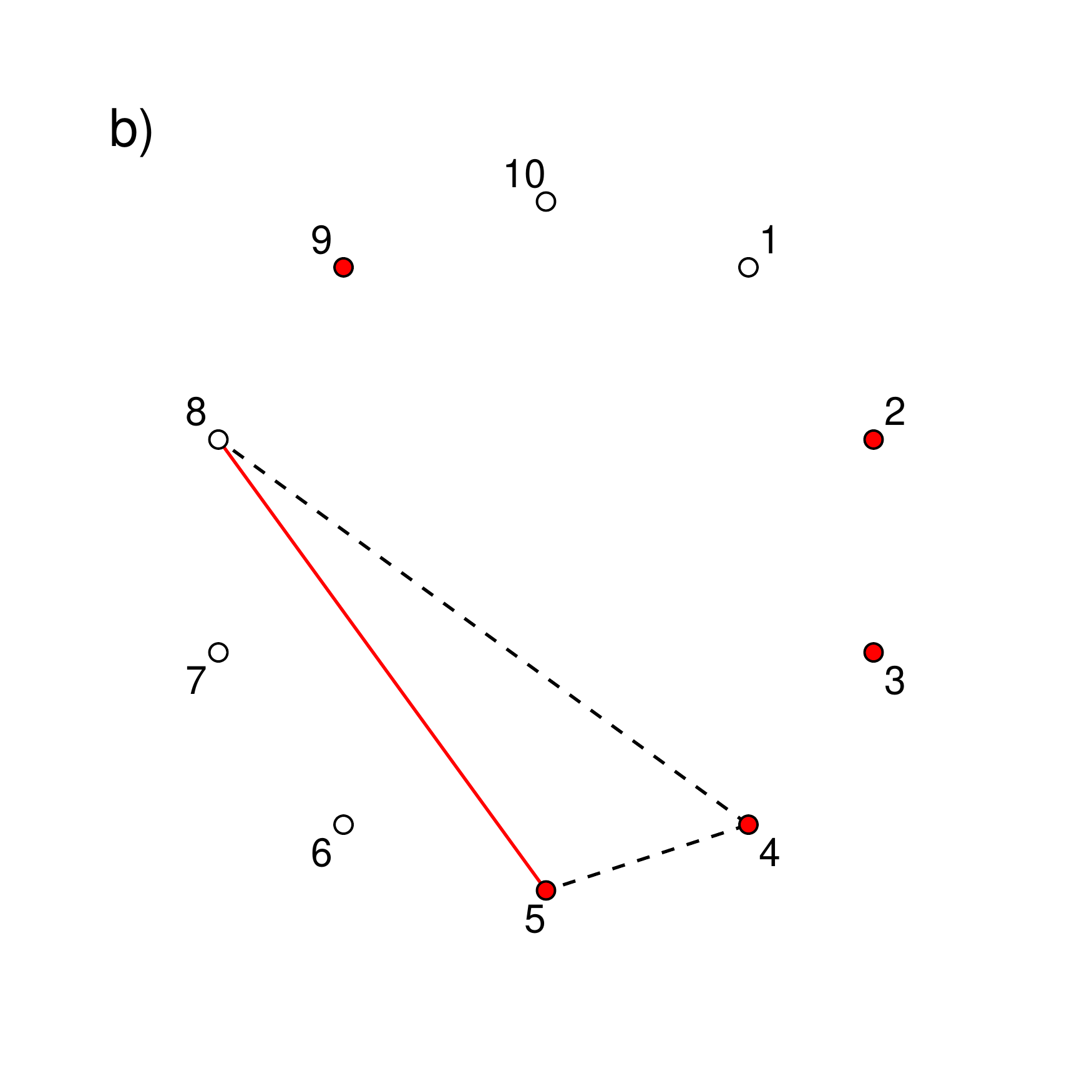}\includegraphics[height=55mm]{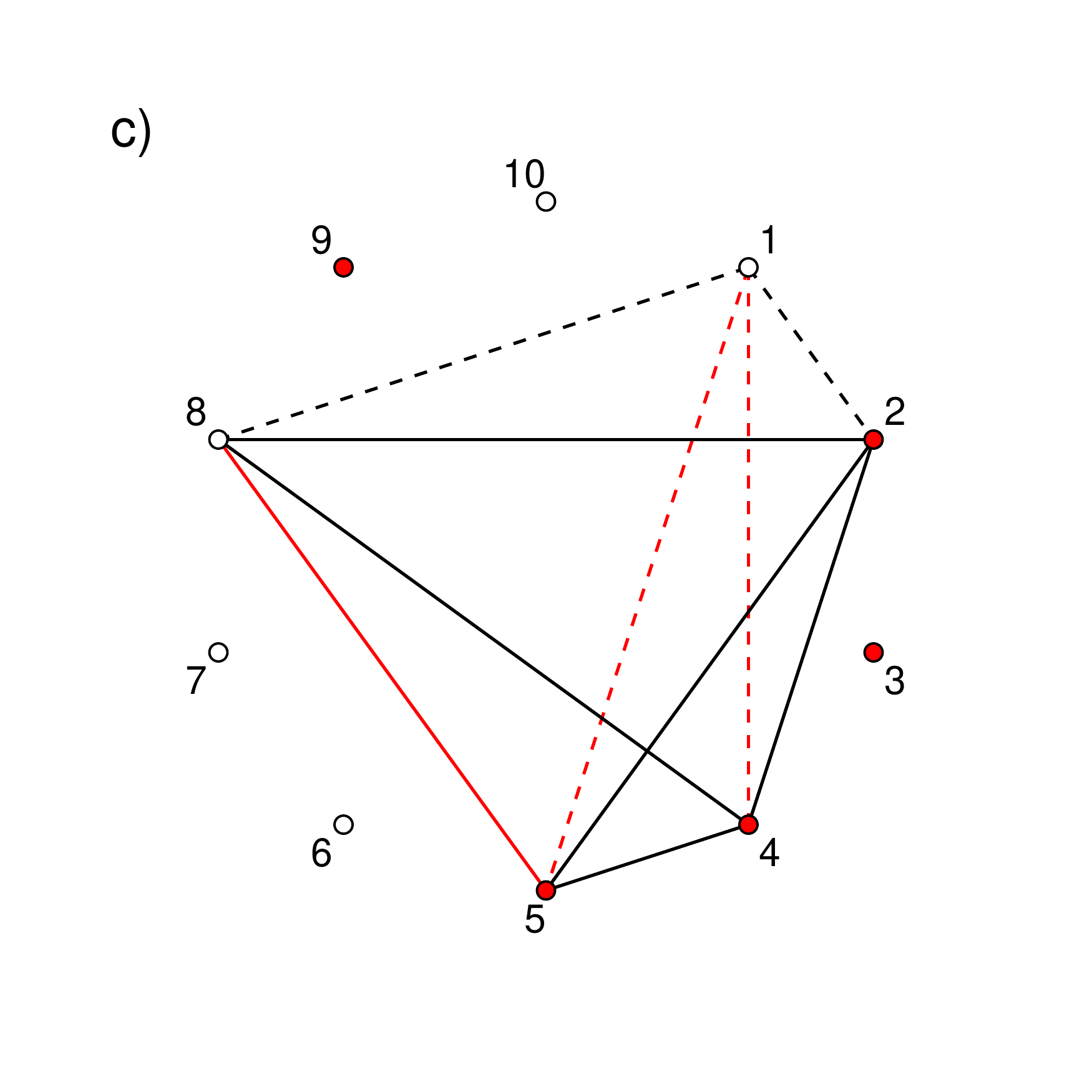}
\includegraphics[height=70mm]{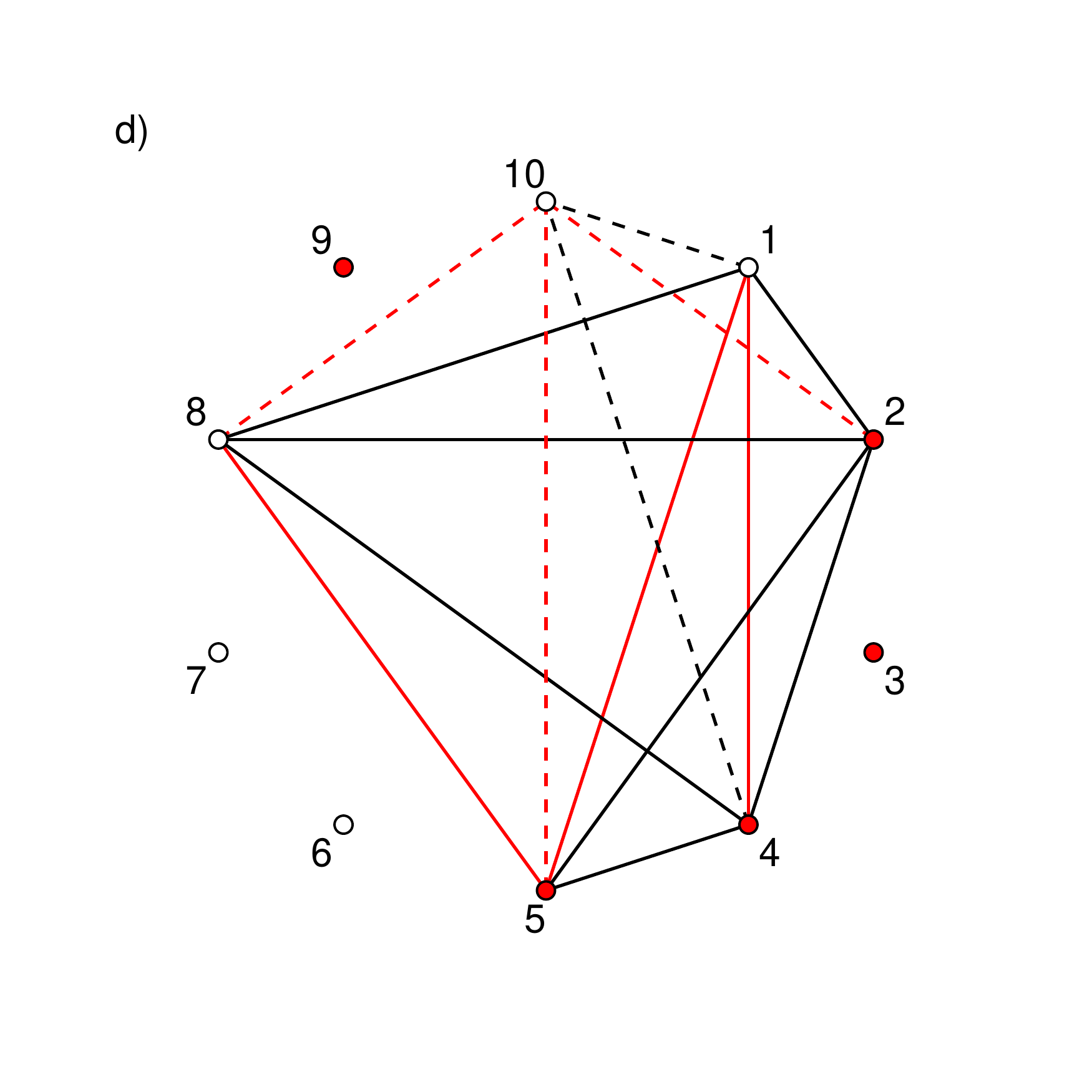}\includegraphics[height=70mm]{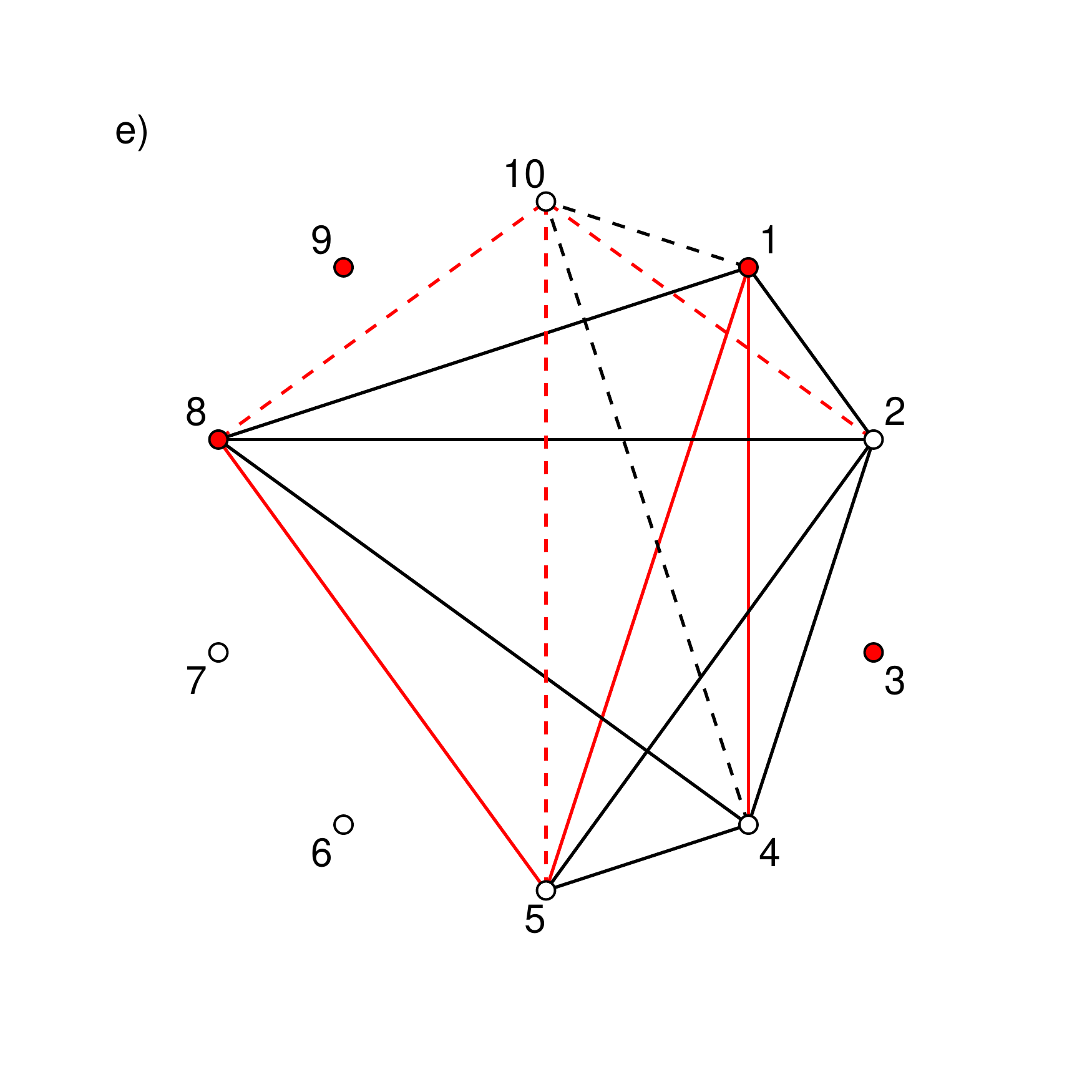}
\caption{Diagram of the cluster algorithm. Forming the cluster (a-c). Flipping clustered nodes (d-e).}\label{b.pic2}
\end{center}
\end{figure}

Above Gibbs sampler to update is one Matropolis-Hastings (MH) step with the Gibbs proposal and acceptance rate equal to one. We can also consider a MH one-step updating, which is more general in Ising model sampling. Denote the current state for $\gamma_j$ as $\gamma_j^0|\bga_{\bar{j}}$ and its flipped state $\gamma_j^*|\bga_{\bar{j}}$, whether or not we move from $\gamma_j^0|\bga_{\bar{j}}$ to $\gamma_j^*|\bga_{\bar{j}}$ depends on the ``energy'' difference $\Delta U=U(\gamma_j^*|\bga_{\bar{j}})-U(\gamma_j^0|\bga_{\bar{j}})$. We prefer the system in lower ``energy'' state since the lower the energy the higher the probability. Thus if $\Delta U\leq0$, the flipped state is accepted with probability 1. We treat the case $\Delta U>0$ probabilistically, that is, with the probability to accept the flipped state as $p(\Delta U)=\exp(-\Delta U)$. These steps can be summarized as that we flip current state to its opposite rather than remaining the current state with probability
\be\label{b.e16}
\min\left(1, \exp(-\Delta U)\right).
\ee
The detailed balance maintains and this MH updating is used in the MCMC Ising model sampling \citep{c21,c22}. In this paper, unless otherwise specified, we adopt this one step MH updating (\ref{b.e16}) with other Gibbs samplers in all cases. Indeed, it is the antithetic updating method discussed by \citet{c22} since $\exp(-\Delta U)$ is the odds of flipping current state with the Gibbs type proposal.

\subsection{Cluster Algorithm}\label{b.sec4.2}
Beyond the single site algorithm, the cluster algorithm is well established for simulating model (\ref{b.e10}) when $J_{ij}$'s and $h_j$'s are fixed. There are abundant literatures available about the cluster algorithm in Ising model \citep{c21,c32,c36}, but it has been introduced to Bayesian variable selection recently only \citep{c22}. In general, a cluster algorithm performances better than the single site updating when $J_{ij}$ is fixed. However, as pointed before, the model (\ref{b.e10}) is difficult in applying the clustering-updating algorithm since there is a random external field $\mathbf h^*$ and the coupling coefficients $J_{ij}$'s follow some unknown distribution and are not independent. Plus the nodes are connected with each other by so called long-range interaction thus the system is a totally disordered complete graph. In this paper, we propose a generalized single-cluster Monte Carlo algorithm which is closel to Wolff's clustering scheme but capable to handle the situation with long-rang random interaction and external field.

In the original SW and Wolff algorithm, clusters are formed through the bonding between paired nodes with positive interactions. Although \citet{c22} proposed an auxiliary variable technique to count the negative coupling between nodes and form clusters including anti-aligned nodes, their method is still based on the single bond between two nodes, which means whether adding a new node to the cluster is determined by the interaction between the new node and ONE node in the cluster. Unlike the usual Ising model on one dimension chain or two/three dimension lattice, the complete graph model of the binary random process is fully connected. This indicates each single node behaves according to the overall effects of all other nodes. Therefore, the clustering dynamics must incorporate this consideration. In other words, the growth of a cluster (adding one new node to the existing cluster) should consider the coupling between the new node and all nodes in the cluster.

Before introduce the cluster algorithm, we specify two types of clusters since the cluster is formed according to the coupling coefficient $J_{ij}$ which can be either positive or negative.
\begin{itemize}
\item a cluster with nodes aligned.
\item a cluster with nodes aligned and anti-aligned.
\end{itemize}
We use $c$ to denote the cluster, and $\bar c$ as the complement of $c$. The single node is considered as special case of the cluster with aligned nodes. So within the second type of cluster, there are two sub clusters anti-aligned to each other. We denote these two sub clusters as $c_1$ and $c_0$ with $\bga_{c_1}=\mathbf1$ and $\bga_{c_0}=\mathbf0$ respectively.

The question then is, given a particularly defined probability $p_a$ of adding a node to the cluster, what is the acceptance ratio that make the flip of the cluster satisfies detailed balance, and how to choose $p_a$ such that the average acceptance ratio is as large as possible? So we derived following generalized Wolff algorithm based on these considerations.
\begin{enumerate}
\item Form the cluster.
\begin{enumerate}
\item Initialize the cluster set $c$ by randomly picking a seed node.
\item Examine the nodes in $\bar c$ one by one, add the  node $j$ in $\bar c$ to the cluster with the probability
\be\label{b.e17}
p_{a,j}=\max\left\{1-\exp\left[\lambda(-1)^{\gamma_j}\left(\sum_{k\in c_1}J_{jk}-\sum_{l\in c_0}J_{jl}\right)\right],0\right\},
\ee
and remove $j$ from $\bar c$ if $j$ added to $c$, where $0\leq\lambda\leq1$. Continue iteratively until no new sites added when each nodes in $\bar c$ has been examined.
\end{enumerate}
\item Flip the nodes in cluster $c$ with probability
\be\begin{split}\label{b.e18}
&\alpha(\bga_c^0\rightarrow\bga_c^*)\\
&=\min\left\{\exp\left[(1-\lambda)\sum_{j\in\bar c}(-1)^{\gamma_j}\left(\sum_{k\in c_1}J_{jk}-\sum_{l\in c_0}J_{jl}\right)+\sum_{j\in c_0}h^*_j-\sum_{j\in c_1}h^*_j\right],1\right\}\\
&=\min\left\{\exp\left[(1-\lambda)(\mathbf1^TJ_{c_0\bar c}-\mathbf1^TJ_{c_1\bar c})(2\bga_{\bar c}-\mathbf1)+\mathbf1^T\mathbf h^*_{c_0}-\mathbf1^T\mathbf h^*_{c_1}\right],1\right\}.
\end{split}
\ee
\item Flip the rest nodes in $\bar c$ (if any left) by single updating method (\ref{b.e16}).
\item Update $\beta_j$'s, $\tau_j$'s and $\phi$.
\end{enumerate}
In (\ref{b.e18}), the last expression is for the convenience of coding using matrix expressions. As we can see, parameter $\lambda$ plays a role of partial clustering similar to \citet{c10}. When $\lambda=1$, all interaction terms in (\ref{b.e18}) are annihilated, which means the coupling of the cluster with its neighbors are totally decoupled. If $\lambda=0$, then no clustering process, the algorithm is reduced to single site algorithm.

The cluster algorithm can be better explained using the diagram in Figure \ref{b.pic2}. Figure \ref{b.pic2} (a-c) demonstrate the clustering process. First we randomly select a seed node, in this diagram, node 8. Then we throw the bond to all neighbors of node 8, and find node $5$ is bonded to $8$ with probability $p_{a,5}$ and forms the cluster (the dashed line is turned into solid lines, meaning 5 is added to the cluster). We scan the remaining nodes again but whether or not a new node should be added is determined by the bonding between the new node and node 5 and 8. For example in Figure \ref{b.pic2} (b), the bond between the new node 4 and the cluster is the overall bonds 4-5 and 4-8. In Figure \ref{b.pic2} (c), after add the last new node 1 into the cluster, we scan all the left nodes and find no new node added to the cluster, then the clustering process stops.

The flipping of the cluster is demonstrated in Figure \ref{b.pic2} (d-e). The cluster formed contains nodes $c=\{8,5,4,2,1\}$. To flip these nodes, we have to cut off the bonding of the cluster with all other nodes in $\bar c$ because in a complete graph the neighbors of a cluster is all other nodes outside of the cluster. For example, the bond between the cluster and node $10$ is demonstrated in Figure \ref{b.pic2} (d), where we can see the bonds between 10 and all nodes in the cluster should be cut off to flip the cluster. Thus to completely flip the cluster, the bonds between all other nodes in $\bar c$ and the nodes in $c$ should be cut off. Similarly, in the reverse process to flip the cluster back, as shown in Figure \ref{b.pic2} (e), all the bonds between the cluster $c$ and the nodes in $\bar c$ must be cut off.

It is easy to show that our algorithm is more general in sense that it is applicable to the complete graph with random interaction. When applied to Ising model on grid with positive fixed interaction $J$ (only interactions among nearest neighbors account), our algorithm evaluate to the original Wolff algorithm: the cluster growth by throwing bonds to nearest neighbors with probability $1-\exp(-J)$.

Following theorem shows the algorithm stated above satisfies the detailed balance and ergodicity.

\noindent{\bf Theorem 1:} {\it With the probability of adding node to the cluster, $p_{a,j}$, and the probability of moving from current configuration $\bga^0_c$ to the flipped configuration $\bga^*_c$, $\alpha(\bga_c^0\rightarrow\bga_c^*)$, as defined as in the generalized Wolff algorithm, the algorithm is detailed balanced and ergodic.}

{\it Proof}: See \ref{app1}.

In this paper, we mainly focus on the noninformative prior for $\bga$. However, since the distribution of $\bga$ given $\bet$ and $\phi$ follows the Boltzman distribution, it is nature to assign a Boltzman prior or Ising prior for $\bga$ if such priori information is available. For example, in some genetic data, the genes form a network that can be descried using special graph model, with this information we can assign a Ising prior with specific interaction matrix that represents the priori graph structure. We will discuss this issue in Section \ref{b.sec6}. Another advantage of the cluster algorithm is it reveals the latent graph structure according to the frequencies of nodes that form a cluster, and this information may help us to distinguish the signals and the noise since the signals and noise should have high frequency to be anti-aligned.

\section{Understanding the Mechanism of Bayesian Variable Selection}\label{b.sec5}
The purpose of this section is to understand how the marginal probability $p(\gamma_j|\mby)$ evaluates under different choice of marginal prior of $\bet$ given the only tuning parameter $b$. Although our Ising model is based on the KM model (\ref{b.e2}), the results of this section is also applicable to SSVS model with the point mass mixture prior for $\bet$. This is because if we integrate out $\bet$, both SSVS and KM models are identical. Note that all the results in this section is based on parametric linear model (\ref{b.e2}) where $\beta_j$ is scalar, but the major results are similar to nonparametric linear model where $\bet_j$ is $M_j\times1$ vector.

Some notations are introduced here. Since $\mbx_j$'s are standarzed, $C=X^TX=[\mathbf c_1,..., \mathbf c_p]$ is the correlation matrix of $\mbx_j$'s and $\mathbf c_j$'s stands for the vector of the correlation between $\mbx_j$ with all predictors. For orthogonal data set, $C=I_n$ or $\mbx^T_i\mbx_j=\delta_{ij}$ and $\mbx^T_j\bel=0, i,j=1,...,p$. The projection of $\mby$ on $\mbx_j$ can be expressed as $\mathbf a=X^T\mby=(\mbx^T_1\mby,..., \mbx^T_p\mby)^T=(a_1,..., a_p)^T$, which are the estimation of the signal $\beta_j$'s under orthogonal design. We may also need notation $\mathbf c_{\bga_{\bar j}}=\bga_{\bar j}\circ\{\mathbf c_j\}_{\bar j}$, where ``$\circ$'' stands for the pointwise product of two vectors, ``$\bar j$'' stands for the $j$th element removed for corresponding vectors and matrices,

\subsection{General Profile of the Marginal Selection Probability}\label{b.sec5.1}
With the hierarchical model defined (\ref{b.e2}) and (\ref{b.e3}) in Section \ref{b.sec2}, the posterior distribution of $\bet$ is multivariate normal given $\bga$ with mean $\bmu$ and variance $\Sigma$ as.
\[\label{b.e19}
\bmu=\phi\Sigma X^T_{\bga}\mby; \Sigma=(\phi X^T_{\bga}X_{\bga}+D)^{-1},
\]
where $D$ is a diagonal matrix with diagonal element $\left\{{\tau_j}\over{b^2}\right\}_{1\le j\le p}$. To understand how $\tau_j$ and $b$ introduce the shrinkage effect, similar to \citet{c3,c25}, it is convenient to introduce the shrinkage coefficient, $\kappa_j=\left({{\tau_j}\over{b^2\phi}}\right)/\left(1+{{\tau_j}\over{b^2\phi}}\right)$. Under the orthogonal design, the posterior mean and variance of $\beta_j$'s corresponding to $j\in\{j: \gamma_j=1\}$ are
\be\begin{split}\label{b.e20}
E(\beta_j|\mby, \phi, b)=[1-E(\kappa_j|\mby, \phi, b)]a_j,\\
Var(\beta_j|\mby, \phi, b)=[1-E(\kappa_j|\mby, \phi, b)]\phi^{-1}.
\end{split}\ee
The coefficient $\kappa_j$'s represent how much shrinkage being placed on the initial estimation of $\beta_j$'s. $\kappa_j\rightarrow0$, yields no shrinkage, and $\kappa_j\rightarrow1$ yields near-total shrinkage. With this definition of $\kappa_j$, it is easy to derive the density function of $\kappa_j$, $p(\kappa_j)$.  Table \ref{t1} lists $p(\kappa_j)$'s based on the three prior settings given $\phi=1$.

In order to compare the performance of different variance mixture priors $p(\tau_j)$ on the marginal selection probability and avoid notation abuses, it is convenient to assume $\phi$ fixed and use a scaleless transformation $\sqrt{\phi}\mby\rightarrow\mby$ such that $a_j\sqrt{\phi}\rightarrow a_j$, $b\sqrt{\phi}\rightarrow b$ and $\sqrt{\phi}C\rightarrow C$. This is equivalent to assume $\phi=1$, but keep in mind that $a_j$'s, $\mathbf c_j$'s and $b$ are scaled by $\sqrt{\phi}$ unless stated otherwise.

\begin{figure}[bth]
\begin{center}
\includegraphics[height=56mm]{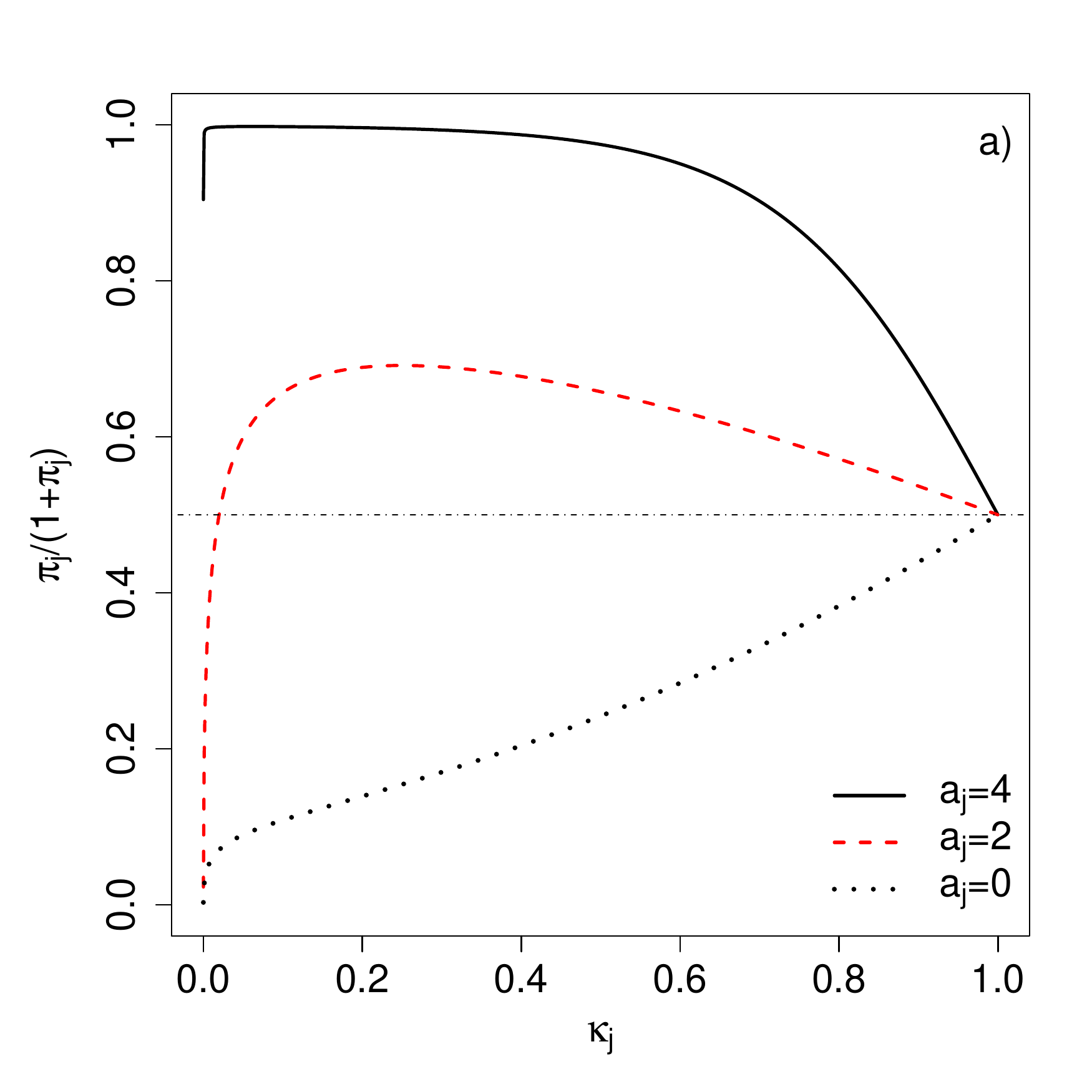}\includegraphics[height=56mm]{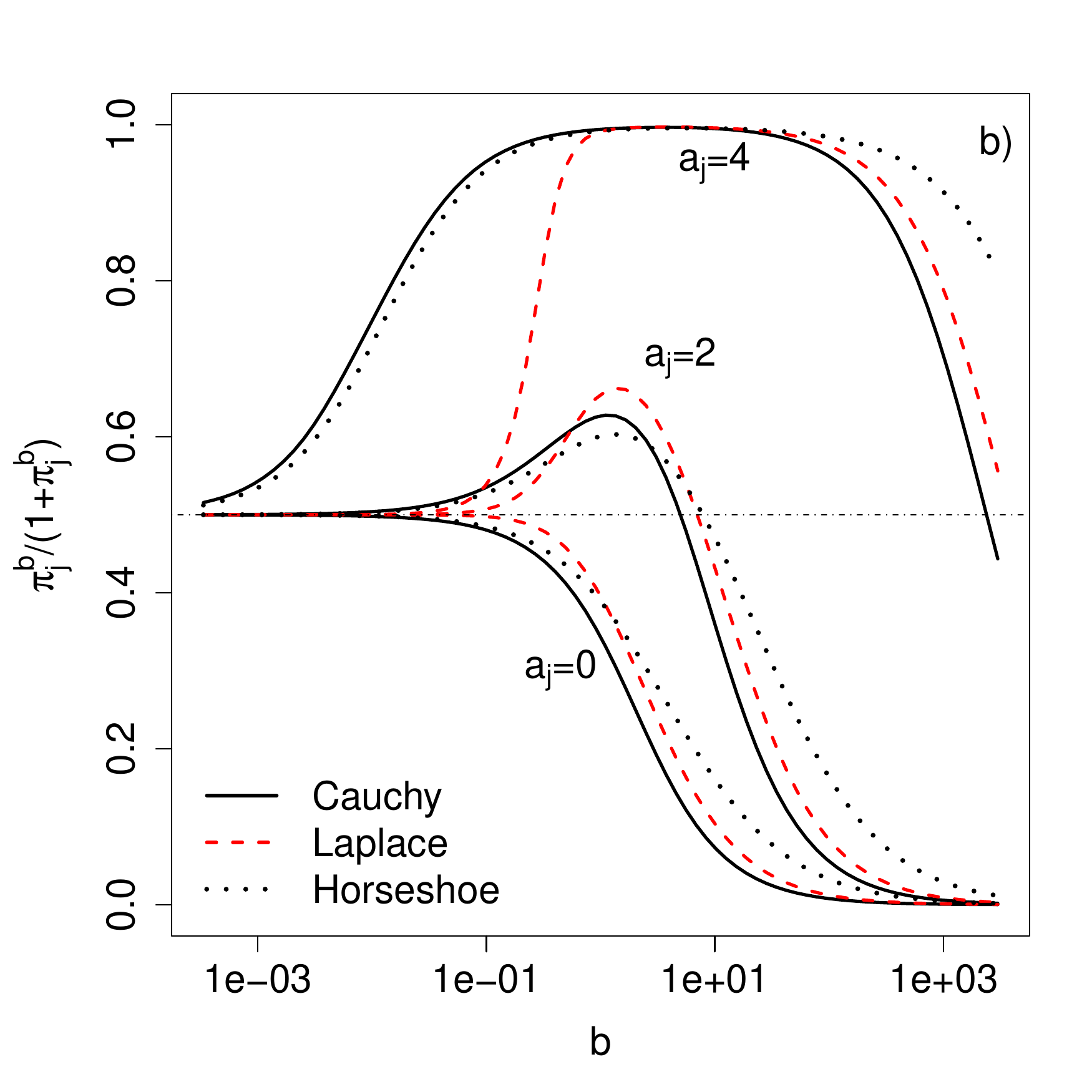}\includegraphics[height=56mm]{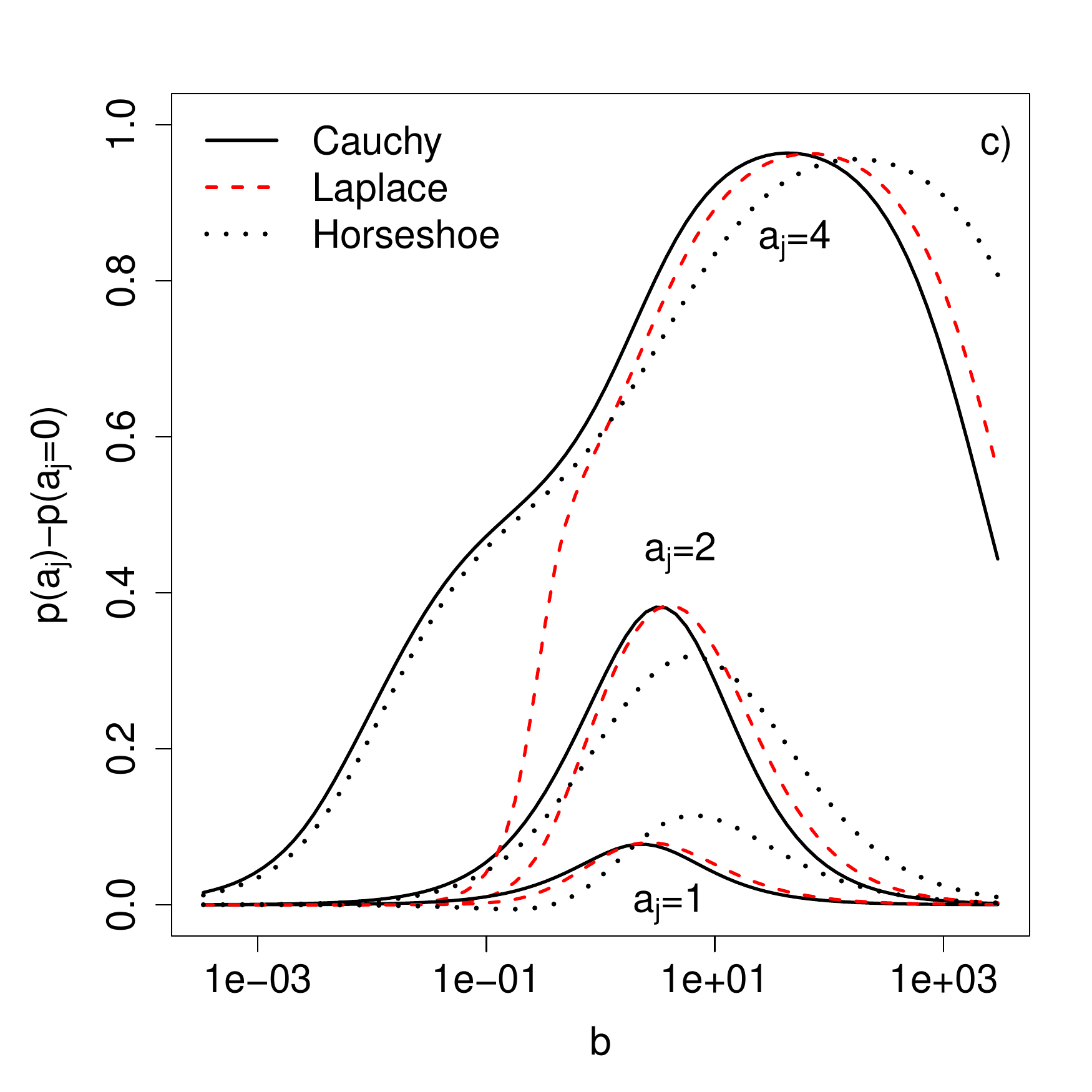}
\caption{The curves of selection probability against $\kappa_j$ (a). The curves of marginal selection probability against global shrinkage parameter $b$ (b). Marginal selection probabilities with baseline subtracted (c). All plots are under orthogonal designs.}\label{b.pic3}
\end{center}
\end{figure}

With these coefficients defined, following theorems connect the marginal odds of $\gamma_j$ given the data with $\kappa_j$'s and $b$. Based on (\ref{b.e2}) and (\ref{b.e3}), the join distribution for $\bga, \bet, \bta$ given $b$ is
\[\label{b.e21}
p(\bga,\bet,\bta|\mby, b)\propto p(\mby|\bga,\bet)p(\bet|\bta,b)p(\bta),
\]
and the marginal probability for $\gamma_j$ is $p(\gamma_j|\mby,b)=\int\sum_{\bga_{\bar j}}p(\gamma_j, \bga_{\bar j},\bet,\bta|\mby, b)d\bet d\bta$. Thus the marginal odds for $\gamma_j=1$ given $b$ is
\[\label{b.e22}
\pi^b_j={{\int\sum_{\bga_{\bar j}}p(\gamma_j=1, \bga_{\bar j},\bet,\bta|\mby, b)d\bet d\bta}\over{\int\sum_{\bga_{\bar j}}p(\gamma_j=0, \bga_{\bar j},\bet,\bta|\mby, b)d\bet d\bta}}.
\]

{\it Theorem 2: For the Bayesian model defined in (\ref{b.e2}) and (\ref{b.e3}), the marginal odds of $\gamma_j$, defined as $\pi^b_j$, has following form
\be\label{b.e23}
\pi^b_j=\int\pi_j\xi_jp(\kappa_j)d\kappa_j,
\ee
where $p(\kappa_j)$ is the density function of $\kappa_j$,
\be\label{b.e24}
\pi_j=\kappa_j^{1\over2}\exp\left[{{a^2_j}\over2}(1-\kappa_j)\right],
\ee
and $\xi_j$ is a positive real function of $\kappa_j$
\be\label{b.e25}
\xi_j={{\int\sum_{\bga_{\bar j}}\xi(\gamma_j=1,\kappa_j,\bga_{\bar j},\bta_{\bar j})p(\bta_{\bar j})d\bta_{\bar j}}\over{\int\sum_{\bga_{\bar j}}\xi(\gamma_j=0,\kappa_j,\bga_{\bar j},\bta_{\bar j})p(\bta_{\bar j})d\bta_{\bar j}}}.
\ee
\begin{enumerate}
\item For general cases
\[\label{b.e26}
\xi(\gamma_j,\kappa_j,\bga_{\bar j},\bta_{\bar j})=\exp\left[{1\over2}\left(\mathbf a_{\bar j}-(1-\kappa_j^{\gamma_j})a_j\mathbf c_{\bga_{\bar j}}\right)^T\Omega_j^{-1}\left(\mathbf a_{\bar j}-(1-\kappa_j^{\gamma_j})a_j\mathbf c_{\bga_{\bar j}}\right)\right]|\Omega_j|^{1/2}|D_{\bar j}|^{1/2}
\]
with $\Omega_j=[D_{\bar j}+X^T_{\bga_{\bar j}}X_{\bga_{\bar j}}-(1-\kappa_j)^{\gamma_j}(\mathbf c_{\bga_{\bar j}}\mathbf c^T_{\bga_{\bar j}})]^{-1}$.

For orthogonal designs, $\xi_j=1$, and
\be\label{b.e27}
\pi^b_j=\int\pi_jp(\kappa_j)d\kappa_j.
\ee
\item  For orthogonal designs, if $\kappa_j\rightarrow 0$, then $\pi_j\rightarrow0$, and if $\kappa_j\rightarrow1$, then $\pi_j\rightarrow1$. Similarly, if $b\rightarrow 0$, then $\pi^b_j\rightarrow1$, and if $b\rightarrow\infty$, then $\pi^b_j\rightarrow0$.
\end{enumerate}

}
{\it Proof:} See \ref{app2}.

From Theorem 2 we can see that in general the marginal odds $\pi^b_j\ne\int\pi_jp(\kappa_j)d\kappa_j$, the marginal odds of the orthogonal design. According to equation (\ref{b.e23}), when the correlation among predictors are not negligible, the odds will be ``blurred'' by the coefficient $\xi_j$, and the marginal selection probability is blurred too. Basically, $\xi_j$ is a complex function of $a_j, \mathbf c_{\bar j}$, and $\tau_k/b^2, k\ne j$ or $\kappa_j$. Furthermore, it is infeasible to calculate $\xi_j$ given large $p$ with more than 2 predictors are correlated. However, we can focus on the orthogonal design to understand the mechanism of marginal selection probability in general since it is much more easier to calculate.

Combining Theorem 2 and Figure \ref{b.pic3}, we can understand the behaviors of $\pi_j$ and $\pi^b_j$ better. Figure \ref{b.pic3} (a) plots the selection probability as a function of $\kappa_j$ according to odds $\pi_j$ (\ref{b.e24}), and Figure \ref{b.pic3} (b) plots the marginal selection probabilities according to $\pi^b_j$ with different prior $p(\kappa_j)$'s. We can see that for the orthogonal design, expression (\ref{b.e24}) and (\ref{b.e27}) indicate that $\pi_j$ and $\pi^b_j$ are monotone functions of $a_j$, this is demonstrated as the different selection probability curves in Figure \ref{b.pic3} too.  In ideal case, all noise predictors will demonstrate the same selection probability since $a_j=0$, which defines the baseline selection probability curve in Figure \ref{b.pic3} (a-b).

Thus ideally, any signals with $a_j\ne0$ are deviated from the baseline curve. However, when correlations among variables do not equal to zero, the situations become complicated. First, even though the correlation among variables are small enough so $\pi_j$ and $\pi^b_j$ are still monotone function of $a_j$, the baseline will be blurred and extended to a band. To see this, consider $k\in\bar{V}^*$ and $j\in V^*$ where $V^*$ is the set of true nodes and its complement is $\bar{V}^*$. Because $c_{kj}=\mbx^T_k\mbx_j\ne0$, $\mbx_k$ will have fake signal: $a_k=\mbx^T_k\mby=c_{kj}a_j$. Thus all the noise predictors will demonstrate false signals as long as they have nonzero correlations with the true signals. This makes separating the true variable with small signals from the noise difficult. Secondly, because of $\xi_j$, even for large signals the selection probability will be distorted by their correlated fake signals. For example in Figure \ref{b.pic3} (b), if $a_j=2$ is the fake signal and $a_j=4$ is the true signal and they are correlated, then the profile curve of $a_j=2$  and $a_j=4$ will show some ``interacting'' behavior at $b\approx1$ where the selection probability of fake signal reaches the maximum (we will show this behavior in the simulation analysis). Thus in general the largest gap that separates $a_j=2$ and $a_j=4$ is not around $b\approx1$, but in two regions around $b\approx0.1$ and $b\approx1000$.

Furthermore, the second result of Theorem 2 states some asymptotic behaviors of $\pi_j$ and $\pi^b_j$ as $\kappa_j\rightarrow0$ or $b\rightarrow\infty$ and $\kappa_j\rightarrow1$ or $b\rightarrow0$. This can be clearly seen in Figure \ref{b.pic3} (a-b) where with small shrinkage ($\kappa_j\rightarrow0$ or $b\rightarrow\infty$), both $\pi_j$ and $\pi^b_j$ approach 0, and with large shrinkage ($\kappa_j\rightarrow1$ or $b\rightarrow0$), they approach 0.5.  However, the dropping rate depends on the magnitude of the signal and the prior $p(\kappa_j)$. For example in Figure \ref{b.pic3} (a) we can see for large signal $a_j=4$, the selection probability maintains at 1 for $\kappa_j\rightarrow0$ till the last point. In Figure \ref{b.pic3} (b), furthermore, we can see the selection probability curves are different for different priors: some drop very fast, such as Laplace prior, some are pretty robust to shrinkage such as horseshoe prior.

So choosing an appropriate prior for $\tau_j$ or $p(\kappa_j)$ is important. Our next question will be how to choose an appropriate prior. Based on Figure \ref{b.pic3} and Theorem 2, there are some guidelines to choose the prior $p(\tau_j)$: (1) The rate of $\pi^b_j$ to increase must be fast when the signal increases, so that the large true signal can be separated from the noise more easily. (2) $\pi^b_j$ drops to 0.5 or 0 slowly when $b\rightarrow0$ or $b\rightarrow\infty$ so we have a wider windows of $b$ where the true signals maintain high selection probability.

Following theorems will further help us to understand the relationship between $\pi^b_j$ and the shrinkage coefficient $\kappa_j$.

{\it Theorem 3
For the Bayesian model (\ref{b.e2}) and (\ref{b.e3}) with orthogonal design, suppose prior $p(\beta_j)$ is a zero mean scale mixture of normals: $[\beta_j|\tau_j,b]\sim N(0,b^2\tau_j^{-1})$, with $\tau_j$ having proper prior $p(\tau_j)$. Define the marginal density $m_j=m(\mby|\gamma_j=1)$ as
\[\label{b.e28}
m_j=\int p(\mby|\beta_j,\gamma_j=1)p(\beta_j|\tau_j)p(\tau_j)d\beta_jd\tau_j.
\]
If $m_j$ is finite for all $\mby$, then
\begin{enumerate}
\item \be\label{b.e29}
E(\beta_j|\mby,\gamma_j=1)=a_j+{1\over{m_j}}\mbx^T_j{\partial\over{\partial\mby}}m_j=a_j+\mbx^T_j{\partial\over{\partial\mby}}\log m_j.
\ee
\item \be\label{b.e30}
E(\beta_j|\mby,\gamma_j=1)={d\over{da_j}}\log\pi^b_j=[1-E(\kappa_j|\mby,\gamma_j=1)]a_j.
\ee
\end{enumerate}
 }
{\it Proof:} See \ref{app3}

The first result of Theorem 3 is well known in Bayesian literature and can be found in \citet{c24} and \citet{c27} for more discussion, here we just simply extend it to the linear model case. We are more interested in the second result which gives the relationship among the expectation of $\beta_j$, the derivative of log $\pi^b_j$ respect to $a_j$, and the shrinkage coefficient. Since we prefer a larger derivative of log marginal odds such that the large signal can be separate from the baseline further. (\ref{b.e30}) indicates that to achieve this purpose, it not only requires a large $a_j$, but also requires the expectation of shrinkage parameter $\kappa_j$ to be small. This is confirmed by Figure \ref{b.pic3} (a), where we see that the largest separation between the signals and the baseline is on the side of $\kappa_j\rightarrow0$. Thus if integrate out $\kappa_j$ to have $\pi^b_j$, we want the density $p(\kappa_j)$ has substantial mass around the region with largest separation. However, we don't want $\kappa_j=0$ since it means exactly no shrinkage and all $\pi_j$'s drop to zero at this point.

Therefore, the general requirement for a $p(\kappa_j)$ based on Theorem 2, 3 and Figure \ref{b.pic3} is to maintain substantial mass around region on the small shrinkage. Surprisedly, (\ref{b.e30}) seems contradict to the usual variable selection strategy that to recover the sparsity in the region of large shrinkage. In fact, it is possible to separate the signal and noise in large shrinkage region and large shrinkage does have some advantages, such as stability, faster mixing, less sensitive to nodes number $p$ and so on. So it is a second choice as long as the signals are robust to large shrinkage, at least for large signals. However, in this paper we focus on the small shrinkage region where the consistency in variable selection seems satisfied more often.

\subsection{Dynamic Properties of the Odds with Different Priors}\label{b.sec5.2}
To explain the different behaviors of the selection probability caused by different priors, we need examine more details about the density distribution of $p(\kappa_j)$. \citet{c3} and \citet{c27} discussed the performance of different types of priors in the Bayesian regularization with difference weight on shrinkage. They focus on the effects on the estimation of the signals. We are looking at those priors from a different point of view in terms of variable selection based on the selection probabilities.

\begin{figure}[bth]
\begin{center}
\includegraphics[height=45mm]{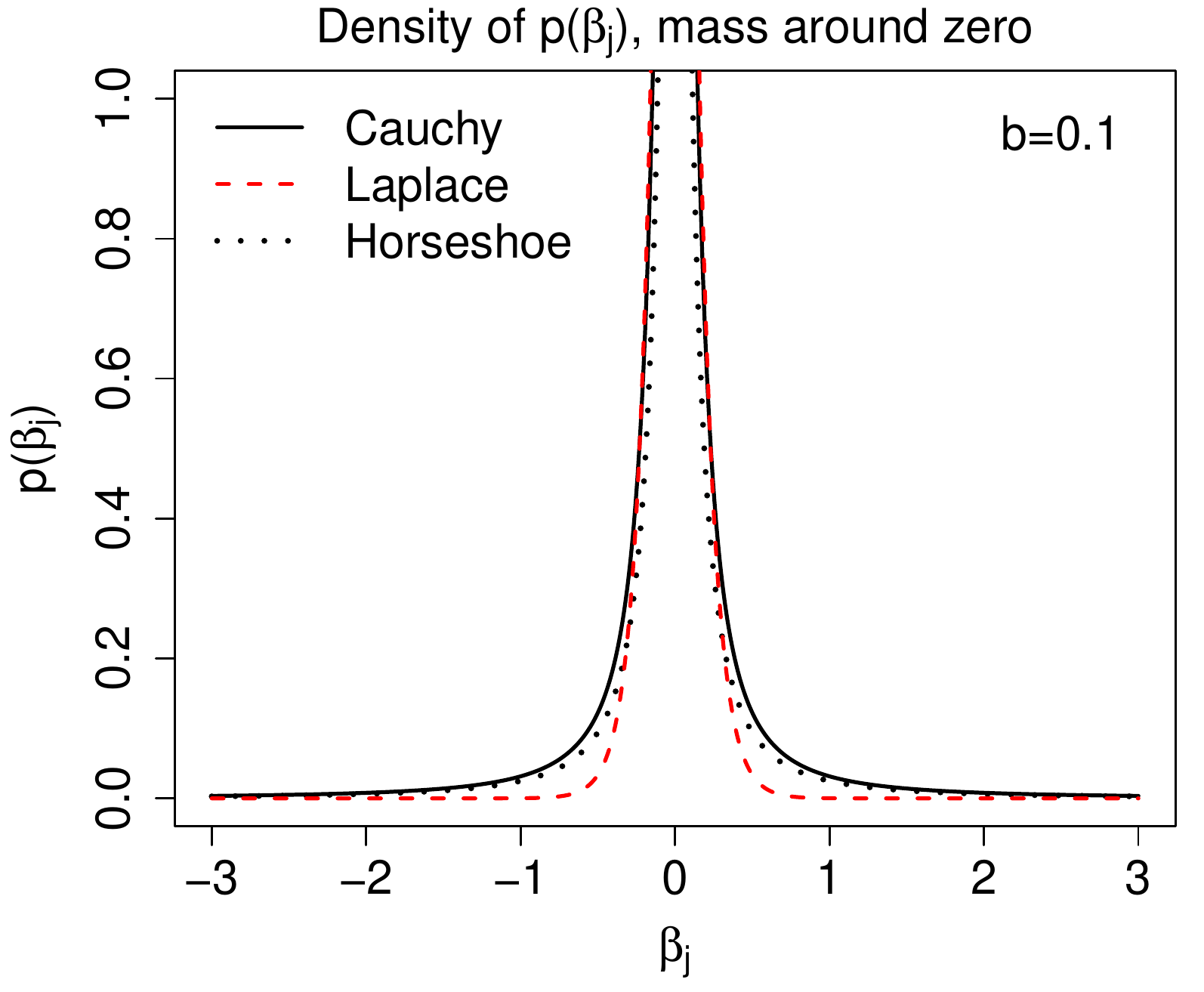}\includegraphics[height=45mm]{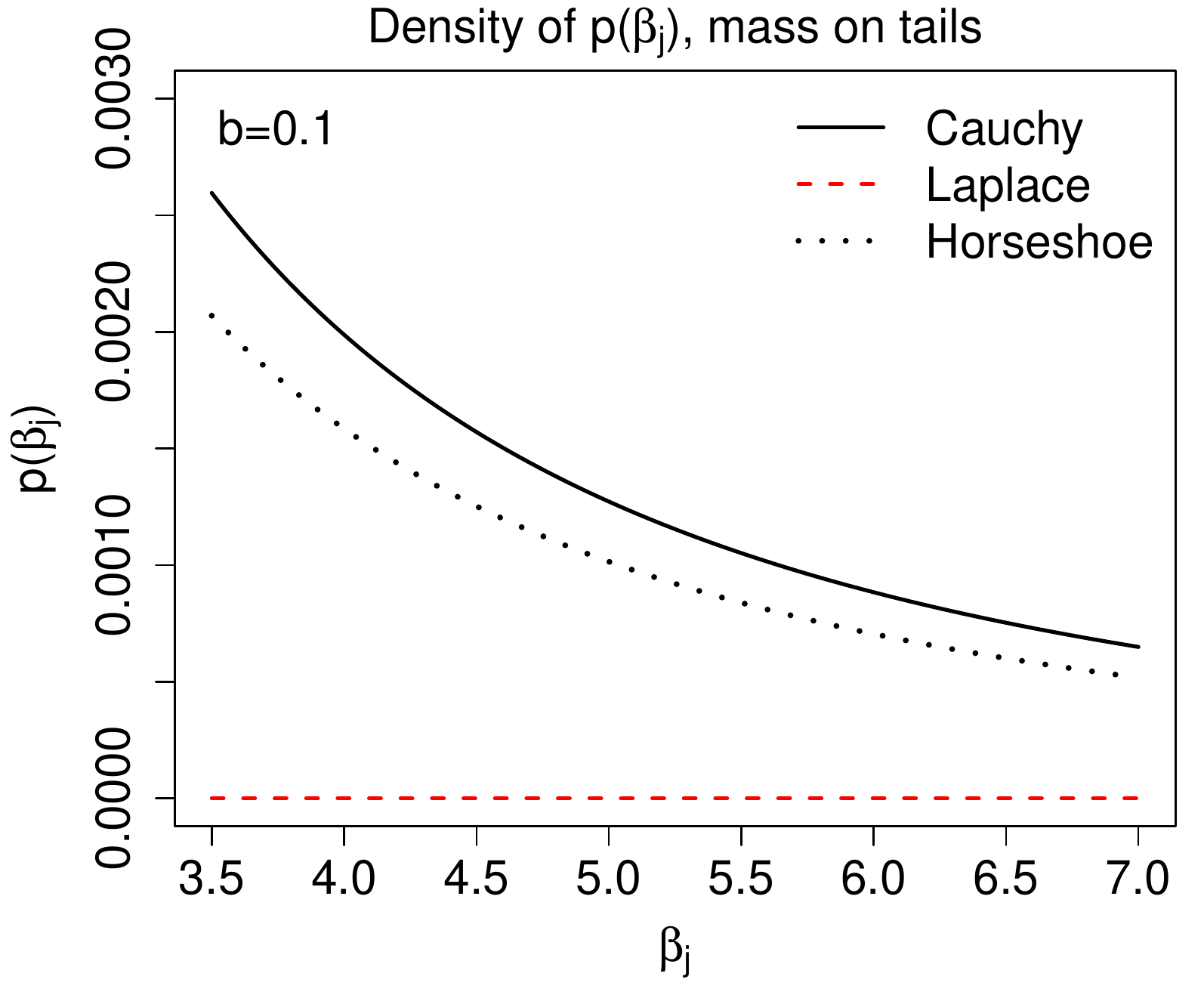}\includegraphics[height=45mm]{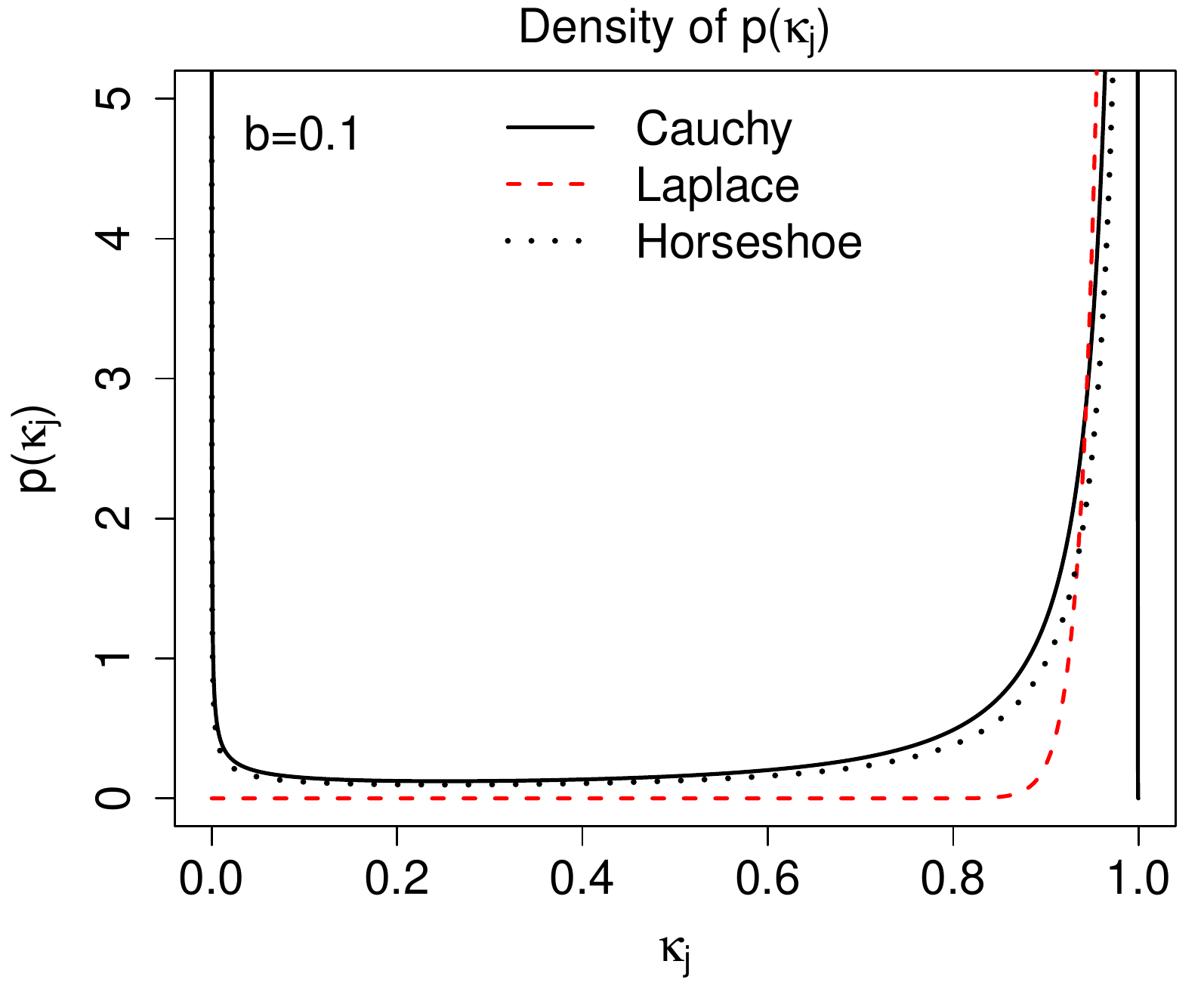}
\includegraphics[height=45mm]{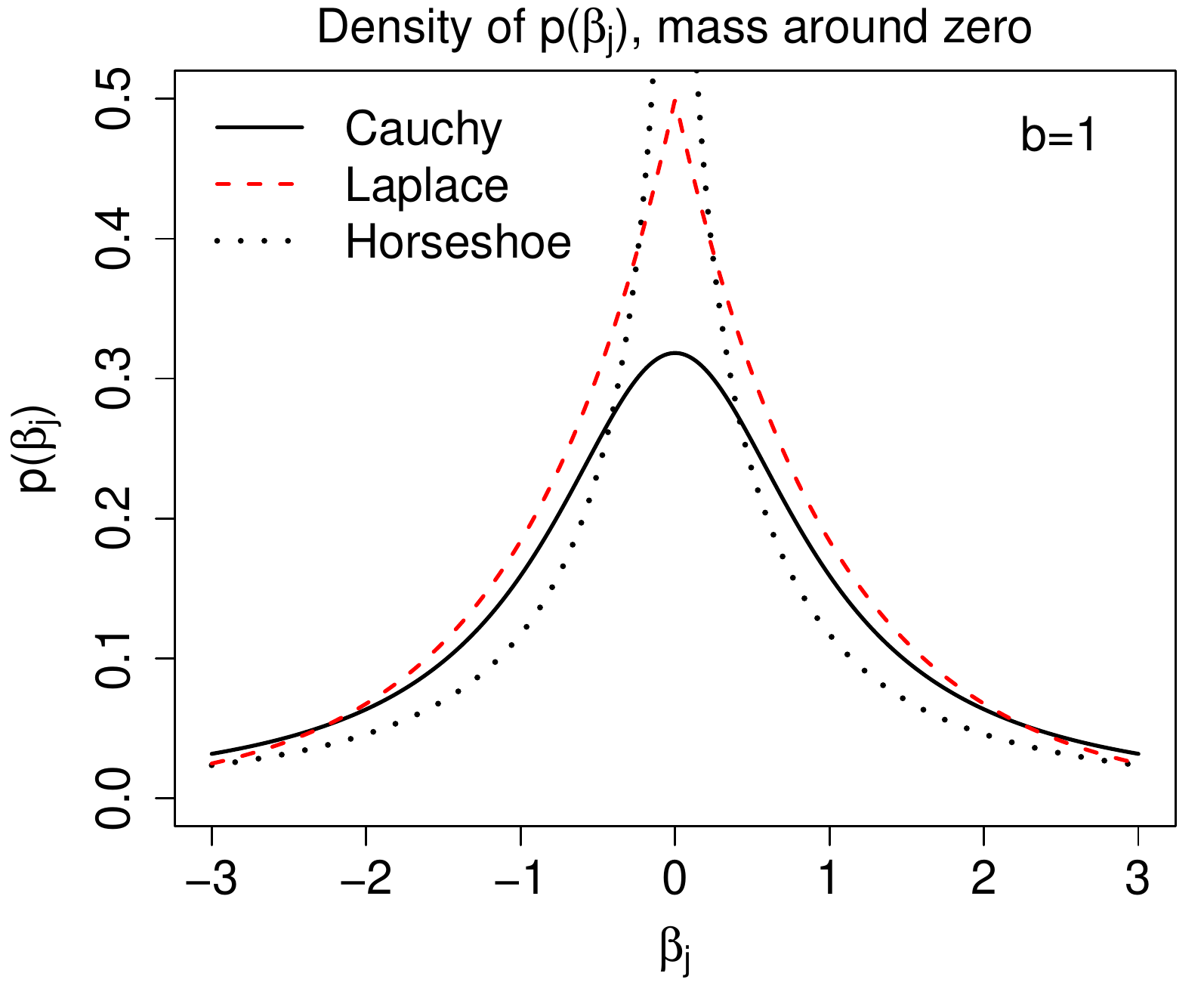}\includegraphics[height=45mm]{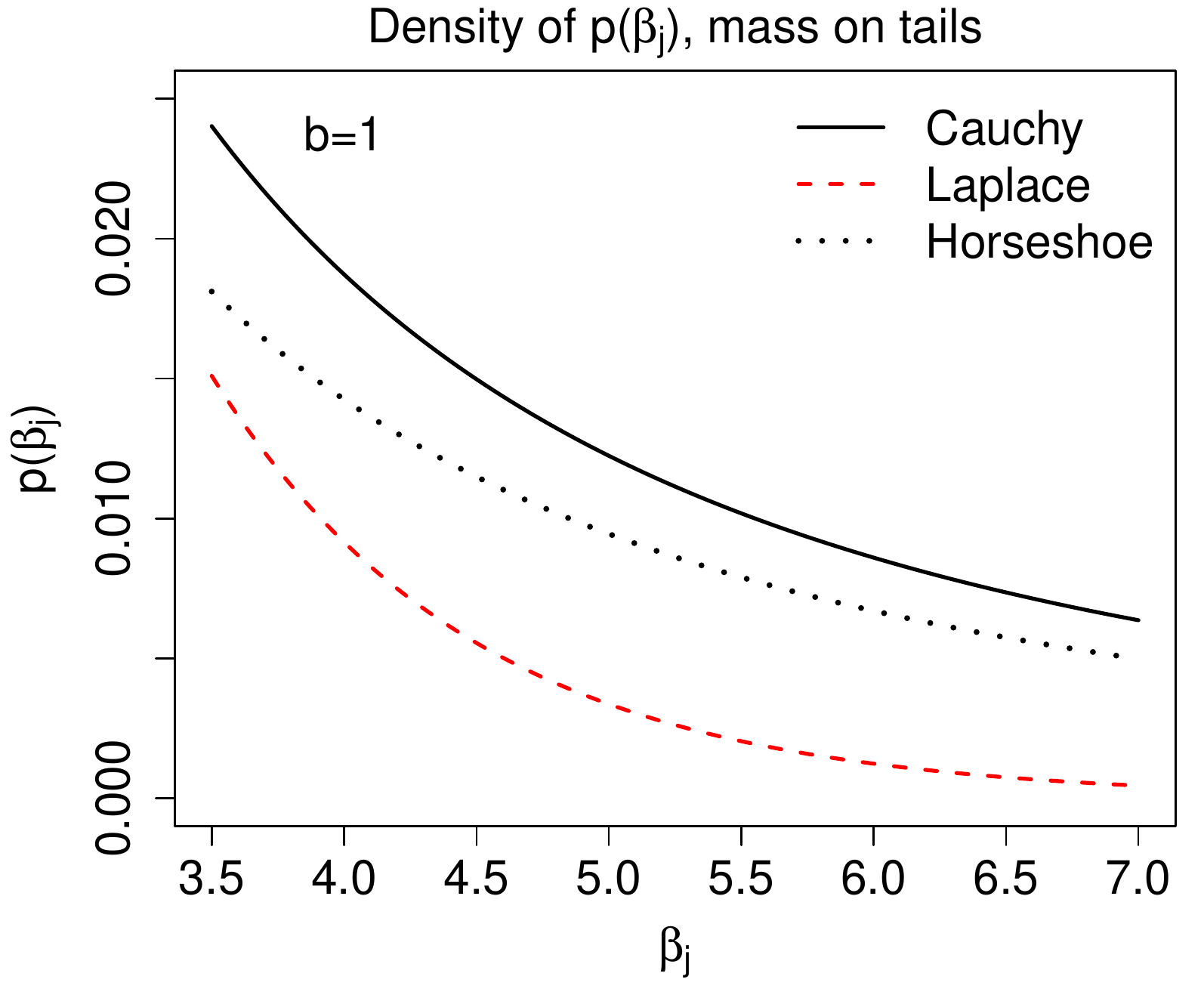}\includegraphics[height=45mm]{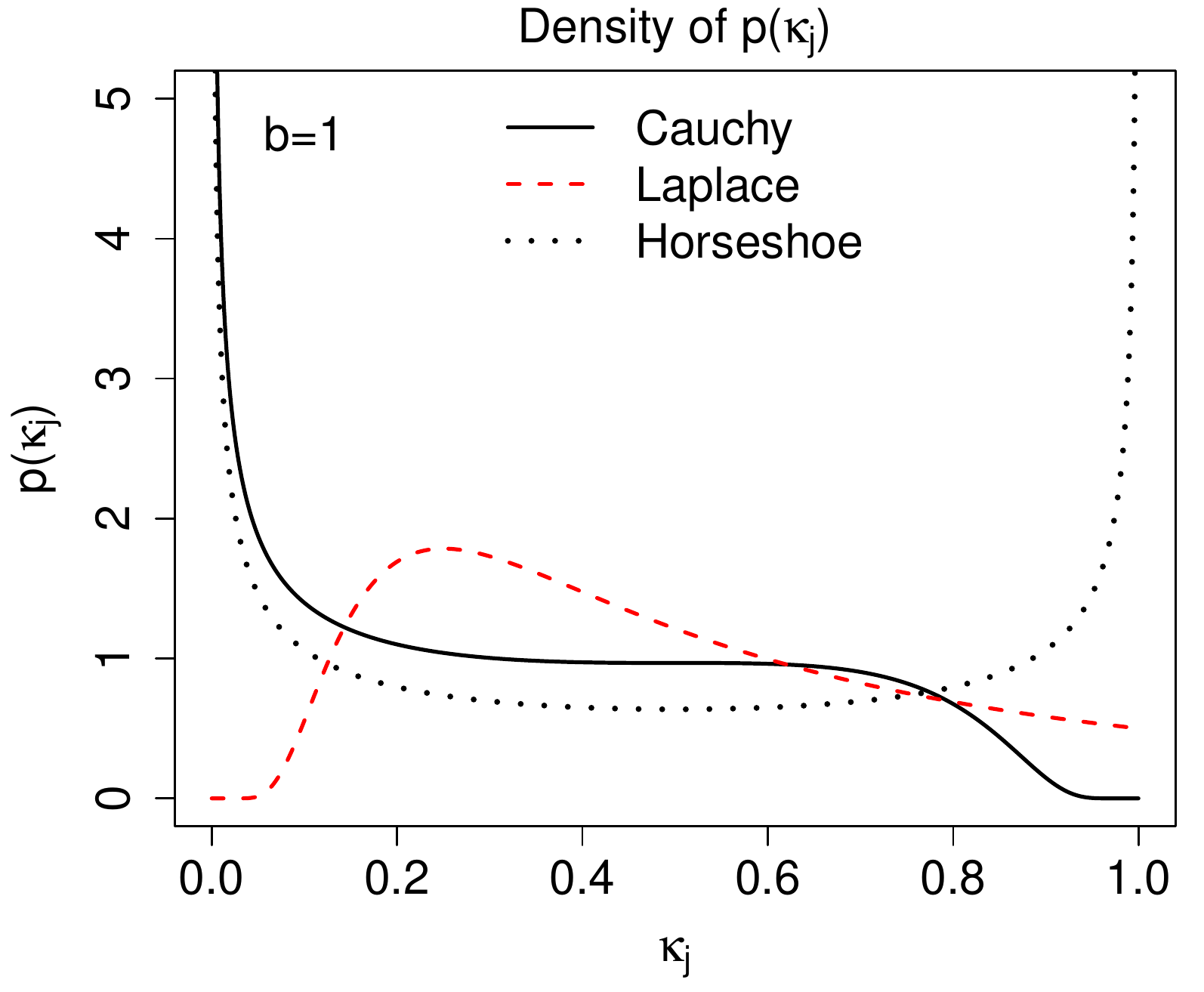}
\includegraphics[height=45mm]{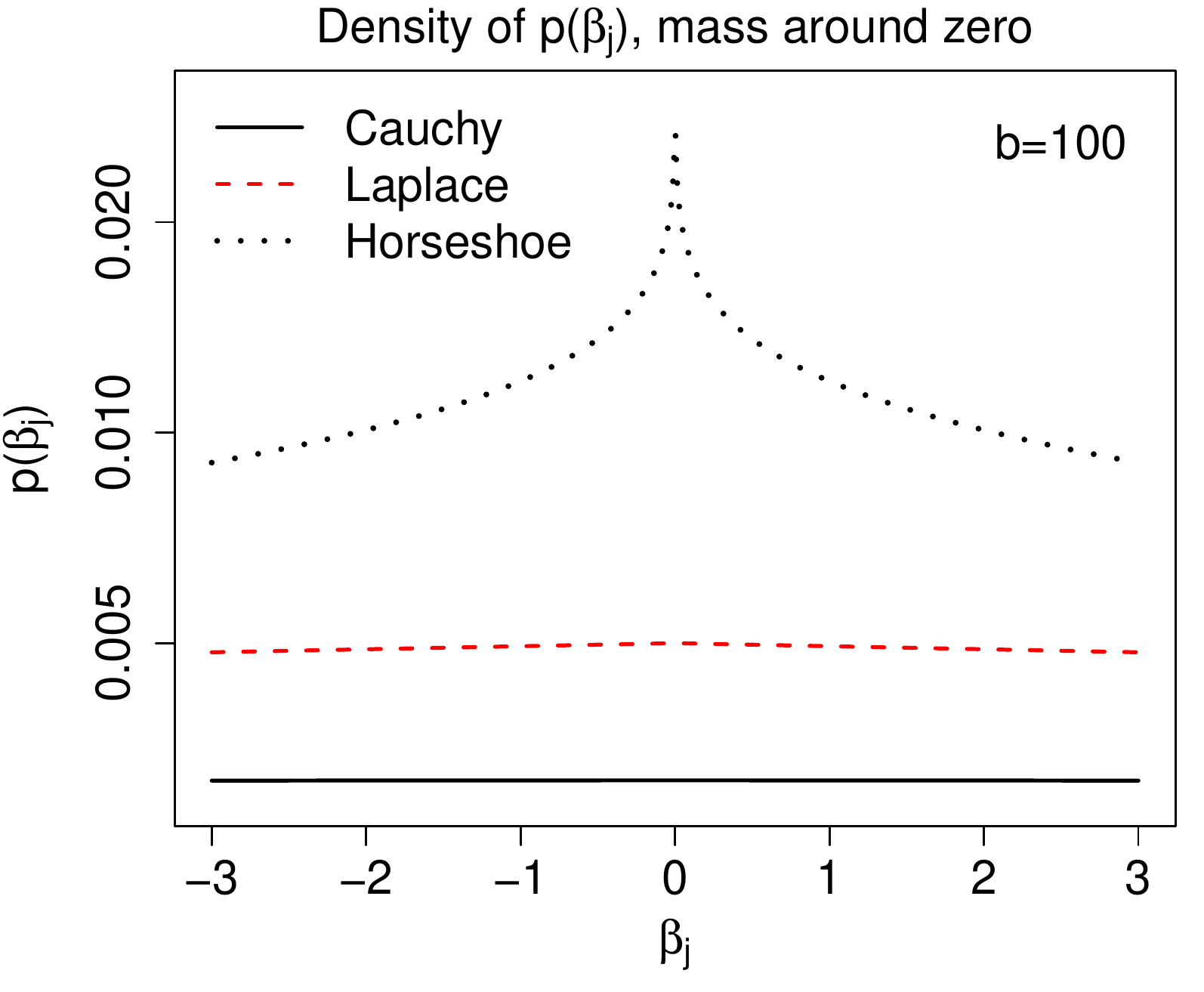}\includegraphics[height=45mm]{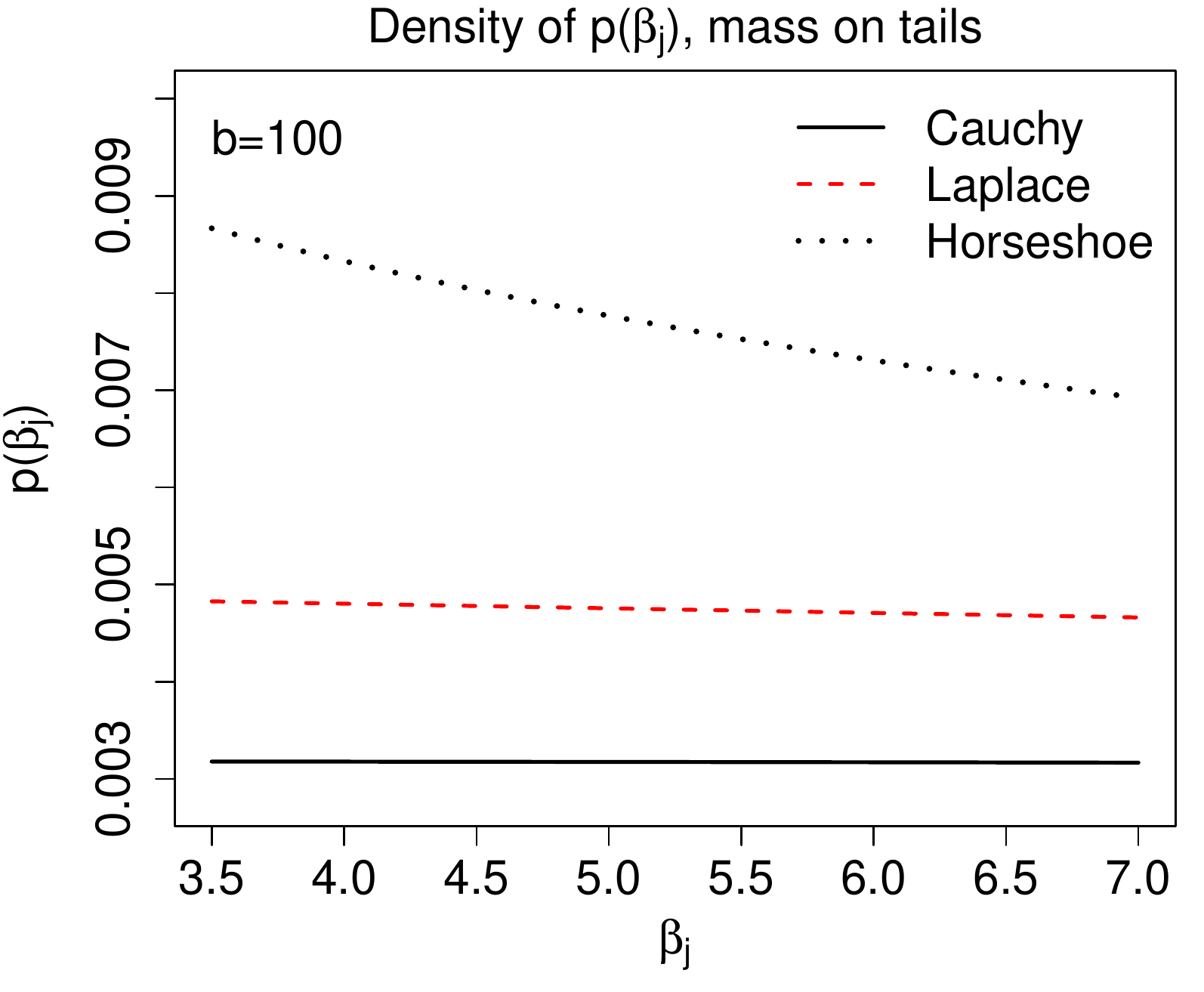}\includegraphics[height=45mm]{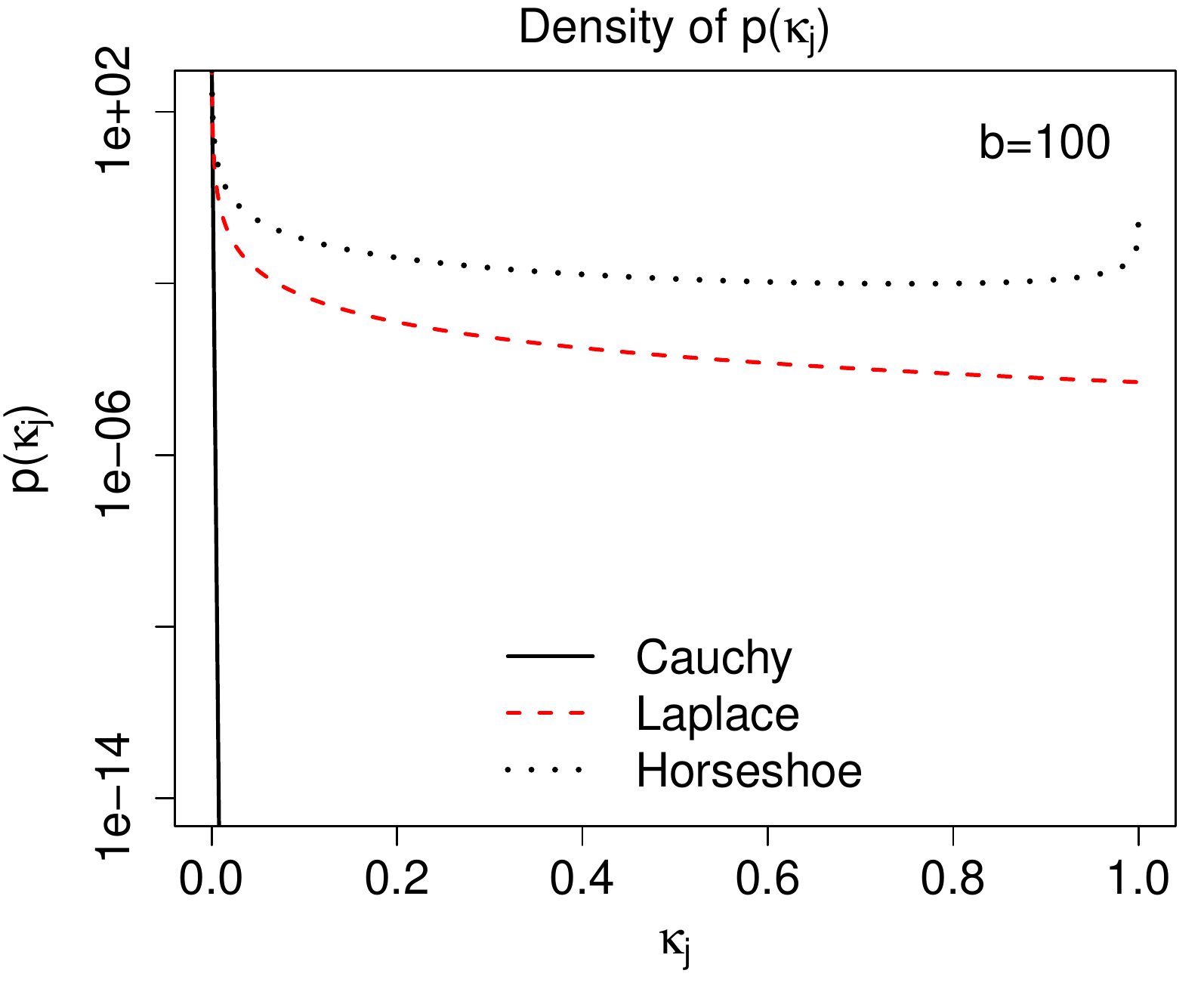}
\caption{Marginal prior density functions of $\beta_j$ and density functions of $\kappa_j$ for different $b$. }\label{b.pic4}
\end{center}
\end{figure}

In the prior for $\beta_j$'s, $b$ is the global parameter. As $b\rightarrow0$, large global shrinkage is applied on all $\beta_j$'s, and as $b\rightarrow\infty$, the global shrinkage effect will be negligible. Table \ref{t1} lists the prior $p(\tau_j)$'s, and corresponding $p(\kappa_j)$'s as well as the marginal prior $p(\beta_j|b)$s. Because of the existing of $b$, how much weight is put on the shrinkage is modified, and for different priors this modification is different.

To see this, Figure \ref{b.pic4} compares density function $p(\beta_j|b)$'s around zero point and on the tails, and density function $p(\kappa_j)$'s given different $b$'s. By examining $p(\kappa_j)$'s in Figure \ref{b.pic4} together with Figure \ref{b.pic3} (c) , we can understand how $b$ effects the selection probability profile through putting different weight on shrinkage. Figure \ref{b.pic3} (c) plots the selection probability profile with the baseline subtracted. It can be seen for orthogonal design, the larger the magnitude of the selection probability, the larger the true signal distinguished from the baseline. In small $b$ ($b\le1$) region, the descendant order of the magnitude is Cauchy, horseshoe and Laplace prior for a given signal, which is consistent with the $p(\kappa_j)$ plots in Figure \ref{b.pic4} at $b=0.1$ where the order of density mass on the small $\kappa_j$ region is Cauchy, horseshoe and Laplace prior. In addition, since Cauchy and horseshoe priors put similar mass around small $\kappa_j$ side, their selection probabilities behave almost identically for small $b$ as shown in Figure \ref{b.pic3} (c). On the other hand, this  order changes for $b=100$ where it becomes horseshoe, Laplace and Cauchy prior in Figure \ref{b.pic4}. Again this is consistent with the selection probability order in Figure \ref{b.pic3} (c) for large $b$ ($b\ge100$). The reason that Cauchy prior becomes worse for large $b$ is because all the mass of $p(\kappa_j)$ is absorbed to $\kappa_j=0$ which is not we expect as mentioned before. For a moderate $b$, such as $b=1$, all priors have substantial mass around small $\kappa_j$ side as shown in Figure \ref{b.pic4}, thus all behave similarly. This is confirmed by Figure \ref{b.pic3} (c) where the selection probabilities for different prior seems similar around $b=1$ at least for large signals.

Above analysis also shows that more weight on large shrinkage is not as important as on small shrinkage in terms of distinguishing the signals. Therefore, horseshoe prior is superior to other two, even though $p(\beta_j|b)$ of horseshoe prior does not have long tail as much as Cauchy prior for small $b$. Horseshoe prior does demonstrate that for a wide range of $b$ $p(\kappa_j)$ maintains substantial mass on the small shrinkage side of $\kappa_j$. On the other hand, Laplace prior has almost zero mass around small $\kappa_j$ side when $b$ is small, and Cauchy prior has all mass abosorbed to $\kappa_j=0$ when $b$ is very large, each deteriorates their performance for those $b$ values respectively. Our argument to evaluate the priors is thus different from \citet{c3} where they argue that the horseshoe prior is superior because it has substantial mass on both small shrinkage and large shrinkage in terms of estimation. Of course, although we prefer small shrinkage in terms of variable selection, large shrinkage does have advantages that some times we must consider. For example, we found in the simulation that with large shrinkage the Gibbs sampler can converge faster even with very large $p$.

To further examine the dynamics of the selection probability profile, following theorem gives some asymptotic behaviors about the derivation of log $\pi^b_j$ respect to $|a_j|$ and $b$, and so it helps us evaluate different priors.

{\it Theorem 4: Consider the inverse of $\tau_j$, $\sigma^2_j=\tau_j^{-1}$, has prior density, $p(\sigma^2_j)$, as $\sigma^2_j\rightarrow\infty$,
 \[\label{b.e31}
 p(\sigma^2_j)\sim {(\sigma^2_j)}^{\alpha-1}e^{-\lambda\sigma^2_j}L(\sigma^2_j)d\sigma^2_j
 \]
for some slowly varying function $L(x)$ such that as $x\rightarrow\infty$ for all $t>0$, $L(tx)/L(x)\rightarrow1$  , then
\begin{enumerate}
\item as $a_j\rightarrow\infty$
\be\label{b.e32}
{d\over{d|a_j|}}\log\pi^b_j\sim\left\{\begin{array}{l  l}
  |a_j|+{{2\alpha-1}\over{|a_j|}}                          &if\;\; \lambda=0  \\
  |a_j|+{{\alpha-1}\over{|a_j|}}-{\sqrt{2\lambda}\over b}  &if\;\; \lambda>0. \end{array} \right.
\ee
\item For large $a_j$, as $b\rightarrow0$
\be\label{b.e33}
{d\over{db}}\log\pi^b_j\sim\left\{\begin{array}{l  l}
  {d\over{db}}\log L^b\left(a_j^2\right)                                  &if\;\; \lambda=0  \\
  {\sqrt{2\lambda}\over b^2}|a_j|+{d\over{db}}\log L^b\left(|a_j|\right)  &if\;\; \lambda>0. \end{array} \right.
\ee
where $L^b(x)$ is a function conditioning on $b$ (see \ref{app4})
\end{enumerate}
}
{\it Proof}: See \ref{app4}

Particulary, a $L^b(x)$ has forms of $b\exp\left(-{b^2\over{2x}}\right), b^{-2}$, and ${b\over{b^2+x^2}}$ for Cauchy, Laplace and horseshoe prior respectively. When $|a_j|$ is large, as Theorem 4 assumes, ${d\over{db}}\log L^b$ is negligible.

Theorem 4 is similar to the tail robustness theorem discussed by \citet{c25} about marginal density $m(\mby|\gamma_j=1)$, which implies that the shrinkage will vanish for any scale mixture normals with $p(\sigma^2_j)$ with heavier tails (such as Cauchy and horseshoe prior), while remain non-diminishing for $p(\sigma^2_j)$ with exponential tails (such as Laplace prior). Combining with (\ref{b.e29}) of Theorem 3, we get the similar conclusion about the estimation of $\beta_j$ that it is robust if estimated by long tail priors. Similar robustness can be found for $\pi^b_j$. The robustness of $\pi^b_j$ means fast change rate of $\pi^b_j$ as signal magnitude increases and small change rate of $\pi^b_j$ as shrinkage increases, which are important since these two characteristics can make distinguishing the signals easier. Large ${d\over{d|a_j|}}\log\pi^b_j$ helps to distinguish the signals from the baseline, while small ${d\over{db}}\log\pi^b_j$ leads to a wide window of $b$ where the selection probability of true signals remain highly.

Expression (\ref{b.e32}) indicates that for priors with exponential tails ($\lambda>0$), the selection probability $\pi_j^b$ increase with a smaller rate when the signal magnitude increases comparing with the heavier tail priors. Meanwhile, expression (\ref{b.e33}) shows that for priors with exponential tails, $\pi_j^b$ drops with much faster rate ($\sim\sqrt{2\lambda}/b^2$) as $b\rightarrow0$. We can also compare the dropping rates of $\log\pi^b_j$ of Cauchy and horseshoe prior as $b\rightarrow0$. Since $L^b(x)=\exp\left(-{b^2\over{2x^2}}\right)b$ and $(1+b^2/x^2)^{-1}b$ for Cauchy and horseshoe prior respectively, it turns out $d\log L^b(x^2)/db\approx1/b$ as $b\rightarrow0$ for both priors. This means the dropping rate as $b\rightarrow0$ is similar for Cauchy prior and horseshoe prior which is confirmed in Figure \ref{b.pic3} (b). Note the second conclusion of Theorem 4 does not apply to $b\rightarrow\infty$ unless $|a_j|\rightarrow\infty$ faster than $b\rightarrow\infty$, i.e., $b/|a_j|\rightarrow0$ so $L^b$ maintains as slowly varying function. However, as shown by the exactly calculation in Figure \ref{b.pic3} (b), horseshoe prior is also the most robust one as $b\rightarrow\infty$.

\begin{figure}[bth]
\begin{center}
\includegraphics[height=80mm]{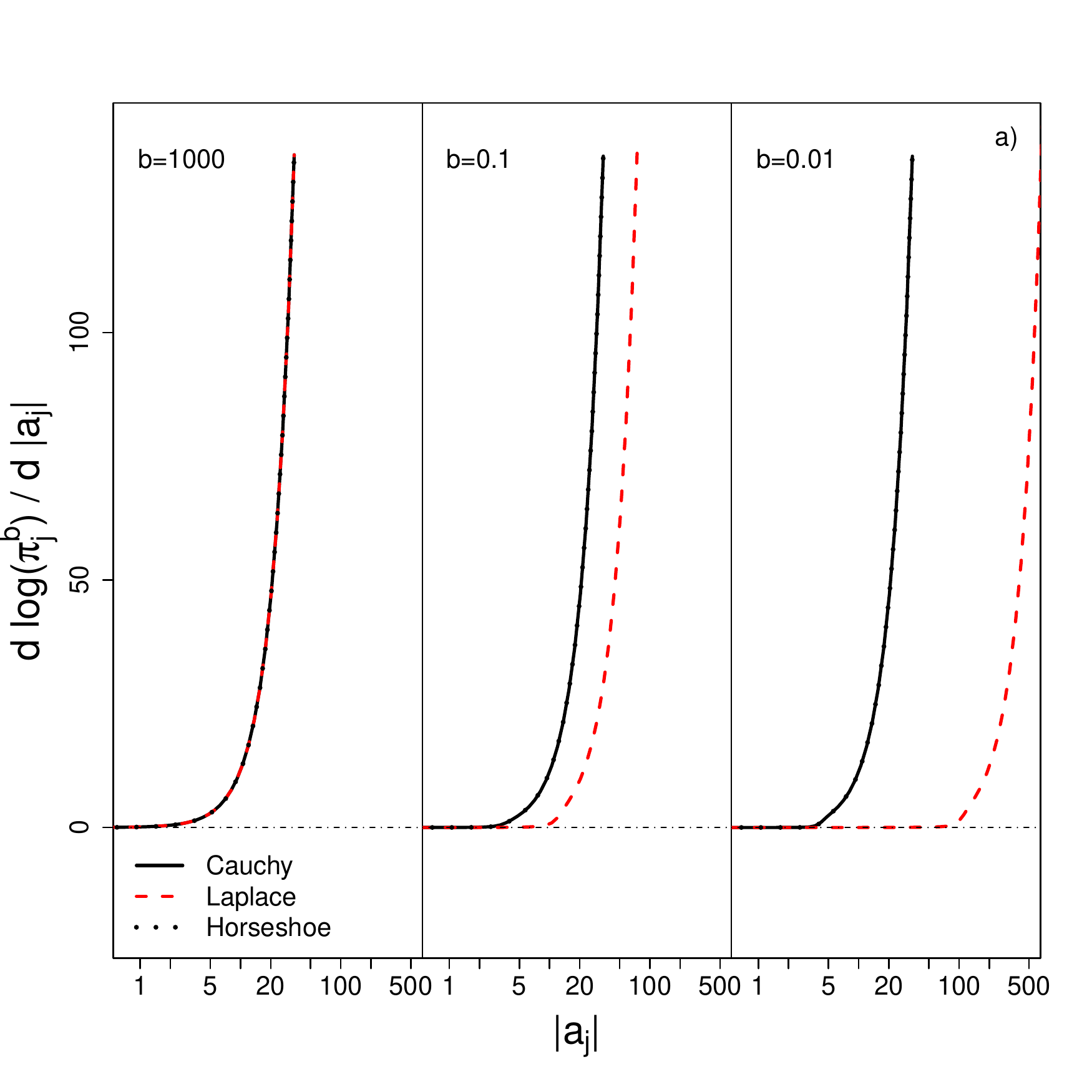}\includegraphics[height=80mm]{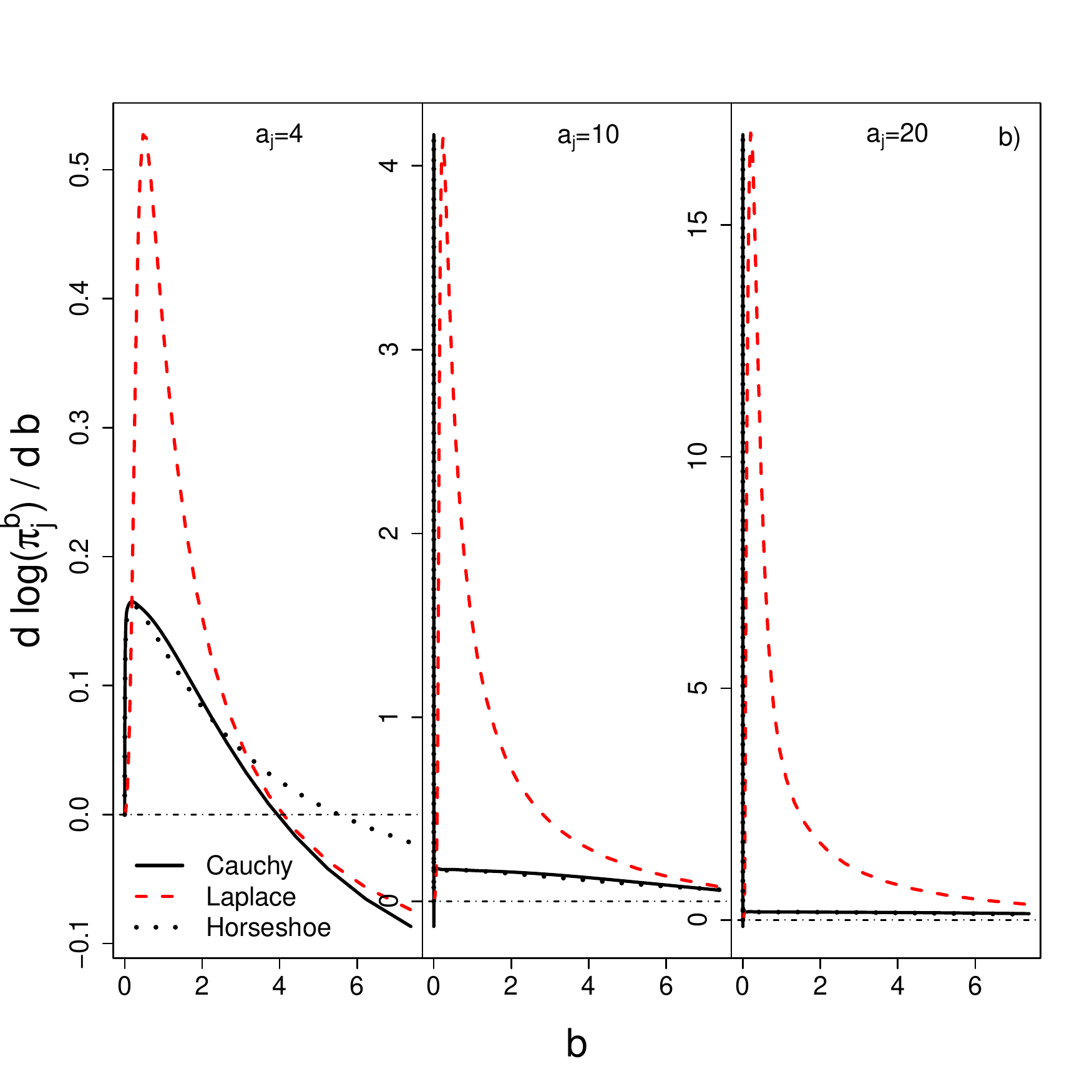}
\caption{The derivative of log odds respect to $|a_j|$ against given different $b$ (a). Note the curves of Cauchy and horseshoe priors are overlapped. The derivative of log odds respect to $b$ given different $a_j$ (b).}\label{b.pic5}
\end{center}
\end{figure}

Figure \ref{b.pic5} gives the exact calculation of $r_a={d\over{d|a_j|}}\log\pi^b_j$ and $r_b={d\over{db}}\log\pi^b_j$ for three priors. In Figure \ref{b.pic5} (a) we can see that $r_a$ is the nearly the same for three priors at large $b$. However, when $b$ is small, $r_a$ is reduced by certain value for Laplace prior, meanwhile, it remains the same for Cauchy and horseshoe prior. So the exact calculation just confirms Theorem 4. Similarly, the exact calculation also confirmes the result of Theorem 4 about $r_b$. As we can see in Figure \ref{b.pic5} (b), $r_b$ increases exponentially as $b\rightarrow0$ for Laplace prior, which means as $b\rightarrow0$, the selection probability by Laplace prior will exponentially drop to 0.5, and this behavior has already been observed in Figure \ref{b.pic3} (b).

Based on discussion in Section \ref{b.sec5.1} and this section, horseshoe prior performs the best in terms of the marginal selection probability, Cauchy prior is in the second place, and Laplace prior is the worst since the selection probability drops too fast as shrinkage increases.

\subsection{Some Expressions for $\pi^b_j$}\label{b.sec5.3}
In above sections, we discussed the properties of the odds $\pi^b_j$ for orthogonal designs, but did not show how to calculate $\pi^b_j$. Those curves are calculated by Monte Carlo simulations which are very precise. In some cases, we may want to calculate $\pi^b_j$ directly. There is no closed form of $\pi^b_j$ for the three different priors. However, we can see at least for Laplace and horseshoe priors, $\pi^b_j$ can be expressed by some special functions.

For Laplace prior,
\be\label{b.e34}
\pi^b_j=\sqrt{\pi\over2}b^{-1}\exp\left[{{(|a_j|-b^{-1})^2}\over2}\right]\left\{\Phi\left(|a_j|-b^{-1}\right)+\exp\left({2|a_j|b^{-1}}\right)\Phi\left(-|a_j|-b^{-1}\right)\right\},
\ee
where $\Phi(\cdot)$ is the CDF of standardized normal distribution. This expression can be directly used to calculated the marginal selection probability given $b$.

For horseshoe prior,
\be\label{b.e35}
\pi^b_j={1\over\pi}b^{-1}Be\left(1,{1\over2}\right)\Phi_1\left({1\over2},1,{3\over2},{a_j^2\over2},1-b^{-2}\right),
\ee
where $Be(\cdots)$ denotes the beta function, and $\Phi_1(\cdots)$ is the degenerate hypergeometric function of two variables \citep{c6,c27}. The calculation of $\Phi_1$ can be employed by using a series of hypergeometric $_2F_1$ functions \citep{c6}.

The derivative of above expressions is shown in \ref{app5}. $\pi^b_j$ of Cauchy prior does not have an analytic form, and its can not be represented by known special functions neither. Hence we simply use the Monte Carlo approach to calculate $\pi^b_j$ for Cauchy prior.

\subsection{Simulated Tempering and Generalization by L\'{e}vy Process}\label{b.sec5.4}
\citet{c15} discussed the difficulty of sampling around phase transition in a SSVS model by assigning a Ising prior for $\bga$. The difficulty is, given a Ising model there is a threshold for the interaction strength, when the interaction magnitude is larger than this threshold, the MCMC sampling will dramatically slow down, resulting in either overwhelming many selected nodes or extremely few ones. It becomes even worse when $J_{ij}$'s and $h_j$'s are all random, such as our model. However, the family of exchange Monte Carlo and simulated tempering algorithm has be developed to handle the slow mixing problem \citep{c5a,c11,c18}. By introducing the scale normal mixture for $\bet$, our model is an special simulated tempering algorithm which thus improves the mixing issue too.

To understand the simulated tempering algorithm, consider the usual Ising model with $U(\bga,J)=\sum_{i<j}J_{ij}\delta_{ij}$ (for simplicity no external field $h^*_i$ included), then the Boltzman distribution is expressed as
\[\label{b.e36}
p(\bga|T,J)={1\over{Z(T)}}\exp\left[-T^{-1}U(\bga, J)\right],
\]
where $T$ represents the temperature (or the scale of variation), and $J_{ij}$ is random and follows some distribution $p(J_{ij})$ such as standard Gaussian distribution. When $T\rightarrow0$, the effective interaction $\tilde J_{ij}=T^{-1} J_{ij}\rightarrow\infty$. Thus if $T$ is lower than some critical temperature, the strong interaction will lead to some non-ergodic behavior such as the slow down of the MCMC and extremely large proportion of $\gamma_j=1$. The reason for this is because the low temperature phase of disordered Ising model generally has numerous local minima which are separated to each other by energy barriers. The characteristic time in which the system escapes from a local minimum, however, increases rapidly as the temperature decreases or the interaction increases. A good review can be found at \citet{c21} about this issue. The family of tempering algorithm treats temperature $T$ as a dynamical variable \citep{c18}, and the joint distribution $p(\bga,T)$ is represented as
\be\label{b.e37}
p(\bga,T, J)\propto p(\bga|T,J)\prod_{i<j}p(J_{ij})p(T),
\ee
where $p(T)$ is the distribution of $T$. The prior information for the $T$ thus represents the range and mass of the temperature to sample the MCMC. With some variable transformation by replace $T_b^{-1/2}J_{ij}$ with $\tilde J_{ij}$, where $T^2=T_b$, the joint distribution (\ref{b.e37}) then becomes
\[\label{b.e38}
p(\bga,T_b, \tilde J)\propto p(\bga|\tilde J)\prod_{i<j}p(\tilde J_{ij}|T_b)T_b^{-1/2}p(T_b),
\]
where $p(\bga|\tilde J)\propto\exp\left[-\sum_{ij}\tilde J_{ij}\delta_{ij}\right]$, $p(\tilde J_{ij}|T_b)\propto p(T_b^{1/2}\tilde J_{ij})T_b^{1/2}$ (with some notation abuse, the later $p(\cdot)$ represents the same density function of $p(J_{ij})$). Clearly, $T_b$ is a global temperature parameter here. If we introduce the local temperature parameter $T_b$ for each interaction $J_{ij}$, then the marginal prior for $\tilde J_{ij}$ is
\[\label{b.e39}
p(\tilde J_{ij})\propto\int_0^\infty p(\tilde J_{ij}|T_b)T_b^{-1/2}p(T_b)dT_b.
\]
If $p(J_{ij})\sim N(0,1)$, above posterior for $\tilde J_{ij}$ is a normal scale-mixture whose mixing measure is expressible in terms of the density of the subprdinator $T_b$. Hence according to the Theorem 3 of \citet{c25}, with the simulated tempering algorithm the interaction of the random Ising model (\ref{b.e37}) can be expressed as a L\'{e}vy process mixture scaled by $T_b$, and  $T_b$ is a nondecreasing pure-jump L\'{e}vy process with marginal density $p(T_b)$ at time $b$.

As another algorithm in the same family,  the exchange monte carlo algorithms (or parallel tempering) is to simultaneously and independently simulate $K\ge2$ replicas of the MCMC trace under different temperatures, and exchange the $\bga$ configurations of the replicas with certain acceptance probability by referring to the energy cost $\Delta U$. The analogy between the exchanged monte carlo and simulated tempering algorithm is clear in terms of the mixture distribution of the interaction $J_{ij}$. For simulated tempering the mixture weight is the continuous prior $p(T_b)$ while the exchanged monte carlo is mixed with weight on a set of discrete temperatures. In both algorithms, the low temperature process can access a representative set of local energy minimums with the accompany of the high temperature process which are generally able to sample large volumes of configuration space to keep the configuration from trapping in some local minimum.

We see that how to understand the simulated tempering algorithm as the Ising model with normal scale-mixture prior mixed by the L\'{e}vy process. On the other hand, BVGM with L\'{e}vy process mixtures can also be understood as an Ising model sampled by simulated tempering algorithm.
To see this, we can generalize both Cauchy and Laplace prior into the framework of normal/generalized inverse Gaussian mixture. The marginal prior of $\beta_j$ for both priors can be expressed using one formula
\be\begin{split}\label{b.e40}
p(\beta_j|u,v,w)&=\int_0^\infty p(\beta_j|\tau_j)\tau_j^{-1}g(\tau_j)d\tau_j=\int_0^{\infty}p(\beta_j|\tau_j)p(\tau_j|u, v, w)d\tau_j\\
&=\int_0^{\infty}{1\over\sqrt{2\pi}}\exp\left(-\beta_j^2\tau_j/2\right)\tau_j^{1/2}{{(u/v)^w}\over{2K_w(uv)}}\tau_j^{w-1}\exp\left[-\left(u^2\tau_j+v^2\tau_j^{-1}\right)/2\right]d\tau_j\\
&={1\over{\sqrt{2\pi}}}\cdot{{(u/v)^w}\over{K_w(uv)}}\cdot{{K_{w+1/2}\left(v\sqrt{\beta_j^2+u^2}\right)}\over{\left(\sqrt{\beta_j^2+u^2}/v\right)^{w+1/2}}},
\end{split}\ee
where $K_w(x)$ denotes the modified Bessel function of the third kind with $w$. $p(\tau_j|u, v, w)$ is the generalized inverse Gaussian ($GING(w,u,v)$) distribution with parameters $u,v$ and $w$ such that $w\in\mathbb{R}$ while $u$ and $v$ are both nonnegative and not simultaneously 0. Note for the two special cases we adopted in this paper, Cauchy and Laplace prior, the values for these there parameters are on the boundary. However, it turns out both the prior $p(\tau_j|u,v,w)$ and the marginal prior $p(\beta_j|u,v,w)$ as the limit exist. For Cauchy prior, $w=1/2$, $u=b$, and $v=0$, thus $p(\tau_j|b,v,1/2)\rightarrow G(1/2, b^2/2)$ as $v\rightarrow0$, where we use identity $K_{-1/2}(x)=\sqrt{\pi/2}x^{-1/2}\exp(-x)$. Hence the distribution of $\beta_j$ reduces to the Cauchy distribution with scale parameter $b$, where we use $\lim_{x\rightarrow0}x^wK_w(x)\rightarrow2^{w-1}\Gamma(w)$ and $\Gamma(w)$ is a Gamma function. For Laplace prior, $w=-1$, $u=0$, and $v=b^{-1}$. the limit of prior $p(\tau_j|u,b^{-1},-1)\rightarrow IG[1,(2b^2)^{-1}]$ as $u\rightarrow0$, where $IG(\cdot)$ stands for inverse gamma distribution, and we used the index symmetry $K_{-w}(x)=K_{w}(x)$.

According to Theorem 3 of \citet{c25}, we see $\tau_j^{-1}g(\tau_j)=p(\tau_j|u, v, w)$ where $g(\tau_j)$ is the density of the subordinator $\tau_j$ at $(u,v,w)$. Analogous to the discussion with the simulated tempering algorithm for Ising model, we can see the temperature parameter in our model is $\tau_j$. However, there are several differences, such as in our model, $J_{ij}\propto\beta_i\beta_j$ and we assign normal mixture prior for $\beta_j$'s, while in the simulated tempering algorithm of regular Ising model, the prior is assigned to $J_{ij}$ directly.

Now we can understand the temperature effect of $\tau_j$. When $\tau_j\rightarrow\infty$, which is equivalent to $\kappa_j\rightarrow1$, the system is in a high temperature state. This means MCMC is exploring the whole configuration space, and no precise sampling for a local energy minimum. This is also equivalent to say for each node, the odds of $\gamma_j=1$ is equal to one since in high temperature every node is heated up and chance to be up and down is even, which means the marginal selection probability for all nodes is $1/2$ as shown in Figure \ref{b.pic3} (a) as $\kappa_j\rightarrow1$.  On the other hand, when $\tau_j\rightarrow0$, which is equivalent to $\kappa_j\rightarrow0$, the system is in low temperature state. Starting from some initial state, all nodes configure will be trapped into their energy minimum (local maximum likelihood) which is $\gamma_j=0$ for most nodes in the orthogonal design, unless the external field $h_j\propto a_j$ is strong enough to force the node in the state $\gamma_j=1$. This is why we see in Figure \ref{b.pic3} (a) for small $\kappa_j$ the selection probability is 0 for nodes with small $a_j$ and remains 1 for nodes with very large $a_j$.

It is easy to understand the role of $b$ too, which is opposite to $\tau_j$ if we look at the prior $p(\beta_j|\tau_j,b)\sim N(0,b^2/\tau_j)$. In fact, we can also understand $b$ by  representing the hierarchical model as $p(\bga|\mby,\bet)\prod p(\beta_j|\tau_j)p(\tau_j|b)$, where $p(\tau_j|b)=\tau_j^{-1/2}\exp(-b^2\tau_j/2)$, $\tau_j^{-2}\exp[-(2b^2\tau_j)^{-1}]$ and $(\tau_j+b^{-2})^{-1}$ for Cauchy, Laplace and horseshoe prior respectively.  Thus we can see that $b$ controls how the local temperature parameter $\tau_j$ distributes. Large $b$ limits the variation of $\tau_j$ and increases the mass around zero, and small $b$ means the range for $\tau_j$ to vary is large. This is also consistent to Figure \ref{b.pic3} (b), where we can see when $b\rightarrow0$, $\tau_j$ can vary widely, thus the system is in high temperature state, and if $b\rightarrow\infty$, $\tau_j$ will be limited around $0$ and the system is in low temperature state.

The generalization of Cauchy prior and Laplace prior into the L\'{e}vy process mixture not only shows that the connection between Bayesian variable selection and the Ising model with tempering algorithm, it also provides flexibility to choose priors with different shrinkage characteristics. We did not discuss generalization of horseshoe prior as a L\'{e}vy process, further discussion can be found in \citet{c26}, but similar conclusion can be drawn for horseshoe prior in terms of shrinkage or tempering.
\section{Incorporating Graph Prior Information}\label{b.sec6}
In this paper, we mainly discuss model (\ref{b.e2}) as a graphical model with noninformative prior for $\bga$, and it works well for $n$ is large enough. However, the priori information about $\bga$ becomes important when $n$ goes small. There are two purposes of incorporating graph prior information for $\bga$. First, it helps to improve the mixing issue so the model works for $n<<p$. Second, it improves the power of detecting the true signals. Since two connected nodes with positive interaction intend to be selected or excluded together, only the prior graph for $\bga$ with positive interaction is meaningful. If we have the information that some selected nodes and their neighbors are all true nodes, then incorporating a graph prior with those nodes connected will improve the power to identify the nodes with small signal. This is because the prior tells us that those nodes with small signal have more chances to be selected together with their neighbors which are true signal. On the other hand, for those nodes that are not true signal, we have more chances to exclude their neighbors too since the prior tells us they should be excluded together. At first glance, assigning a prior graph for $\bga$ seems like manipulating the weight to select which nodes and their neighbors, but if the prior information is true, then assigning such a prior is reasonable. Even though the prior information is not exactly correct, it will help if the prior graph contains the true graph about which nodes are networked. For example, given a true model, $\mby=\sum_{j\in S}\mbx_j\beta_j$ where $S={1,...,k}$ is sequential index up to $k$ and $k<p$. Obviously there are some information about the true variables such that there are $k$ sequential nodes are true nodes, and $p-k$ sequential nodes are not in the true model. Therefore, a Ising prior with one dimensional linear chain will be a very efficient prior since this prior reflects the information that sequential nodes are selected or excluded together. Another example is the genetic pathway data within which different sets of genes function together.  Some gene sets are related to the phenotype diseases, some are not. Therefore the prior with this pathway graph helps distinguishing different set of genes in the pathway since among those genes if one node is selected then its connected neighbors have high chance to be selected too. Further example about incorporating prior graph information can be seen in \citet{c15,c20,c31,c33}.

Since we are only interested in the network prior information, we only apply a graph prior for $\bga$ with the interaction matrix $W=\{W_{ij}\}$ without the external field:
\[\label{b.e41}
p(\bga)\propto\exp\left(\sum_{i<j}W_{ij}\delta_{ij}\right),
\]
where $W_{ij}$ represents the prior coupling information between node $i$ and $j$. For simplicity, considering $W=w\Lambda$, where $w$ is a small positive interaction parameter, and $\Lambda=\{\lambda_{ij}\}$ is the adjacency matrix with $\lambda_{ij}=1$ if node $i$ and $j$ are connected and $\lambda_{ij}=0$ if $i$ and $j$ are independent. With this prior, the posterior distribution for $\bga$ is modified as:
\be\begin{split}\label{b.e42}
p(\bga|\mby,\bet,\phi)&\propto p(\mby|\bga,\bet,\phi)p(\bga)\\
&\propto\exp\left(\sum_{i<j}J^*_{ij}\delta_{ij}+\sum_jh^*_j\gamma_j\right),
\end{split}\ee
where $J^*_{ij}=(J_{ij}+W_{ij})$. $J_{ij}$ and $h^*_j$ are defined in (\ref{b.e11}) and (\ref{b.e12}).

Correspondingly, the two expressions for the cluster algorithm are modified as
\be\label{b.e43}
p_{a,j}=\max\left\{1-\exp\left[(-1)^{\gamma_j}\left(\sum_{k\in c_1}\lambda_{ik}J^*_{jk}-\sum_{l\in c_0}\lambda_{jl}J^*_{jl}\right)\right],0\right\},
\ee

\be\begin{split}\label{b.e44}
&\alpha(\bga_c^0\rightarrow\bga_c^*)\\
&=\min\left\{\exp\left[\sum_{j\in\bar c}(-1)^{\gamma_j}\left(\sum_{k\in c_1}(1-\lambda_{jk})J^*_{jk}-\sum_{l\in c_0}(1-\lambda_{jl})J^*_{jl}\right)+\sum_{j\in c_0}h^*_j-\sum_{j\in c_1}h^*_j\right],1\right\}.
\end{split}
\ee
Above two expressions tell us that $p_{a,j}$ and $\alpha(\bga_c^0\rightarrow\bga_c^*)$ are also conditional on $\Lambda$.

\section{Extension to Nonparametric Regression Models:\\*Bayesian Sparse Additive Model (BSAM)}\label{b.sec7}
Although the BVGM is based on the parametric linear regression model (\ref{b.e2}), it is easy to be extended to nonparametric regression models. Some similar approaches have been suggested, such as nonparametric regression using Bayesian variable selection \citep{c30} and Bayesian Smoothing Spline ANOVA models \citep{c28}, both use the spline techniques. In the former case, the binary random variable  is applied to each knots of spline function in stead of each predictor, thus the model is capable to select the knots of each nonparametric function. In the second paper, second order interactions are included by using function ANOVA. In this paper, we only employ BSAM to demonstrate how easy it is to extend the parametric regression model based on BVGM.

Extending the multiple parametric linear regression model (\ref{b.e2}) to an additive model is straightforward. In Bayesian point of view, there is no strict difference between parametric and nonparametric additive regression model in sense of that both assign prior to the basis coefficients. In general, both choosing a basis to express the marginal regression predictor $f_j(\mbx_j)$. For linear parametric regression, $f_j(\mbx_j)=\beta_j\mbx_j$, where the predictor $\mbx_j$ itself can be considered as the basis to represent $f_j$ and $\beta_j$ is a univariate random variable. This is just a special case of nonparametric regression model considering $f_j(\mbx_j)=Z_j\bet_j$, where $Z_j$ is some basis matrix for $j$th predictor and $\bet_j$ is multivariate random variable, where the basis length $M_j\geq1$ can vary for different predictor. Despite the variation of the basis chosen, each predictor is corresponding to a univariate random vector $\mbr_j=f(\mbx_j)=Z_j\bet_j$. Then the generalized additive model can be expressed as
\[\label{b.e45}
\mby=\mu+\sum_{j=1}^{p}\gamma_{j}f_j(\mbx_j)+\bel.
\]
For this model, similarly, we can consider following prior for $\bet_j$'s
\be\begin{split}\label{b.e46}
[\bet_j|\tau_j]&\sim N(\mathbf0,b^2\tau^{-1}_jI),\\
[\tau_j]&\sim p(\tau_j).
\end{split}\ee
where $N$ is multivariate $M_j$ dimensional normal distribution, and $p(\tau_j)$ is some priors similar to previous discussions, such as $G(1/2,1/2), IG(1,1/2)$ or $C^+(0,1)$. For some special basis, the multivariate normal prior of $\bet_j$ may have two variance components such as LS basis (see \ref{app6}). Note that because the dimension of $\bet_j$ is changed, if we integrate out $\tau_j$ by assigning the same $p(\tau_j)$ as in parametric linear models, the marginal prior $p(\bet_j|b)$ is no longer Cauchy, Laplace, or horseshoe prior any more, but it shares the similar properties as linear parametric case.

Similarly we can define matrix $R=[\mbr_1,...,\mbr_p]$, design matrices $Z=[Z_1,...,Z_p]$, $Z_{\bga}=[\gamma_1Z_1,...,\gamma_pZ_p]$ and the coefficients vector $\bet=(\bet_1^T,...,\bet^T_p)^T$, but here we should treat $\bet$ and $Z$ as blocks. The total dimension for the design matrix $Z$ is $n\times M$ and $M\times1$for $\bet$, where $M=\sum_jM_j$. Without any confusion, we can use the same posterior distribution expressions in (\ref{b.e4}) and (\ref{b.e6}) to update $\bet_c$ and $\phi$ except we need keep in mind $\bet$ and $Z$ are in blocks and $D_c$ is diagonal block matrix with block $\tau_j/b^2I_{M_j}, j\in c$ in the diagonal, where $I_{M_j}$ is $M_j$ dimensional identity matrix. $\Sigma_c$ and $\boldsymbol \mu_c$ are expressed as
\be\begin{split}\label{b.e47}
\Sigma_c&=\left(\phi Z_c^TZ_c+D_c\right)^{-1},\\
\bmu_c&=\phi\Sigma_cZ_c^T\left(\mby-Z_{\bga_{\bar{c}}}\bet_{\bar{c}}\right).
\end{split}\ee
The calculation for $J_{ij}$'s and $h_j$'s is exactly the same as (\ref{b.e11}) since those formulas involve $R$ which is a $n\times p$ matrix for both cases. In \ref{app6} we will introduce a specific additive model with the natural cubic spline represented by Lancaster and \v{S}alkauskas (LS) basis. Of course, other spline basis to define $Z$ is possible.

\section{Simulation Study}\label{b.sec8}
\subsection{Case One: Comparison of Three Priors}\label{b.sec8.1}
The first simulation study will examine a simple linear regression model with a general form
\be\label{b.e48}
y=\sum_{j\in S}\beta_jx_j+\epsilon,
\ee
where $S=\{2,3,5,10\}$, sample size $n=50$ and $p=100$, $x_j\sim N(0,1), j=1,...,p$ and $\epsilon\sim N(0,1)$. Particularly, we will consider one large signal set and one small signal set: $\{\beta_2,\beta_3,\beta_5,\beta_{10}\}=\{-4,2,-1,2.5\}$ and $\{-0.9,0.7,-0.6,0.8\}$.

\begin{figure}[bth]
\begin{center}
\includegraphics[height=55mm]{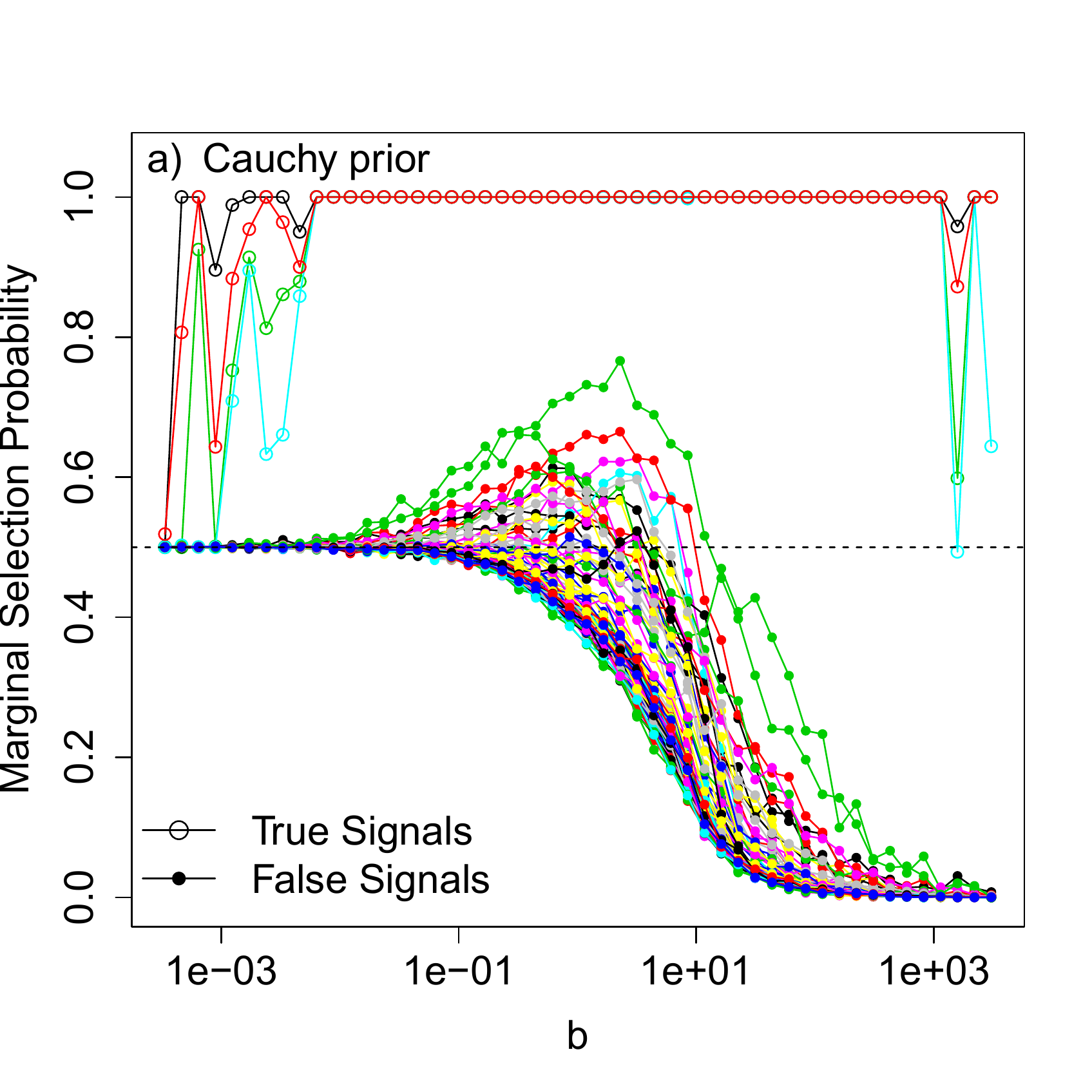}\includegraphics[height=55mm]{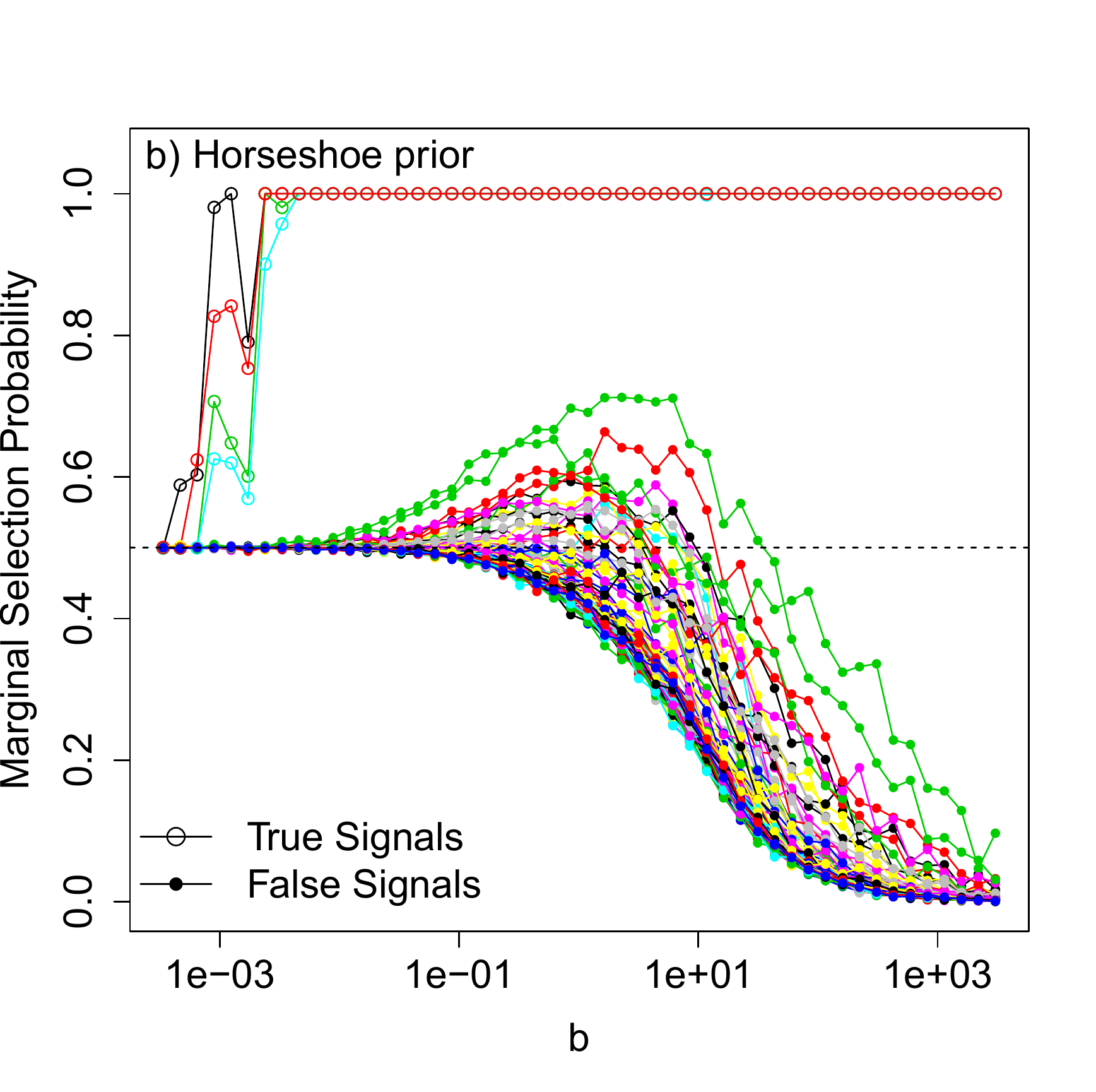}\includegraphics[height=55mm]{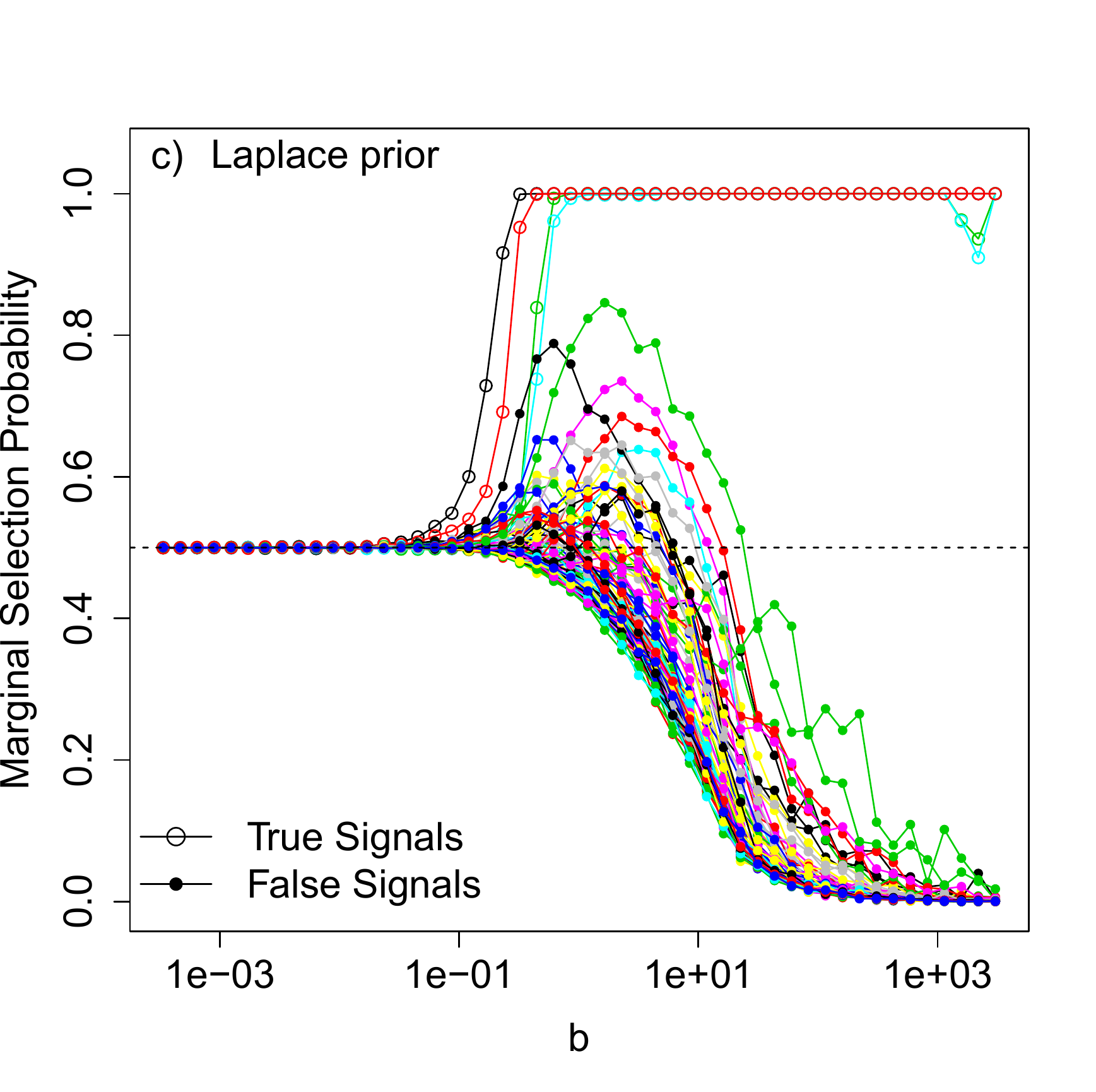}
\includegraphics[height=55mm]{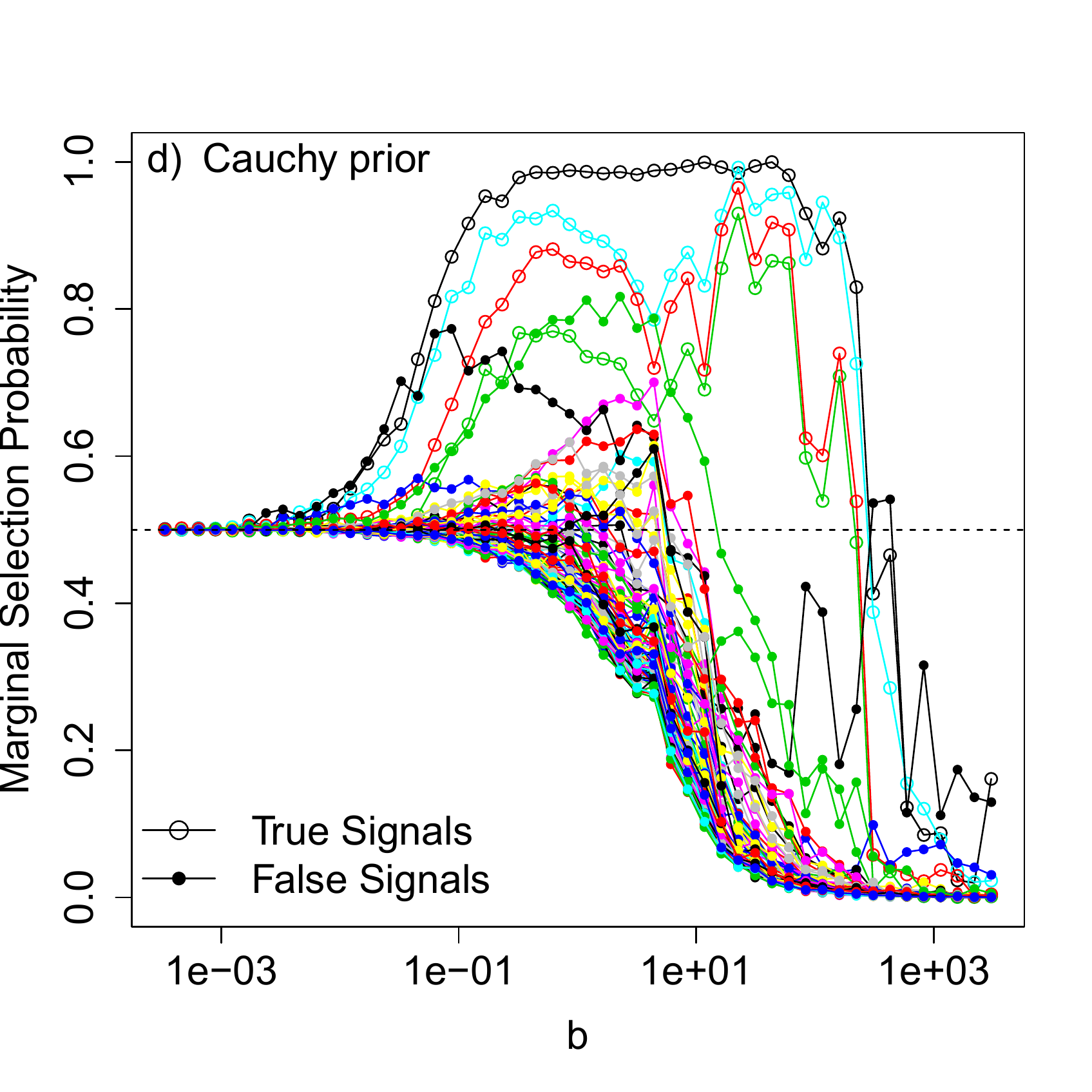}\includegraphics[height=55mm]{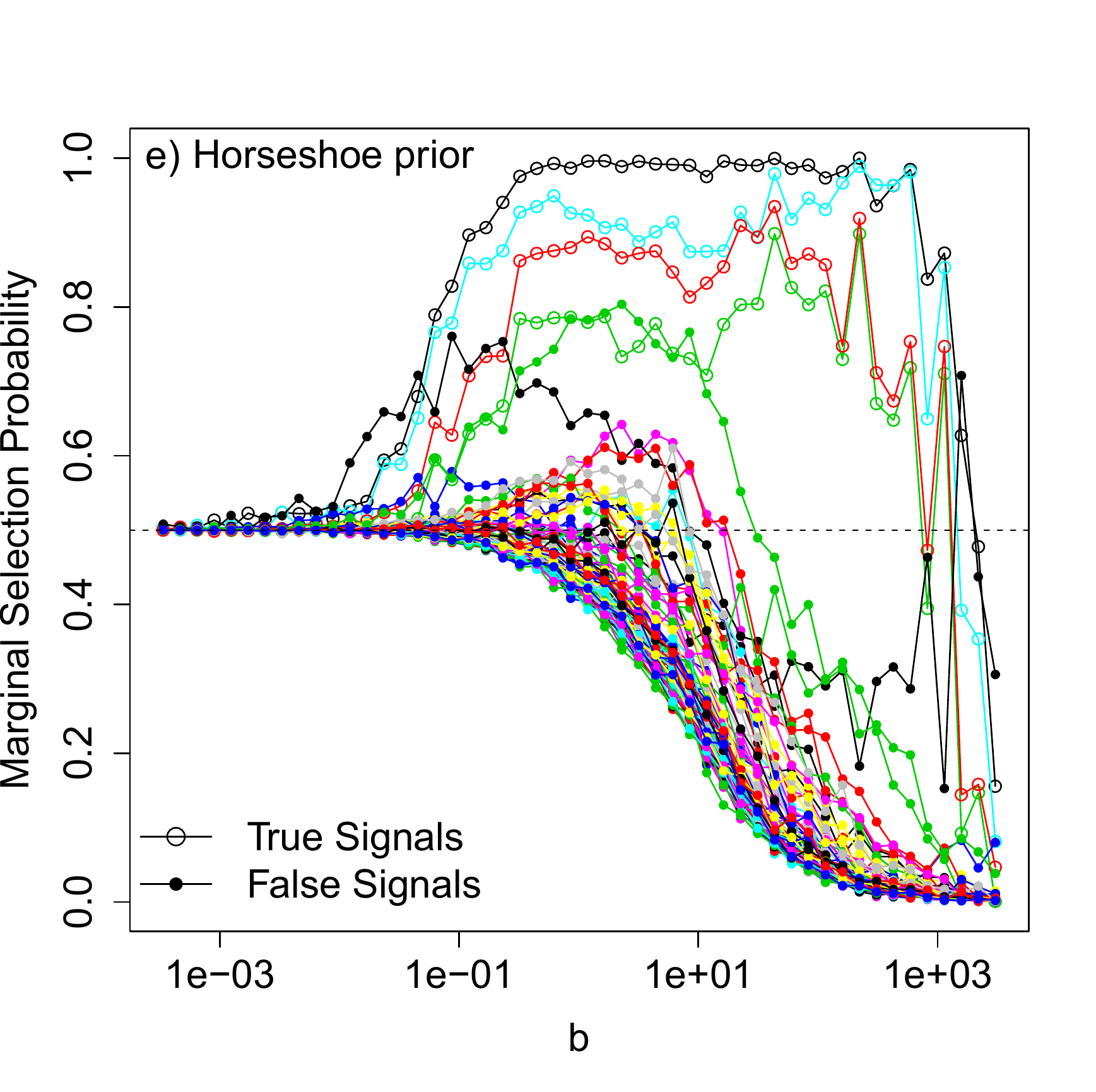}\includegraphics[height=55mm]{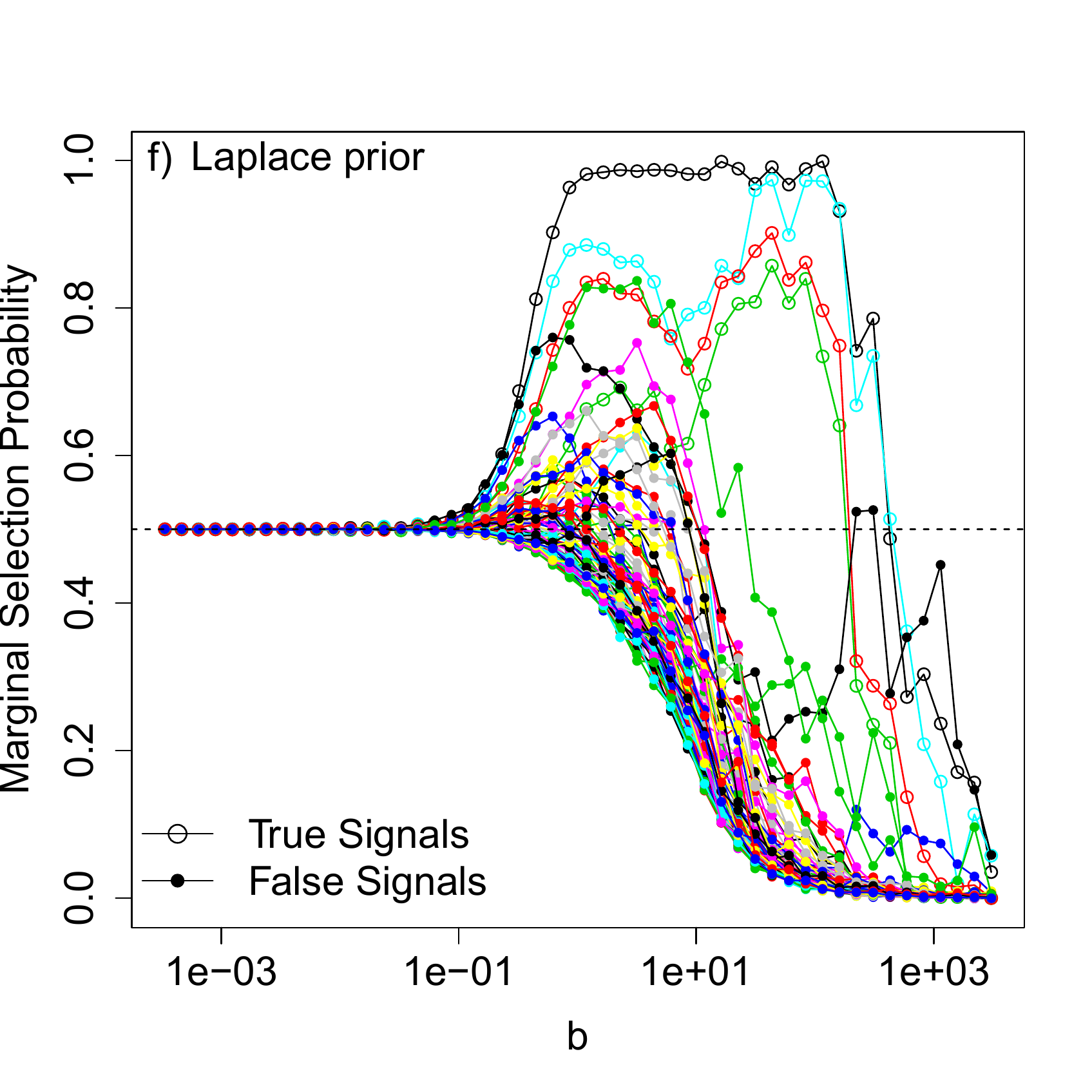}
\caption{The profile curves of the selection probability of simulation model (\ref{b.e48}) with different priors for large signal setting (a-c), and small signal setting (d-f).}\label{b.pic6}
\end{center}
\end{figure}

In this simulation, we performed the single site updating with total 6000 iterations for each settings and discarded the first 2000 iterations as burn-in, then calculated the average $\gamma_j$'s over total $N=4000$ iterations as the marginal selection probabilities. Figure \ref{b.pic6} plots the marginal selection probability of all variables against the global shrinkage parameter $b$. For large signals, as shown in the upper row of Figure \ref{b.pic6}, horseshoe and Cauchy priors perform similarly and show the robustness of large signals, i.e., as $b$ decreases, the selection probability of true signals maintains 1 till very small $b$ and drops to $0.5$. Horseshoe prior also shows better robustness than Cauchy prior on the large $b$ side. Both priors have a wide window of $b$ in which the true signals are well separated from the noise signals. On the other hand, Laplace prior does not demonstrate such robustness for large signals: as the $b\rightarrow0$, the selection probability of true signals drops to 0.5 very fast. Around $b=0.1$, all signals reach  the  $0.5$ line for Laplace prior. On the large $b$ side, Laplace prior seems perform a little better than Cauchy prior. Recall the exact calculation of the marginal selection probabilities in Figure \ref{b.pic3}, we can see that the conclusion made from the simulation about the performance of the three priors is exactly the same.

The bottom row of Figure \ref{b.pic6} is the simulation results for small signals. In general, the window for true signals maintaining high selection probability gets narrower for all three priors.  The selection probability of true signal for all priors starts to drop to $0.5$ around $b=0.1$, and around $b=1000$, they drop to 0. However, the drop rate on both side of $b$ is different for three priors, resulting in different width of the working widow of $b$. Horseshoe prior has the widest window, Laplace prior gets the narrowest one. Again, the conclusion we made from the simulation is exactly the same as the calculation in Figure \ref{b.pic3}.

Other observations can be found from Figure \ref{b.pic6}, especially for small signals. First we can see around $b=10$, the selection probability of true signals first drops a little and then grows up again. This happens right above the peak of the selection probability of the noise, which indicates potential interaction between the noise and the true signals. This is easy to understand since when sample size is small, the correlation between true signal and noise is large, meaning the parameter $\xi_j$ in (\ref{b.e23}) is large such that the profile curve is distorted. The second observation from Figure \ref{b.pic6} is, although the overall performance of three priors is different in sense of different width of the working window, we can select a right value of $b$ so that for all priors the noise and signals are well distinguishable.  For instance, for all priors, with a cut-off probability 0.5 all true signals are separated from the noise at some fixed $b$  between 10 and 1000.

\subsection{Case Two: Three Regions of Global Shrinkage Parameter $b$}\label{b.sec8.2}
Based on the case study one, horseshoe prior has the largest working window, thus in the rest of this paper, we employ horseshoe prior only unless stated otherwise. In this simulation we will examine the case when $p$ is large, say $p=1000$ or $p=500$. The linear model still has the form (\ref{b.e48}) with $x_j\sim N(0,1), j=1,...,p$ and $\epsilon\sim N(0,1)$, but we consider following specific models and settings
\begin{enumerate}[label={\it Model} \Roman*,align=left,leftmargin=*]
\item\begin{enumerate}[label=\Alph*.,align=left,leftmargin=*]\label{st1}
\item $p=1000, n=200$, $\beta_j=0.8$ if $j$ is odd; $\beta_j=1.0$ if $j$ is even. $S=\{31,91,...,931\}\bigcup\{60,120,...,960\}$
\item $p=1000, n=500$, $\beta_j=0.8$ if $j$ is odd; $\beta_j=1.0$ if $j$ is even. $S=\{31,91,...,931\}\bigcup\{60,120,...,960\}$
\end{enumerate}
\item\begin{enumerate}[label=\Alph*.,align=left,leftmargin=*]\label{st2}
\item $p=500, n=100$, $\beta_j=-0.8$ if $j$ is odd; $\beta_j=0.8$ if $j$ is even. $S=\{31,91,...,451\}\bigcup\{60,120,...,480\}$
\item $p=500, n=500$, $\beta_j=-0.8$ if $j$ is odd; $\beta_j=0.8$ if $j$ is even. $S=\{31,91,...,451\}\bigcup\{60,120,...,480\}$
\end{enumerate}
\end{enumerate}
Thus for \ref{st1} the number of true $\beta_j$'s are the cardinality $|S|=32$, and $|S|=16$ for \ref{st2}. For each setting, we performed the single site updating with total 8000 iterations and discarded the first 3000 as burn-in, then calculated the average $\gamma_j$'s over total $N=5000$ iterations as the marginal selection probabilities.

\begin{figure}[bth]
\begin{center}
\includegraphics[height=80mm]{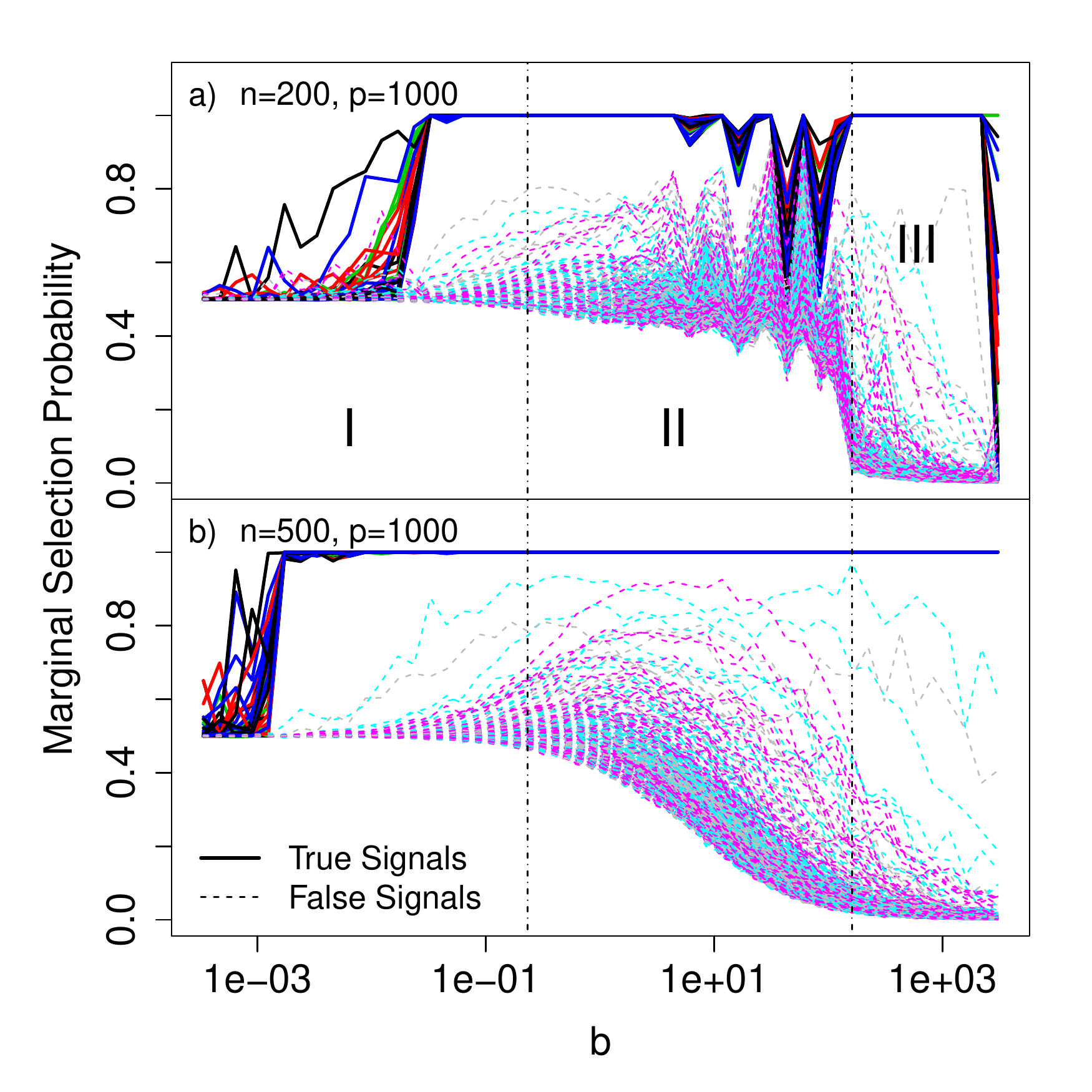}\includegraphics[height=80mm]{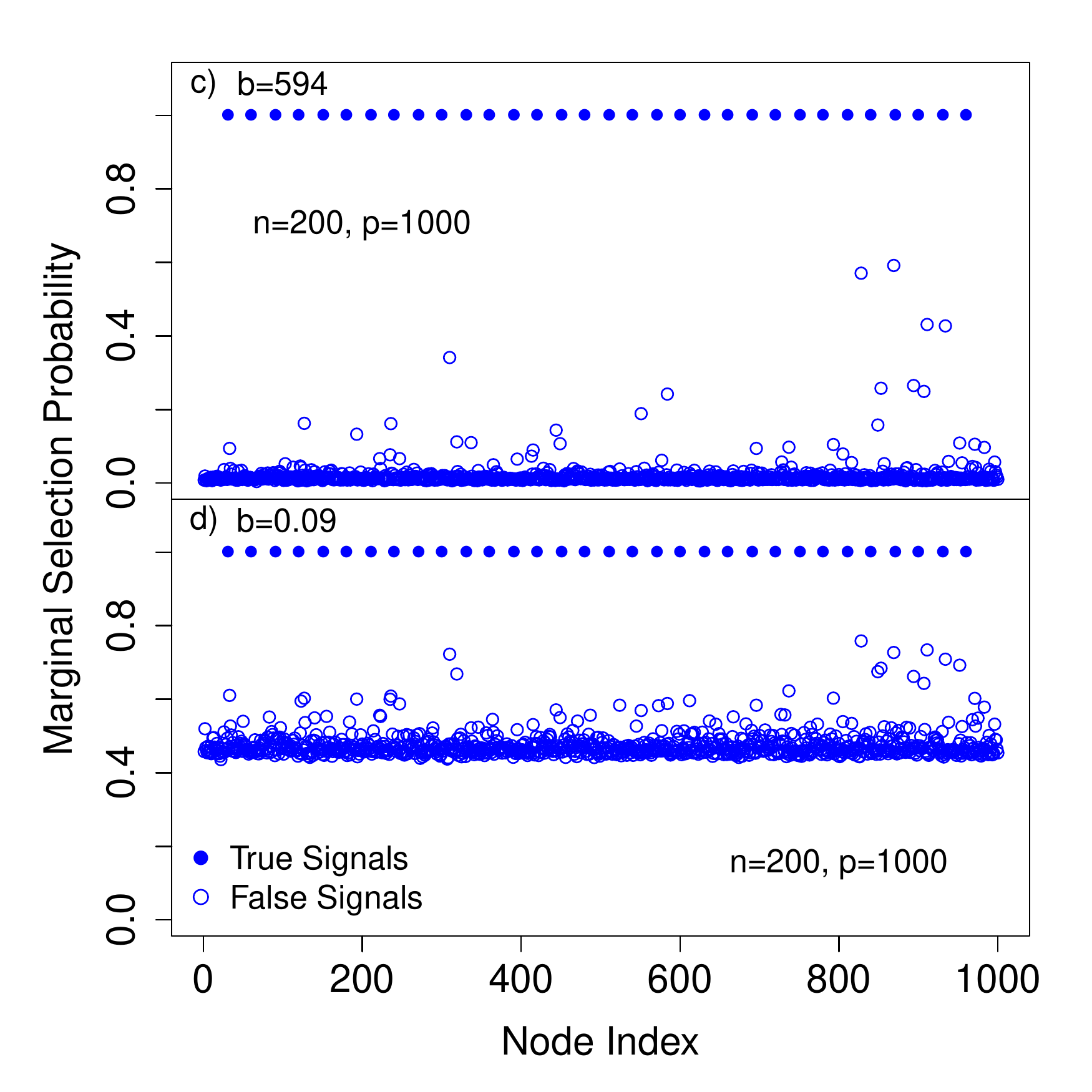}
\caption{The profile curves of the selection probability of \ref{st1} A and B (a-b). Selection probability at two $b$ values for \ref{st1} A (c-d).}\label{b.pic7}
\end{center}
\end{figure}

\begin{figure}[bth]
\begin{center}
\includegraphics[height=80mm]{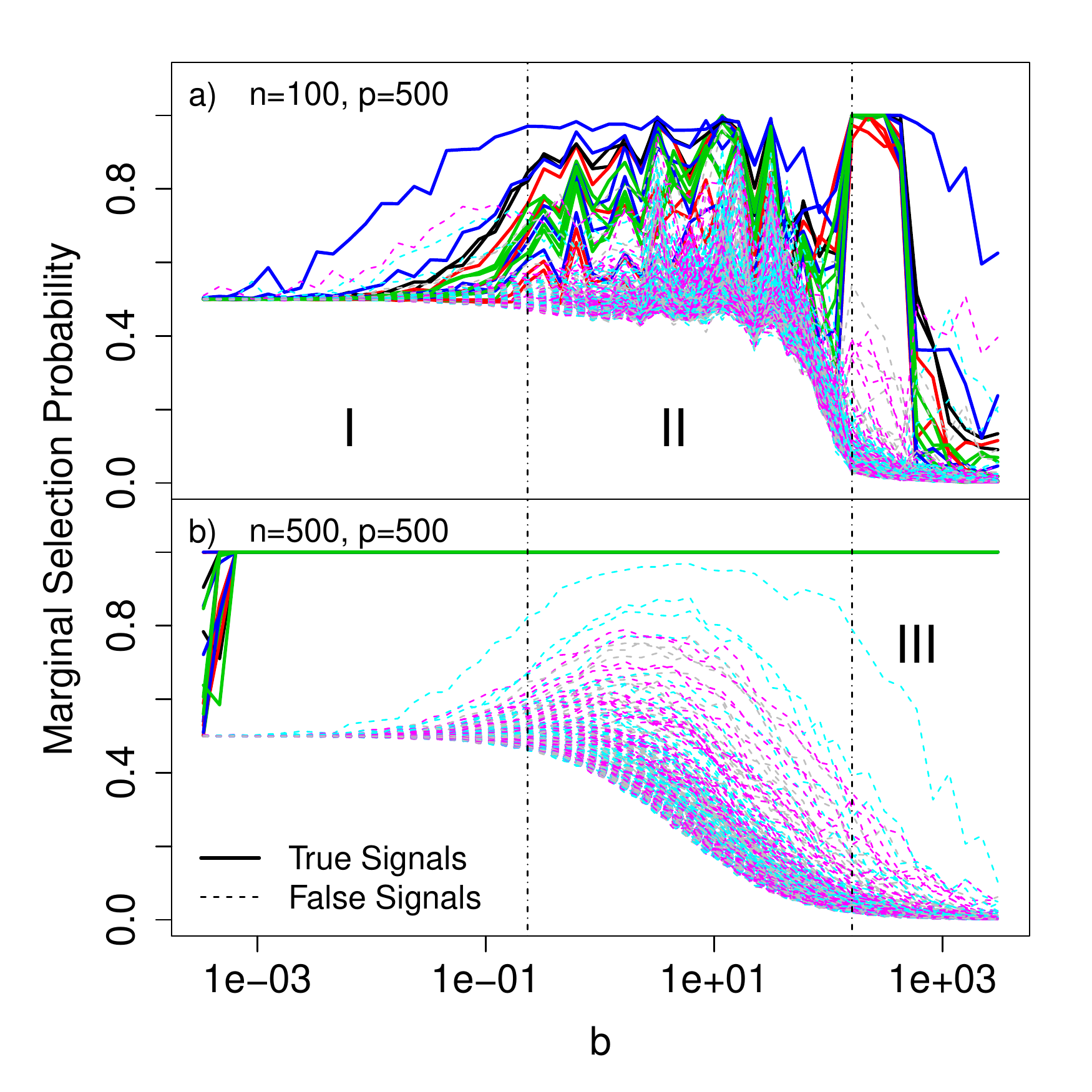}\includegraphics[height=80mm]{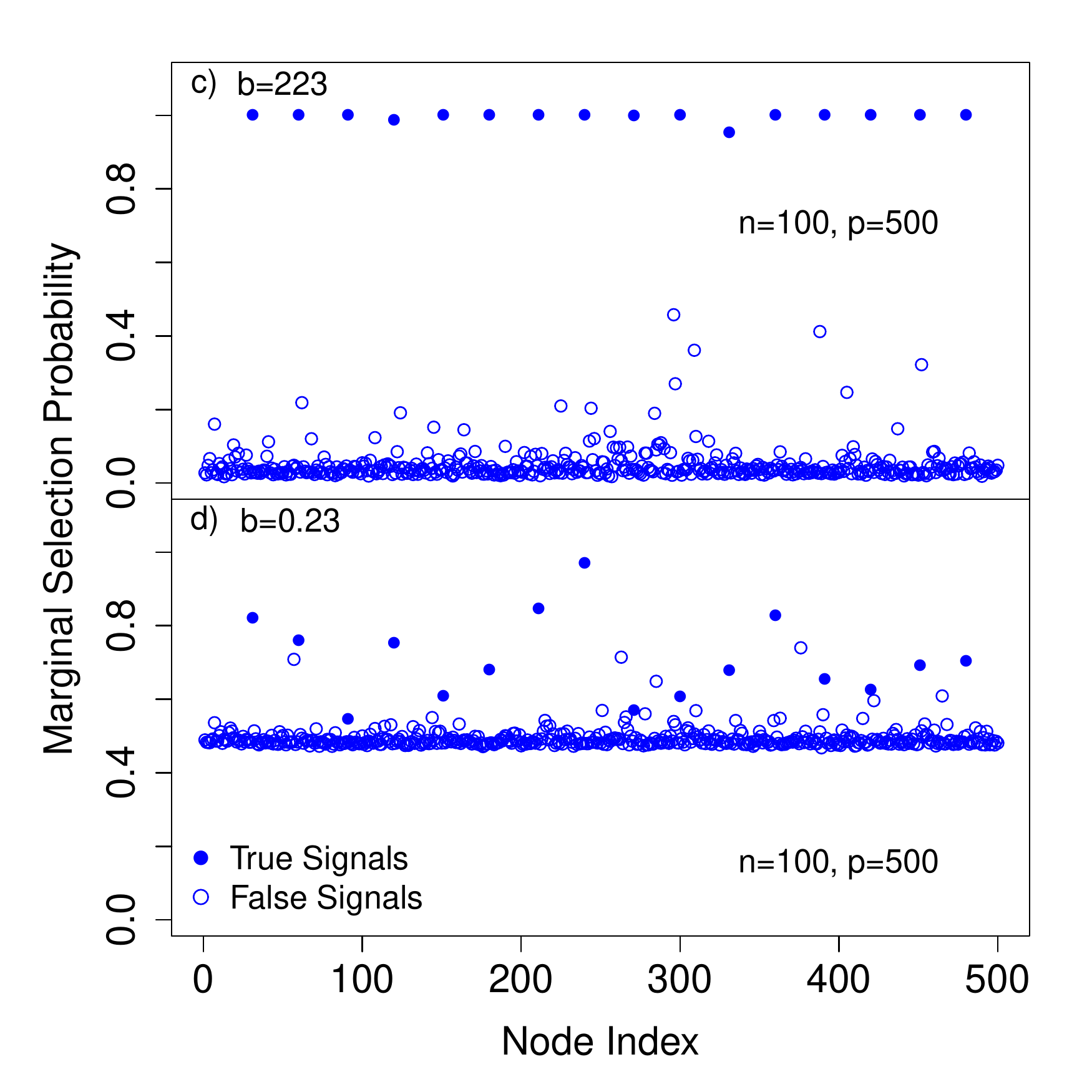}
\caption{The profile curves of the selection probability of \ref{st2} A and B (a-b). Selection probability at two $b$ values for \ref{st2} A (c-d).}\label{b.pic8}
\end{center}
\end{figure}

Figure \ref{b.pic7} (a-b) and Figure \ref{b.pic8} (a-b) plot the marginal selection probability of all variables against $b$ for all settings, so we can have a overall view about all possible global shrinkage. In this simulation it is easier to examine how the working window of $b$ suitable for variable selection changes. For example, in Figure \ref{b.pic7} (a), the working window is from $b\approx0.01$ to $b\approx2000$, and from $b\approx0.001$ to $b\approx3000$ in Figure \ref{b.pic7} (b). Within this window, we can see for both setting A and B, $b$ can be further divided into three regions I, II and III. In Figure \ref{b.pic7} (a) Region I represents the high temperature or large shrinkage area, where $b$ is around $0.1$ or smaller. Region II is a moderate shrinkage area with $b$ between 1 and 100, and the last Region III is around 1000 varying from several hundreds to several thousands. the widths of these three regions also change for different signal strength. We can see in most cases, the signals are well separated from the noise in Region I and III. On the other hand, in Region II, if the signal is not large enough or the sample size $n$ is small, some oscillations or strong interactions occur between signal and noise on the profile curves, resulting in a total mixture up of noise and signal. Thus if to suggest the appropriate value of $b$, it must be selected to avoid Region II.

Another interesting observation from this simulation is although in general, both Region I and III both can be used to detect signals, the performance of the MCMC sampling may have different properties in these two regions. We have discussed that Region I has large shrinkage property, but it may not have the sparse consistency. This can be understood from the point of view of the oracle properties of the estimation given the shrinkage parameter. Fan and Li 2001 shows in general the sparse consistency requires $\lambda\rightarrow0$ or small shrinkage ($\lambda\propto b^{-2}$ is the shrinkage parameter in their paper). Region III, representing small shrinkage area, hence may maintain sparse consistency while Region I loses it. This phenomena is shown in Figure \ref{b.pic8} (a) for \ref{st2} A where the best value of $b$ is in Region III with which all signals are distinguishable from the noise. On the other hand the noise and signals mix up in Region I. This can be seen more clearly in Figure \ref{b.pic8} (c-d) for two specific $b$ values: for $b=223$, most of the true signals have selection probability 1 and distinguished from the noise, while for $b=0.23$, some the true signals have smaller selection probability than some noise.

How to determine the parameter $b$ is a interesting topic. Some authors suggest assigning another prior for $b$, such as horseshoe prior \citet{c25}. Unfortunately, because $b$ is a global parameter, when $p$ is large, the posterior distribution of $b$ will be forced to some value that is not in Region III where we prefer. Therefore in this paper, we will not consider assigning a prior for $b$, instead, we consider it as a tuning parameter. A practical way to select $b$ is to try several $b$ values and choose the one we see the largest gap in selection probability and $b$ is usually between ten to over thousands.

\subsection{Case Three: Comparison of Cluster and Single Site Algorithm}\label{b.sec8.3}
As discussed in Section \ref{b.sec5.4}, assigning the scale normal mixture prior for $\beta_j$'s with shrinkage parameter $b$ is also a tempering algorithm, which means our model already makes improvement in the mixing issue. So there may be no much space left from improving the mixing with a cluster algorithm. We will show that, the performances of cluster and single site algorithms both are $b$-dependent. In some region of $b$, one may outperform the other but performs worse in other region.

To demonstrate this, we consider the simple simulation with the same model as (\ref{b.e48}) with large signals $\{\beta_2,\beta_3,\beta_5,\beta_{10}\}=\{-4,2,-1,2.5\}$, $n=200$ and we vary $p$ from 50 to 1500. We run the simulation with four representative $b$'s, two are large and two are small, so we can compare the difference behavior of two algorithm with different shrinkage paramters.

\begin{figure}[bth]
\begin{center}
\includegraphics[height=80mm]{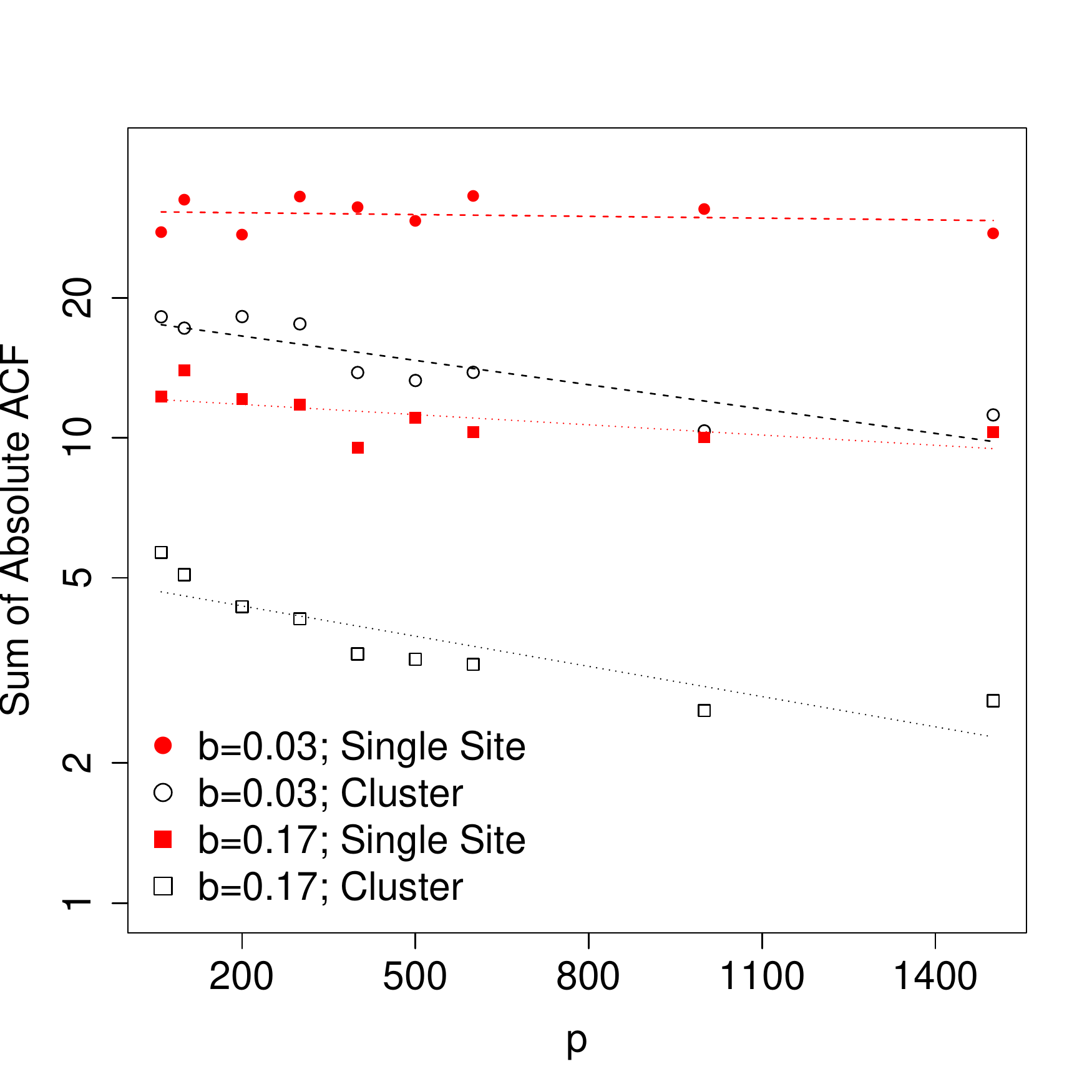}\includegraphics[height=80mm]{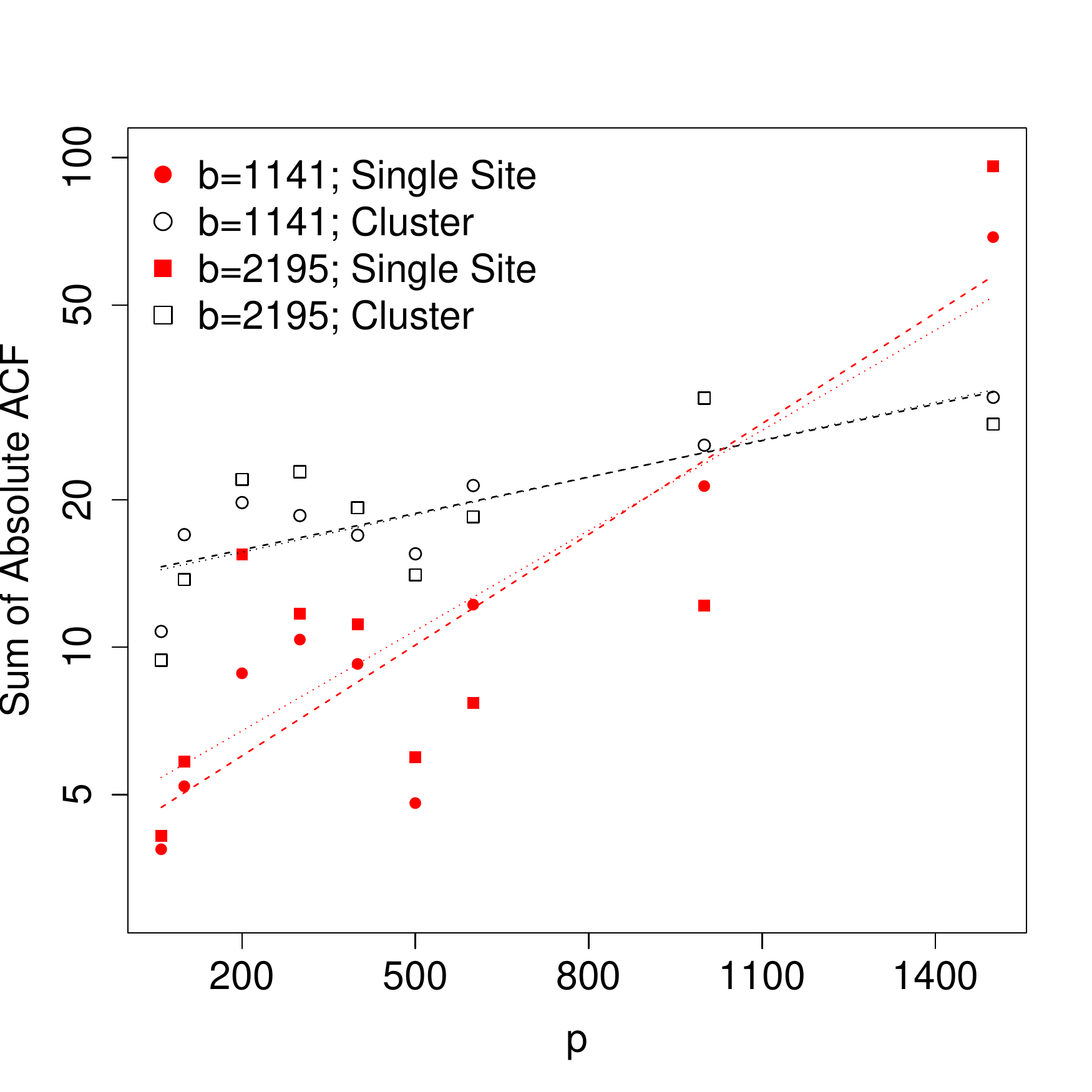}
\caption{The sum of absolute ACF against variable number $p$ for cluster algorithm and single site algorithm at different $b$ values.}\label{b.pic9}
\end{center}
\end{figure}

To measure the mixing or correlation time, it is convenient to define the ``magnetization'', $M^{(i)}$, which represents the average value of the binary random variable $\gamma_j$'s at $i$th sweep of the MCMC iteration.
\[\label{b.e49}
M^{(i)}={1\over p}\sum_{j=1}^p\gamma^{(i)}_j.
\]
Thus the mixing time of the MCMC iteration can be measured using the time-delayed autocorrelation function (ACF) of the Monte Carlo chain of ``magnetization'',
\[\label{b.e50}
C{(t)}={{\sum_{i=1}^{N-t}(M^{(i)}-\bar M)(M^{(i+t)}-\bar M)}\over{\sum_{i=1}^N(M^{(i)}-\bar M)^2}},
\]
where $t$ is the lag or the iteration time from the origin, measured in Monte Carlo sweeps (MCS), and $\bar M$ is the average magnetization over total $N$ iterations. We assume the absolute value of $C(t)$ decays exponentially, i.e., $|C(t)|\approx C_0\exp(-t/\tau)$, where $C_0$ is some positive constant, and $\tau$ is defined as the exponential correlation time. Therefore, we can use $\tau$ to measure how fast the chain converges or mixes. The smaller the $\tau$, the faster the system mixes up. Another way to measure the mixing time is simply using the summation of the autocorrelation time, $\sum_{t=0}^L|C(t)|$, where $L$ is the maximum lag calculated. \\

For each $p$ we performed 15000 iterations or sweeps for each setting and discarded the first 5000. From the remaining $N=10000$ sweeps we calculated the autocorrelation function $C(t)$ up to $L=100$ lags. Figure \ref{b.pic9} shows the summation of absolute ACF time against the nodes size $p$ for $b=0.03, 0.17,1141$ and $2195$.

From Figure \ref{b.pic9} we can see the different behavior of cluster algorithm and single site algorithm. In large shrinkage region, $b=0.03$ or $0.17$, the cluster algorithm has mixing time uniformly smaller than the single site algorithm. Note that as the node size increases, the mixing time for all algorithm first decreases slightly and then stabilizes. It may goes up when $p$ goes further. This profile is not well understood yet. Probably because in large shrinkage area, the effect of large node size is pressed by the shrinkage when the number of true nodes is fixed and small. Nevertheless, in this region, we can conclude that cluster algorithm is uniformly outperform the single site algorithm in terms of fast mixing time, and the mixing time of cluster algorithm is at least two times shorter.

In the small shrinkage region where $b=1141$ and $2195$, as shown in Figure \ref{b.pic9} (b), we see different characteristics. First, the measured mixing time is much more noisy than in Figure \ref{b.pic9} (a), but the trend against $p$ is clear.  Secondly, unlike large shrinkage area, here we see for both algorithms the mixing time increases as $p$ increase. Furthermore, when $p$ is small, the single site algorithm has shorter mixing time, but slow down very fast as $p$ increases. For example, when $p=60$, the summation of autocorrelation function is only several MCS, but reaches almost 100 MCS when $p$ is large than 1500, which means extremely slowing down for the MCMC process. On the other hand, although the cluster algorithm is about two times slower when $p$ is small and it also slows down with $p$ increases, the mixing time increases with smaller rate and reaches no more than 50 when $p=1500$.

Hence in general, we can see cluster algorithm outperforms single site algorithm in terms of mixing time. However, which algorithm should be used depends on the data. Single site algorithm is much less time consuming since the cluster algorithm spends time in forming the cluster. The overall computational time for cluster algorithm is expensive when $p>1000$. Plus, in many situations, the mixing time may not be so worse for single site algorithm. Thus we prefer using single site algorithm to achieve the results quickly.
\subsection{Case Three: Bayesian Sparse Additive Model}\label{b.sec8.4}
In this section, we demonstrate variable selection on following Bayesian sparse additive model:
\be\label{b.e51}
y=f_1(x_1)+f_2(x_2)+f_3(x_3)+f_4(x_4)+\epsilon,
\ee
where $x_j=(w_j+tu)/(1+t), j=1,...,p$ and $w_1,...w_p$ and $u$ are iid from Uniform (0,1), and $\epsilon\sim N(0,1.74)$. Therefore $\Cor(\mbx_i,\mbx_j)=t^2/(1+t^2)$ for $i\ne j$. We consider $t=0$ and $t=1$. The later one gives the correlation between two predictors around 0.5. This simulation is similar to Example 1 in \citet{c16a} but with $p=10, 80$ and 150. We also consider sample size $n=100$. Functions $f_j$'s have following forms.
\be\begin{split}\label{b.e52}
f_1(x)&=x,\\
f_2(x)&=(2x-1)^2,\\
f_3(x)&=\sin(2\pi x)/[2-\sin(2\pi x)],\\
f_4(x)&=0.1\sin(2\pi x)+0.2\cos(2\pi x)+0.3\sin^2(2\pi x)+0.4\cos^3(2\pi x)+0.5\sin^3(2\pi x).
\end{split}\ee
As shown in \ref{app6}, for each $\mbx_j$, the LS basis employs two precision parameters $\tau_{ej}=\sigma^{-2}_{ej}$ and $\tau_{dj}=\sigma^{-2}_{dj}$. We treat all set of $\{\tau_{ej}, \tau_{dj}: j=1,...p\}$ independently. Similarly, we can still assign $G(1/2, 1/2)$, $IG(1,1/2)$, or $C^+(1)$ prior for them. However, since $\bet_j$ is the $M_j\times 1$ vector and for each node we have two variance components, the marginal prior for $\beta_j$ given $b$ is no longer simple Cauchy, Laplace or horseshoe prior any more, but it will share the similar properties to its counterpart in linear parametric model. In this simulation, we employ the independent $G(1/2,1/2)$ prior for each $\tau_{ej}$ and $\tau_{dj}$ only. For the number of knots of the LS basis, we may consider each predictor has different number of knots, but it turns out a fixed number for all $M_j$'s, say $M_j=6$, will give good enough results. Therefore, we fix $M_j=6, j=1,...,p$ in this simulation. Totally $6000$ iterations have been employed by the single site algorithm and first $2000$ ones are discarded for all settings.

\begin{figure}[bth]
\begin{center}
\includegraphics[height=80mm]{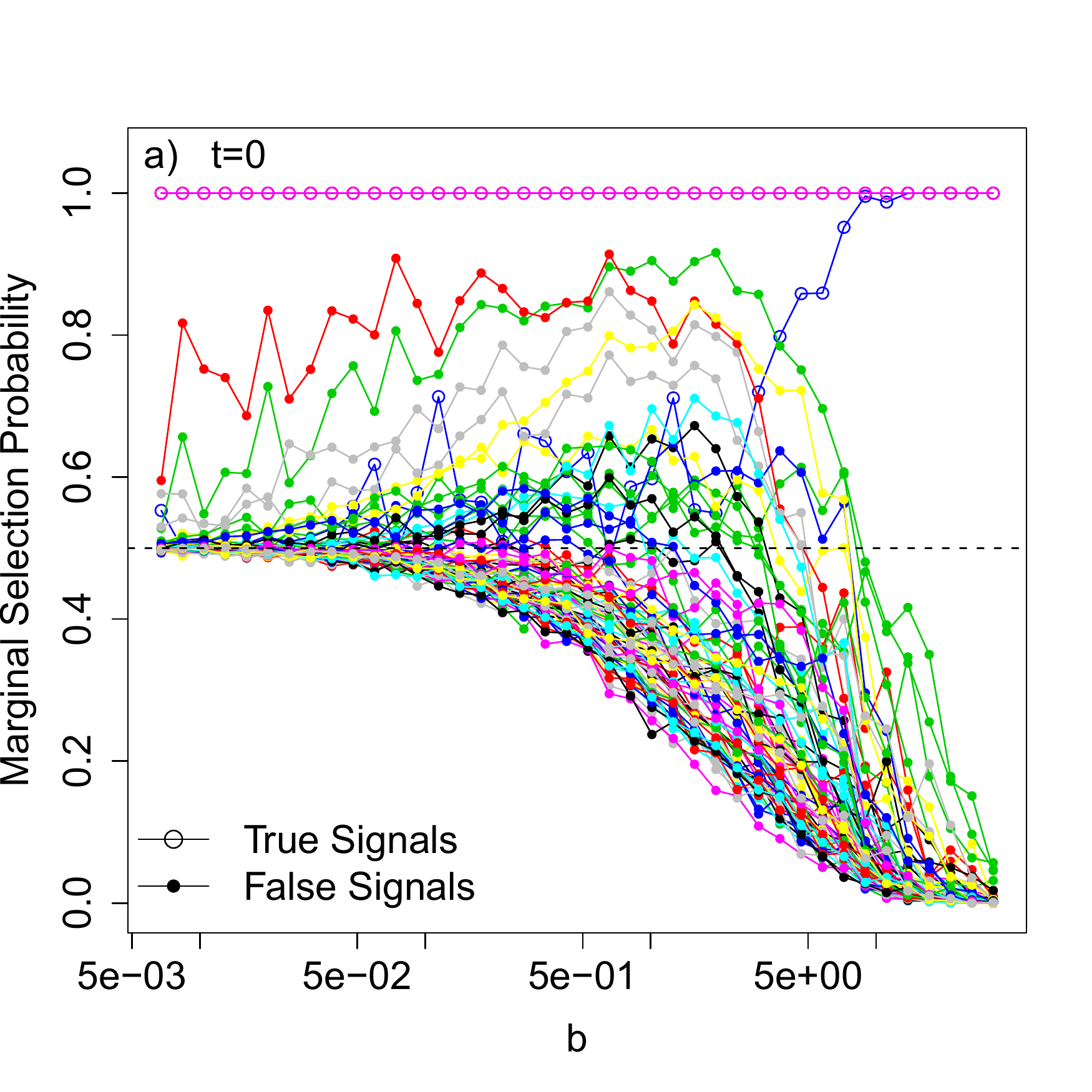}\includegraphics[height=80mm]{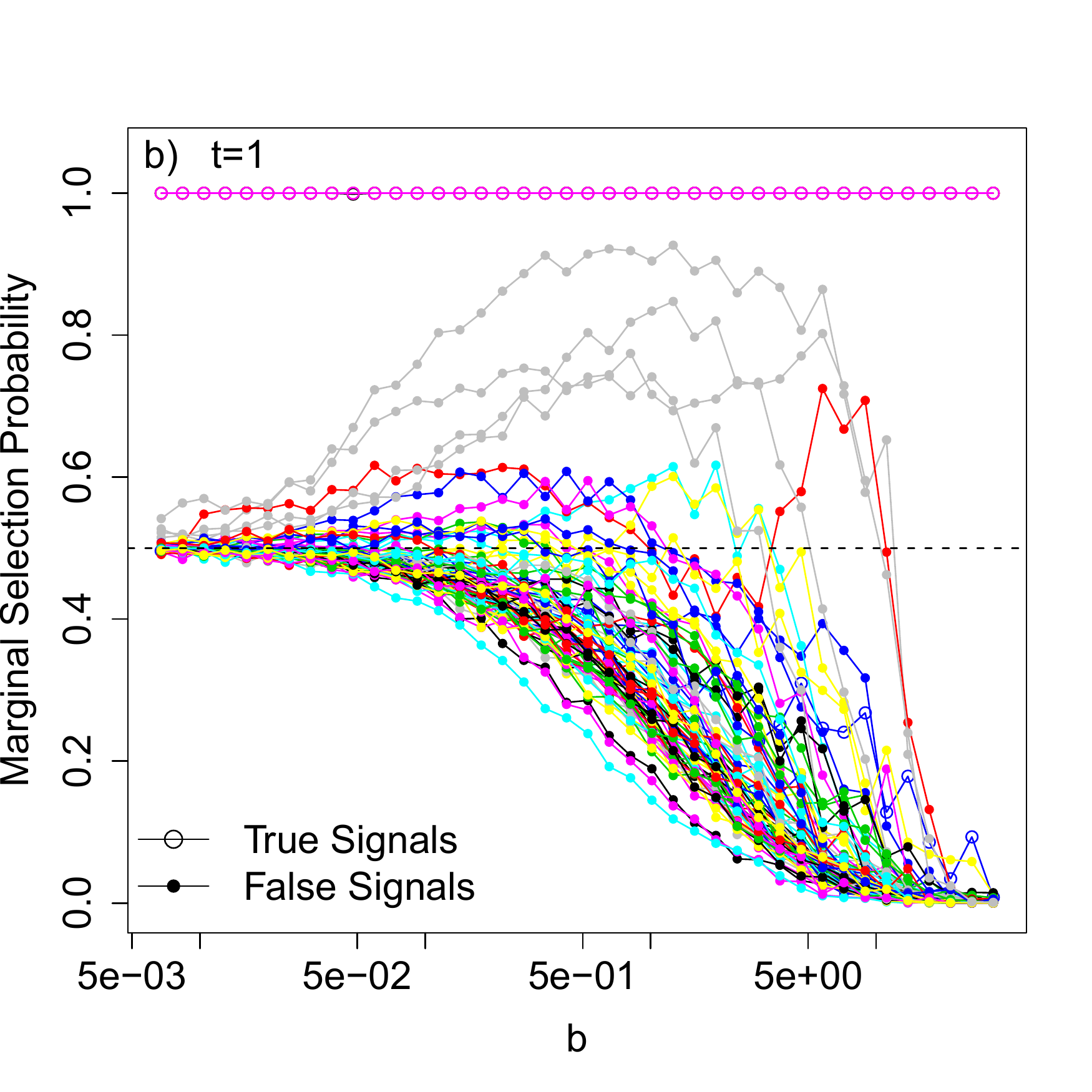}
\caption{The profile curves of the selection probability of simulation model (\ref{b.e51}) with $p=80, n=100$ for independent setting (t=0) (a), and correlated setting (t=1) (b).}\label{b.pic10}
\end{center}
\end{figure}
\begin{figure}[bth]
\begin{center}
\includegraphics[height=100mm]{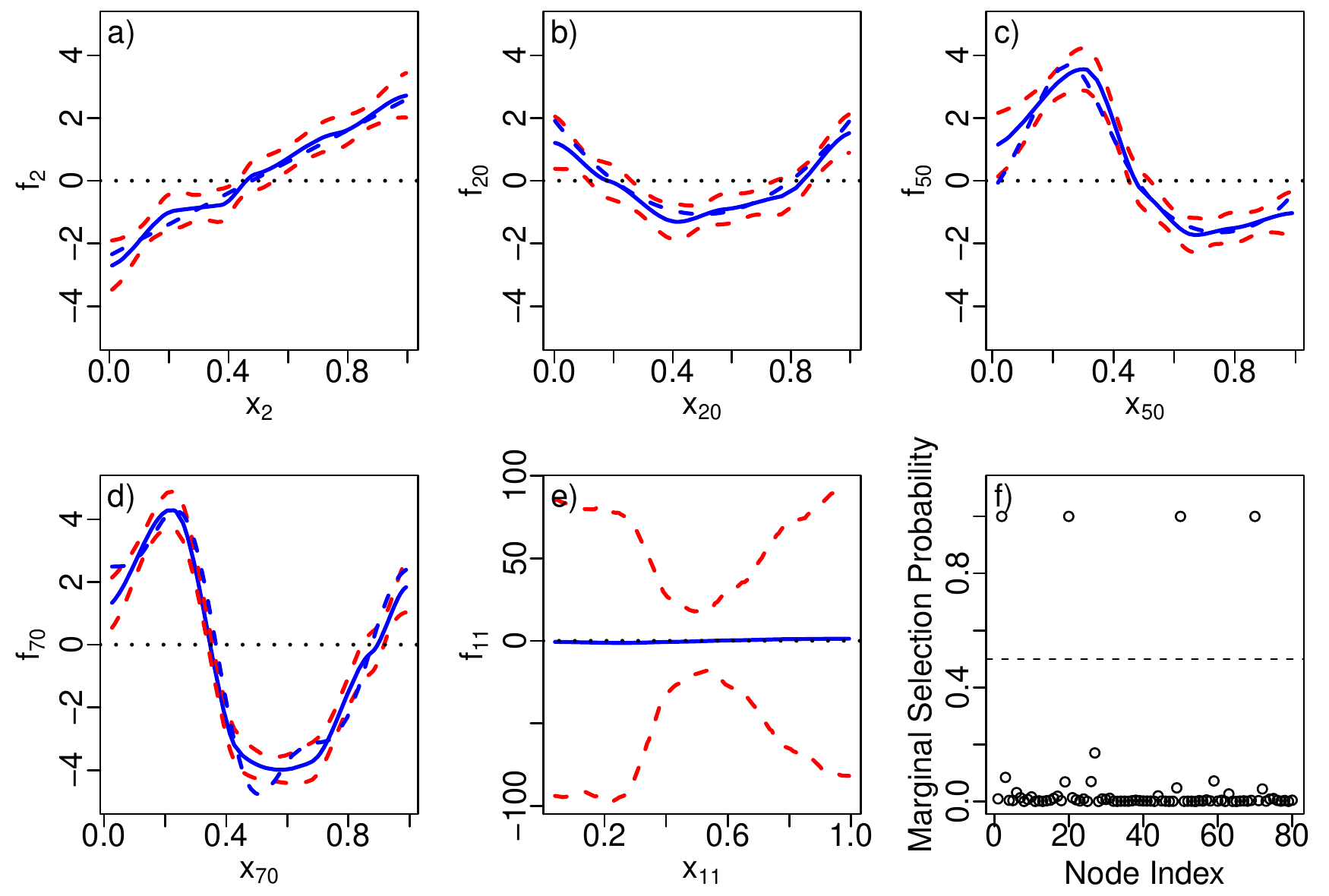}
\caption{True function $f_j$'s (blue dashed lines) and estimated function $\hat f_j$'s (blue solid lines) with $95\%$ credible interval (red dashed lines) for the 4 true nodes (a-d) and a noise node (e) of a run of simulation model (\ref{b.e51}) with independent setting $t=0$ and $p=80$. The marginal selection probability at $b=26$ (f). Note we reordered the first 4 true nodes number to (2, 20, 50, 70) for a better view.}\label{b.pic11}
\end{center}
\end{figure}

\begin{figure}[bth]
\begin{center}
\includegraphics[height=100mm]{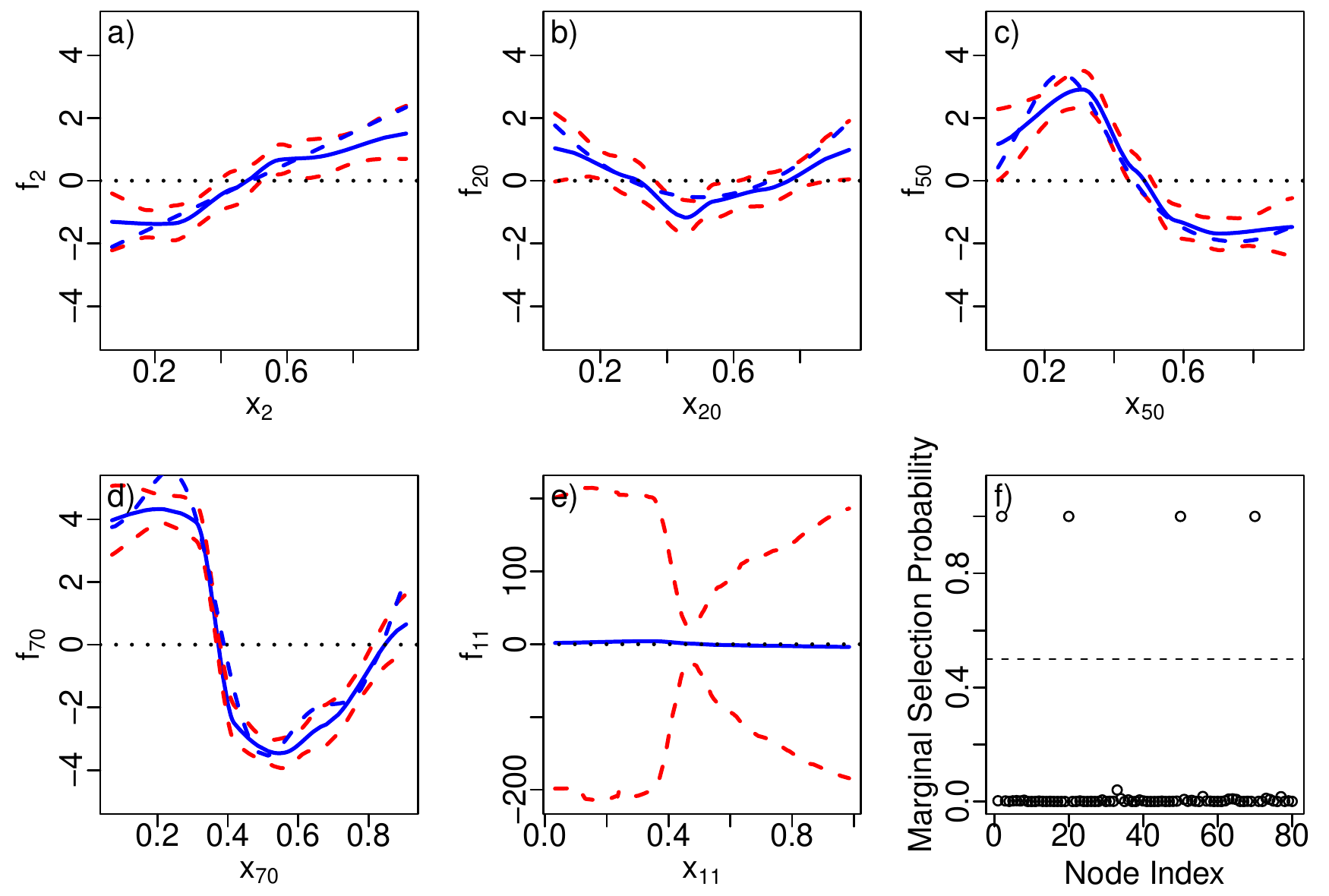}
\caption{True function $f_j$'s (blue dashed lines) and estimated function $\hat f_j$'s (blue solid lines) with $95\%$ credible interval (red dashed lines) for the 4 true nodes (a-d) and a noise node (e) of a run of simulation model (\ref{b.e51}) with independent setting $t=1$ and $p=80$. The marginal selection probability at $b=26$ (f). Note we reordered the first 4 true nodes number to (2, 20, 50, 70) for a better view.}\label{b.pic12}
\end{center}
\end{figure}

\begin{table}[h]
\scriptsize
\centering
\caption{Simulation results of sparse additive model (\ref{b.e51}) for 500 runs.}
\scriptsize
\begin{tabular}{lllccccccc}
\hline\hline
                       &       $t$          &  $p$   & FP-rate    & FN-rate     &  MS         & $f_1$SE   & $f_2$SE   & $f_3$SE    & $f_4$SE         \\
\hline
\multirow{5}{*}{BVGM}  &\multirow{3}{*}{$0$}& $10$   & 0.00(0.02) & 0.00(0.03)  & 3.99(0.17)  & 0.07(0.05)& 0.16(0.06)& 0.18(0.08) & 0.74(0.27) \\
                       &                    & $80$   & 0.00(0.01) & 0.00(0.03)  & 4.16(0.48)  & 0.07(0.05)& 0.15(0.06)& 0.18(0.08) & 0.70(0.27) \\
                       &                    & $150$  & 0.00(0.02) & 0.01(0.04)  & 4.81(5.52)  & 0.08(0.09)& 0.16(0.07)& 0.18(0.11) & 0.73(0.44) \\\cline{2-10}
                       &\multirow{3}{*}{$1$}& $10$   & 0.00(0.01) & 0.09(0.10)  & 3.56(0.54)  & 0.08(0.07)& 0.18(0.09)& 0.16(0.08) & 0.79(0.40) \\
                       &                    & $80$   & 0.00(0.01) & 0.09(0.11)  & 3.68(0.66)  & 0.09(0.08)& 0.18(0.08)& 0.16(0.07) & 0.77(0.39) \\
                       &                    & $150$  & 0.01(0.03) & 0.11(0.11)  & 4.48(7.17)  & 0.10(0.10)& 0.18(0.09)& 0.18(0.10) & 0.80(0.40) \\
\hline
\multirow{5}{*}{COSSO} &\multirow{2}{*}{$0$}& $10$   & 0.00(0.01) & 0.00(0.00)  & 4.00(0.06)  & 0.07(0.04)& 0.05(0.04)& 0.11(0.06) & 0.32(0.13) \\
                       &                    & $80$   & 0.07(0.08) & 0.18(0.07)  & 9.85(11.2)  & 0.17(0.28)& 0.79(0.11)& 1.55(0.32) & 5.28(0.58) \\\cline{2-10}
                       &\multirow{2}{*}{$1$}& $10$   & 0.01(0.03) & 0.04(0.08)  & 3.86(0.48)  & 0.07(0.07)& 0.26(0.10)& 0.14(0.10) & 2.00(1.00) \\
                       &                    & $80$   & 0.10(0.09) & 0.19(0.10)  & 12.5(14.0)  & 0.36(0.40)& 0.37(0.16)& 1.05(0.44) & 4.67(0.54) \\
\hline
\hline
\end{tabular}
\label{t2}
\normalsize
\end{table}
Figure \ref{b.pic10} (a-b) show us how the selection probabilities changes for a range of $b$ with $t=0$ and $t=1$ for simulation setting $p=80$. Note that in Figure \ref{b.pic10} (a), one true signal is buried in the noise till $b=1$, while the same signal is always mixed with noise in Figure \ref{b.pic10} (b). As shown in Figure \ref{b.pic10}, when $b\approx26$ all false signals go to 0 and we achieve the largest gap between signals and the noise.

One feature of our BVGM is the capability to select the important variables as well as estimate the selected function components at the same time. The four true functions and one noise, the corresponding estimated functions, and the selection probability for all nodes at $b=26$ of a simulation run with $p=80$ for $t=0$ and $t=1$ are shown in Figure \ref{b.pic11} and Figure \ref{b.pic12} respectively. Based on the LS basis, the function components estimated are always centered, so we also centered the true functions. In this simulation run, the four true nodes are selected exactly, and the estimated functions of them are calculated by $\hat f_j=Z_jE(\bet_j|\gamma_j=1)$, where the expectation is based on the $N=4000$ iterations, and the $95\%$ credible intervals are plotted as well. As shown in Figure \ref{b.pic11} and \ref{b.pic12}, for both $t=0$ and $t=1$ the estimated functions are very close to the true functions.  Note for the noise function, $f_{11}$, the selection probability is close to zero, thus the estimated function is calculated by $\hat f_j=Z_jE(\bet_j)$, a expectation over all $N$ iterations. This is why we see a very wide credible interval for $f_{11}$ because when $\gamma_j=0$ the posterior of $\bet_j$ is multiple normal with large variance. Also note that to have a better view of the selection probability, we reordered the nodes such that the 4 true nodes are $(2,20,50,70)$.

To further examine the performance of variable selection and estimation accuracy of our method, 500 simulation runs have been employed for $p=10, 80$ and $100$ respectively. We calculated seven statistics: ``False Positive Rate (FP-rate)'', ``False Negative Rate (FN-rate)'', ``Model Size (MS)'', and ``Squared Error (SE)'' of 4 true functions, where $\hbox{FP-rate}={\#{False\,Positive}\over{\#False\,Positive + \#True\,Negative}}$, $\hbox{FN-rate}={{\#False\,Negative}\over{\#False\,Negative + \#True\,Positive}}$, and $\hbox{SE}=\sum_i^n(f_{j,i}-\hat f_{j,i})^2/n$. The estimated function is calculated by $\hat f_j=Z_jE(\bet_j|\gamma_j=1), j\in\hbox{\{true nodes\}}$. Since it can happen that $p(\gamma_j=1|\mby)=0$ for any true function components, we simply estimate $f_j$ by $\hat f_j=0$ for the 4 true nodes if $p(\gamma_j=1|\mby)=0$ in each run. Statistics SE can be used to assess the accuracy of the estimation of the nonlinear function $f_j$ because the smaller the SE the closer the estimation $\hat f_j$ to true function $f_j$. The average and standard deviation of those statistics over 500 runs are reported in Table \ref{t2} and compared with Component Selection and Smoothing Operator (COSSO) \citep{c16a}.

As shown in Table \ref{t2}, the results for our method is pretty robust to $p$. For each $t$, all statistics are similar for different $p$ except a little increase in the mean and standardized deviation of those statistics. For different $t$, our method is also pretty robust, except the increase in the values of FN-rate and SE's. On the other hand, we can see COSSO only performs well for small $p$. When $p=80$ (COSSO can not work for $n<p$ case, so no result for $p=150$), all the statistics of COSSO increase, especially for those four true function components, SE's are very large meaning COSSO can not estimate those function components correctly.  In general, we can see our method works very well for BSAM even for large $p$ and large correlation cases in both variable selection and function component estimation.
\subsection{Case Four: Linear Chain Prior}
Again, we consider the same form of model (\ref{b.e48}) but with setting  $p=100, n=100$, $\beta_j=0.4$ if $j$ is odd, $\beta_j=0.8$ if $j$ is even, and $S=\{1,2,...,15\}$. This example is special in sense of its true predictor set $S$ and false signal set $\bar S$ both are continuous in their node index. Obviously, the simplest prior network information is a linear chain: any node's two neighbors are most likely to aligned to this node. Although this is not true for neighbored node 15 and 16, but this discontinuity has small effect on the whole system. With this knowledge, we would consider the linear chain prior for the nodes. $W=\{w\lambda_{ij}\}$, $\lambda_{ij}=1$ for $|i-j|\le1$ and $\lambda_{ij}=0$ otherwise. In order to have a exchangeable prior, two boundary nodes 1 and $p$ can be treated as neighbors, i.e., $\lambda_{ij}=1$ if $|i-j|=p-1$. To fully use this prior information, we employ the cluster algorithm and using this adjacency matrix $\Lambda={\lambda_{ij}}$ to form the cluster. We also compare the results with noninformative prior (employed by single site algorithm).

\begin{figure}[bth]
\begin{center}
\includegraphics[height=50mm]{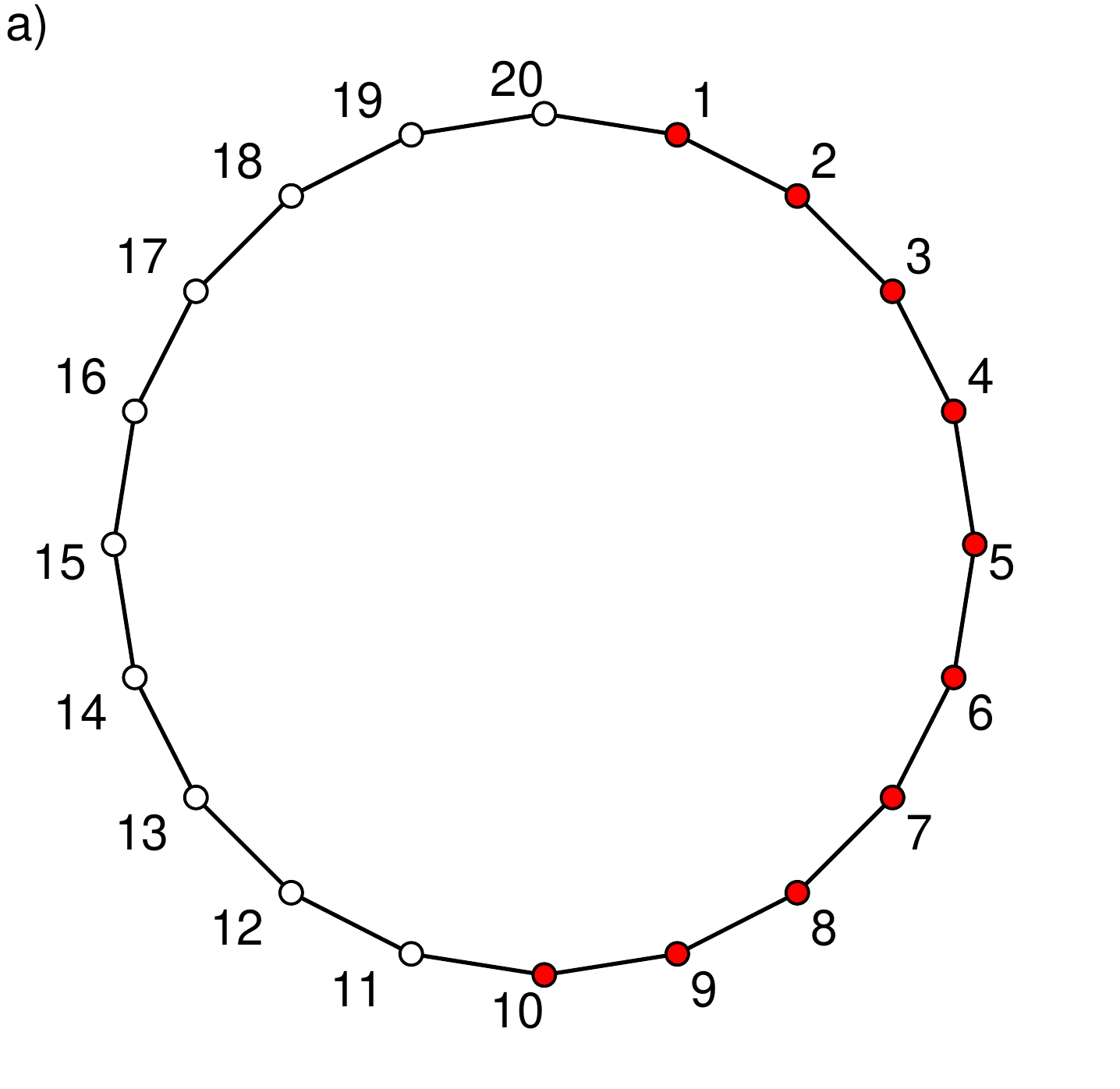}\includegraphics[height=50mm]{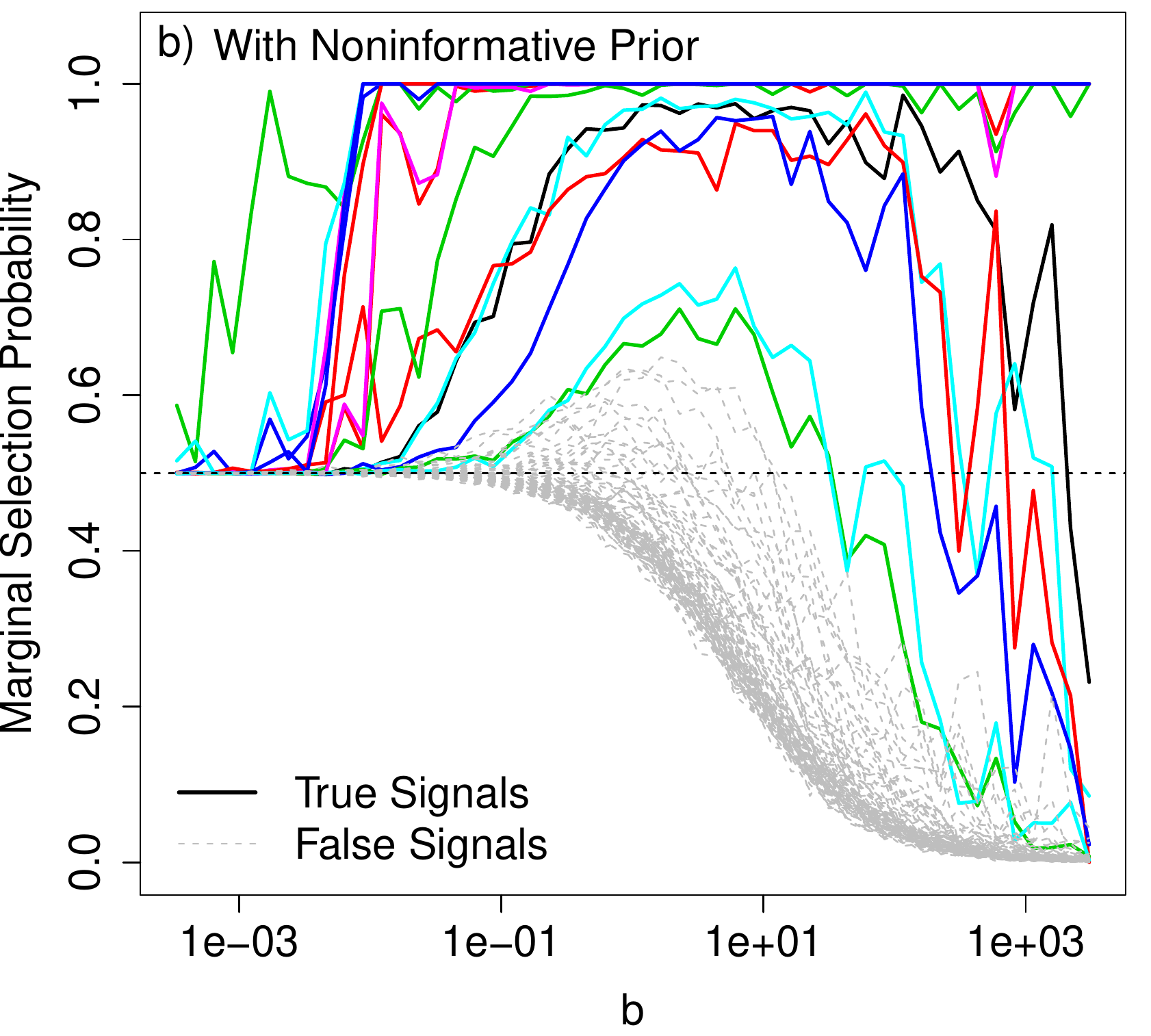}\includegraphics[height=50mm]{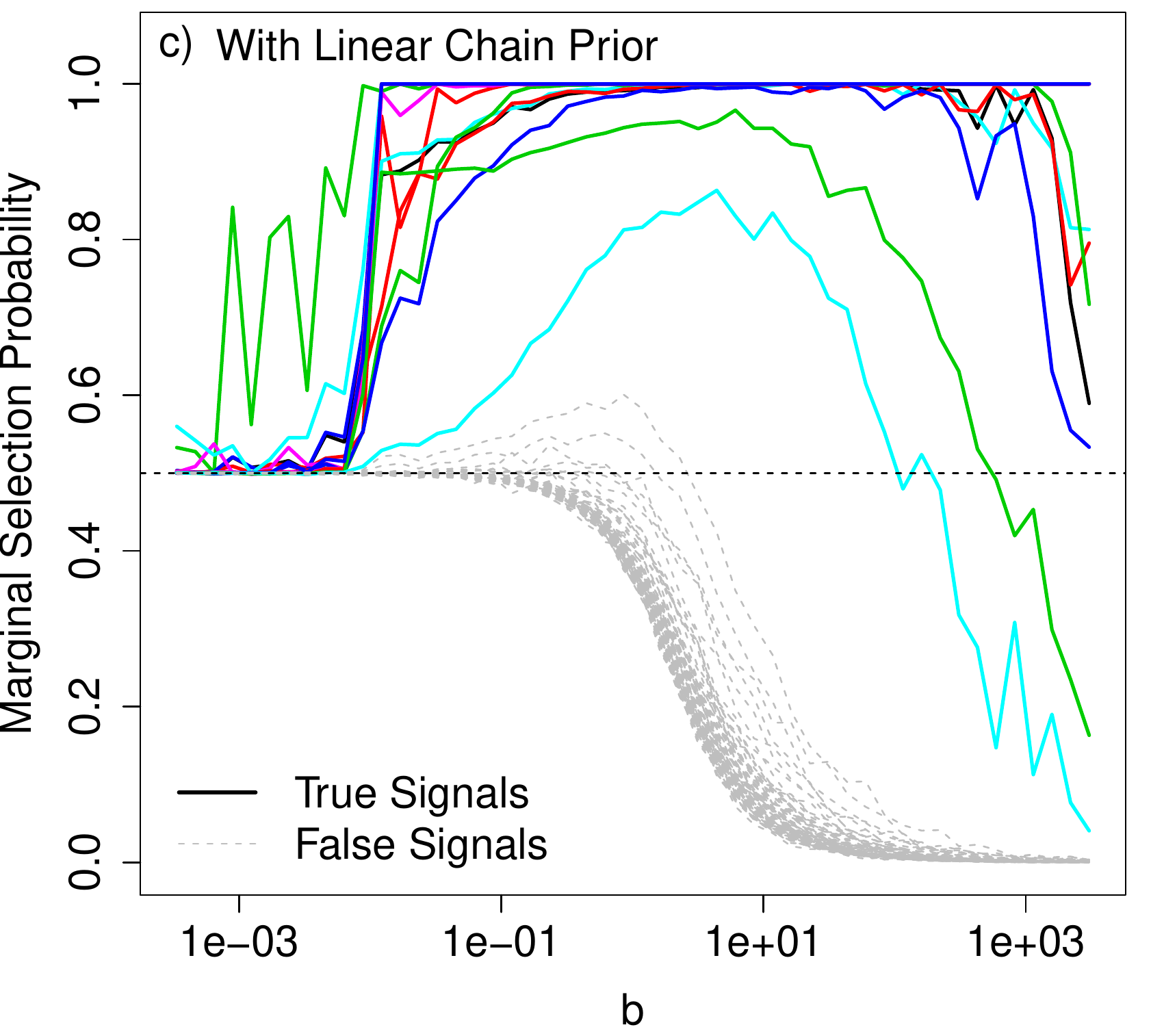}
\caption{The graph of a linear chain prior with 20 nodes with 1 through 10 nodes ``in'' (a). The profile curves of the selection probability of case four model calculated by the cluster algorithm with noninformative prior (b), and with the linear chain prior \label{b.pic13} for $\bga$ (c).}
\end{center}
\end{figure}

Figure \ref{b.pic12} (a) shows a example graph of a linear chain with $p=20$ nodes, note how the two end nodes connected. Figure \ref{b.pic13} (b-c) are the selection probability profile plot with noninformative prior and linear chain prior. For each $b$ in both plots, total 6000 iterations have been employed with first 2000 burn-in. For the linear chain prior we take $v=\Phi(\log(b))$ where $\Phi$ is the standard normal CDF such that the interaction strength vanishes for large shrinkage and maintains at 1 for small shrinkage. The difference of two plots is obviously: with the noninformative prior, two true signals are very close to the noise for all range of $b$ and hard to be separated from the false signals, while with the linear chain prior, we can see for a large range of $b$ all the true signals are well distinguishable from the noise.

\section{Real Data Analysis}\label{b.sec9}
\subsection{Ozone Data}\label{b.sec9.1}
As an illustration of BSAM implemented by BVGM, we consider an example, the ozone data analyzed by \citet{c16a}. The ozone data is available in R package  \verb"cosso" or  \verb"gss". In the Ozone data, the interesting response variable is the daily maximum one-houraverage ozone concentration and eight meteorological variables were recorded in the Los Angeles area for 330 days in 1976. The sample size $n=330$, and the 8 variables are Height (Hgt), Wind Speed (WS), Humidity (Hum), Temperature (Temp), Inversion Base Height (InvHt) , Pressure (Press), Inversion Base Temperature (InvTp),  and Visibility (Vis). All predictors were standardized and the response was transformed using logarithm to have normal distributed response. We applied the Bayesian graph model described in Section \ref{b.sec5.4} with $M_j=6$ for all predictors. Total 20000 iterations have been employed with single site algorithm and half of them were discarded as burn in. By quickly examining a series of $b$, we see the selection probability profile curves are all well defined due to small variable number (not shown). So it is more appropriate to choose a modest shrinkage, which is $b=1.6$ in this case such that all selected predictors reach their highest selection probability. At $b=1.6$, two predictors have selection probability less than 0.5.
\begin{figure}[bth]
\begin{center}
\includegraphics[height=100mm]{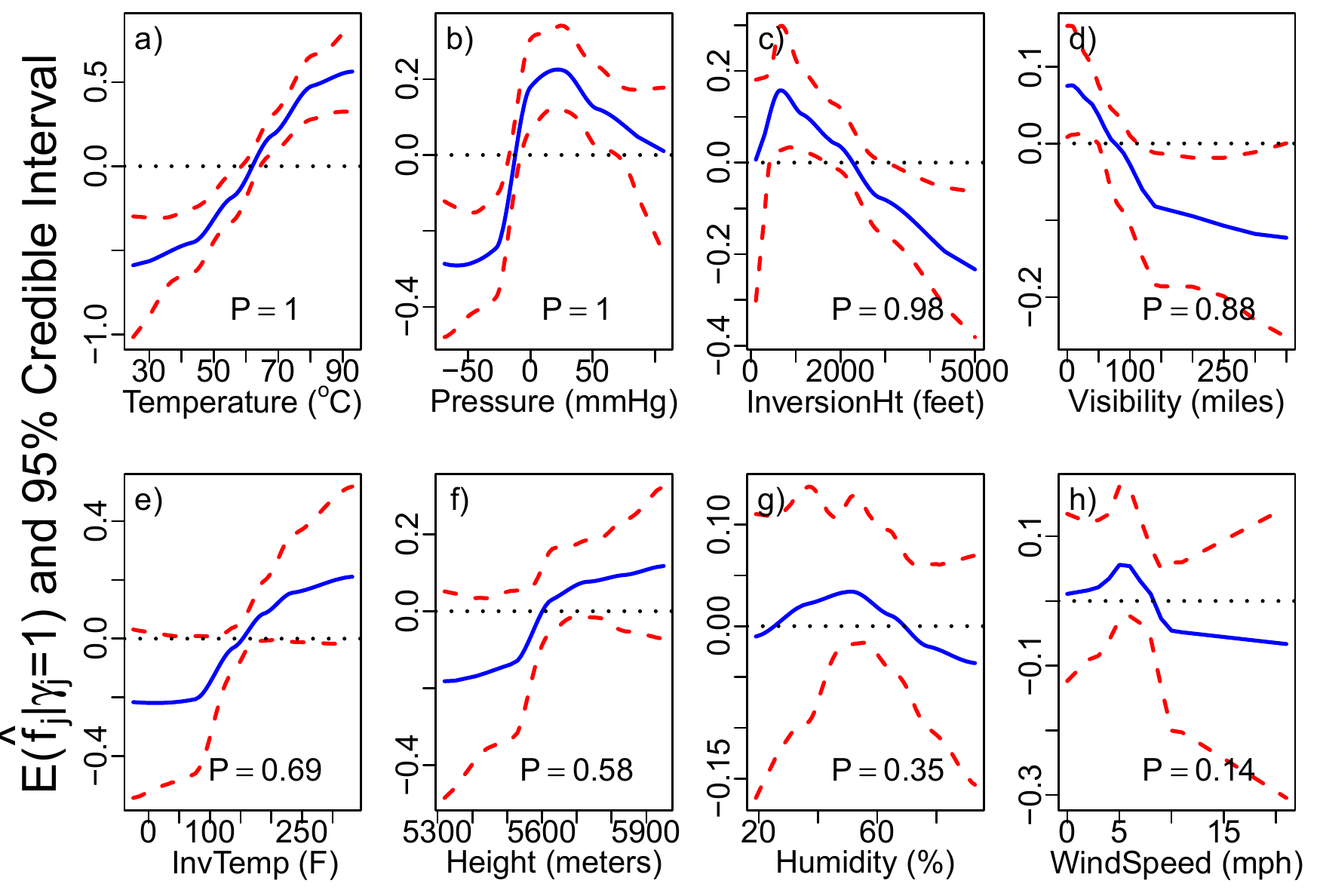}
\caption{Estimated function $\hat f_j$ (blue solid lines) with $95\%$ credible interval (red dashed lines) for the 8 predictors of ozone data labeled by the marginal selection probability $P=p(\gamma_j=1|\mby)$ at $b=1.6$. }\label{b.pic14}
\end{center}
\end{figure}

The estimated results for $\bet_j$'s are summarized in Figure \ref{b.pic14}, where the additive function components, $\hat f_j=Z_jE(\beta_j|\gamma_j=1)$'s, are plotted with 95\% credible interval. Because the smallest $P=p(\gamma_j|\mby)$ is at least 0.14, we have enough iterations for all $\gamma_j=1$ to estimate all $\hat f_j$. The marginal selection probability, $P$, for each variable is labeled in each plot. We can then identify three groups of the variables. The first group has ate least $P=0.88$ including Temp, Press, InvHt and Vis. The second group includes InvTp and Hgt with $P=0.69$ and $0.58$. The last group contains Hum and WS with $P$ smaller than 0.5. In variable selection point of view, we will select all variables in the first group surely, and we will not select the last group since their selection probabilities are very close to the baseline. Because of the small variable number and nearly independence of each variable, we can consider the second group as true variables with small signals. In the point view of function components estimation, the selection probability is consistent to it signal estimation. As shown in Figure \ref{b.pic16}, for the first group of variables the credible intervals only cover a small part of the zero line, while for the third group of variables, the zero line is almost in the center of the credible interval. Although the credible intervals of the second group of variables cover the zero line totally, the zero line is close to the edges of the credible interval.

We also report the summary statistics for all variance components (their inverses) and the intercept in Table \ref{t4}. In this example we include the intercept term $\mu$ in model (\ref{b.e51}) and assign a prior for $\mu$: $[\mu]\sim N(0,\tau_{\mu}^{-1})$, and a Gamma prior for $\tau_{\mu}$: $[\tau_{\mu}]\sim G(4,2)$. The full conditional distributions for $\mu$ and $\tau_{\mu}$ are easy to derive (not shown here). Note how the posterior means of these parameters adapt to the data. Especially for $\tau_{ej}$'s and $\tau_{dj}$'s, all start with the same prior, but the posterior means are different. For the first group variables, their posterior means for $\tau_{ej}$'s and $\tau_{dj}$'s are obviously different from their priors.

\begin{table}[h]
\scriptsize
\centering
\caption{Parameter estimation of ozone data under Bayesian sparse additive model at $b=1.6$.}
\scriptsize
\begin{tabular}{lllcccccc}
\hline\hline
 \multirow{2}{*}{Parameter}& \multicolumn{2}{c}{Prior}&&  \multicolumn{5}{c}{Posterior}               \\\cline{2-3}\cline{5-9}
                     &Mean   &Std.Dev.         && Mean  &Std.Dev.  & Median &Lower 2.5\%&Upper 2.5\%   \\
\hline
$\tau_e^{Temp}$      &1.000  &1.414            && 0.671 & 1.192    & 0.150  & 0.003     & 4.199      \\
$\tau_e^{Press}$     &1.000  &1.414            && 0.787 & 1.270    & 0.282  & 0.008     & 4.570      \\
$\tau_e^{InvHt}$     &1.000  &1.414            && 0.198 & 0.597    & 0.021  & 0.000     & 1.700      \\
$\tau_e^{Vis}$       &1.000  &1.414            && 1.316 & 1.517    & 0.805  & 0.021     & 5.485      \\
$\tau_e^{InvTp}$     &1.000  &1.414            && 1.010 & 1.365    & 0.500  & 0.010     & 4.884      \\
$\tau_e^{Hgt}$       &1.000  &1.414            && 1.062 & 1.392    & 0.547  & 0.005     & 5.032      \\
$\tau_e^{Hum}$       &1.000  &1.414            && 1.050 & 1.431    & 0.509  & 0.001     & 5.129      \\
$\tau_e^{WS}$        &1.000  &1.414            && 1.104 & 1.505    & 0.545  & 0.001     & 5.334      \\
$\tau_d^{Temp}$      &1.000  &1.414            && 0.226 & 0.724    & 0.026  & 0.002     & 2.212      \\
$\tau_d^{Press}$     &1.000  &1.414            && 0.033 & 0.047    & 0.021  & 0.002     & 0.125      \\
$\tau_d^{InvHt}$     &1.000  &1.414            && 0.889 & 1.352    & 0.349  & 0.004     & 4.731      \\
$\tau_d^{Vis}$       &1.000  &1.414            && 0.751 & 1.235    & 0.227  & 0.004     & 4.226      \\
$\tau_d^{InvTp}$     &1.000  &1.414            && 1.122 & 1.413    & 0.611  & 0.005     & 5.227      \\
$\tau_d^{Hgt}$       &1.000  &1.414            && 1.045 & 1.429    & 0.521  & 0.009     & 5.140      \\
$\tau_d^{Hum}$       &1.000  &1.414            && 1.061 & 1.474    & 0.513  & 0.001     & 5.048      \\
$\tau_d^{WS}$        &1.000  &1.414            && 0.949 & 1.383    & 0.404  & 0.000     & 4.985      \\
$\phi$               &-      &-                && 6.602 & 0.542    & 6.585  & 5.599     & 7.681     \\
$\mu$                &0.000&$\sqrt{\tau_{\mu}}$&& 2.143 & 0.066    & 2.145  & 2.011     & 2.265     \\
$\tau_{\mu}$         &2.000  &1.000            && 1.043 & 0.495    & 0.971  & 0.316     & 2.205     \\
\hline
\hline
\end{tabular}
\label{t4}
\normalsize
\end{table}
\subsection{Gene Selection in Pathway Data}\label{b.sec9.2}
\citet{b21} presented an pathway based analysis  to test a priori defined pathways for association with the diabetes disease. A pathway is a predefined set of genes that serve a particular cellular or physiological function. Therefore a genetic pathway can be expressed by a graph to prrsent the gene network within this pathway. \citet{b21} identified several significant pathways among which ``Oxidative phosphorylation'', ``Alanine-and-aspartate metabolism'' et al. are interesting ones. However, even with those significant pathways identified,  gene selection in microarray data analysis is still difficult because alterations in gene expression are modest due to the large number of genes, small sample sizes and variability between subjects. \citet{c31} provide a Bayesian technique to incorporate biological information into linear models to select genes and pathways. Similar to \citet{c31}, we also incorporate the pathway network information into our graph model, and apply it to gene selection of the diabetes data from \citet{b21}. However, in our method, we use the gene network information in the pathway as the prior for $\bga$, and we don't select pathways. The data contains gene expressions from $n=35$ subjects, 17 normal and 18 Type II diabetes patients. We merged three interesting pathways, ``Oxidative phosphorylation'', ``Alanine-and-aspartate metabolism'' and ``Glutamate-metabolism'' into one graph with total $p=173$ nodes (some nodes are different probe sets of the same gene, so the gene names are identical) which is a subgraph of the corresponding merged graph obtained from KEGG database. The response $\mby$ is the continuous glucose level.
\begin{figure}[bth]
\begin{center}
\includegraphics[height=80mm]{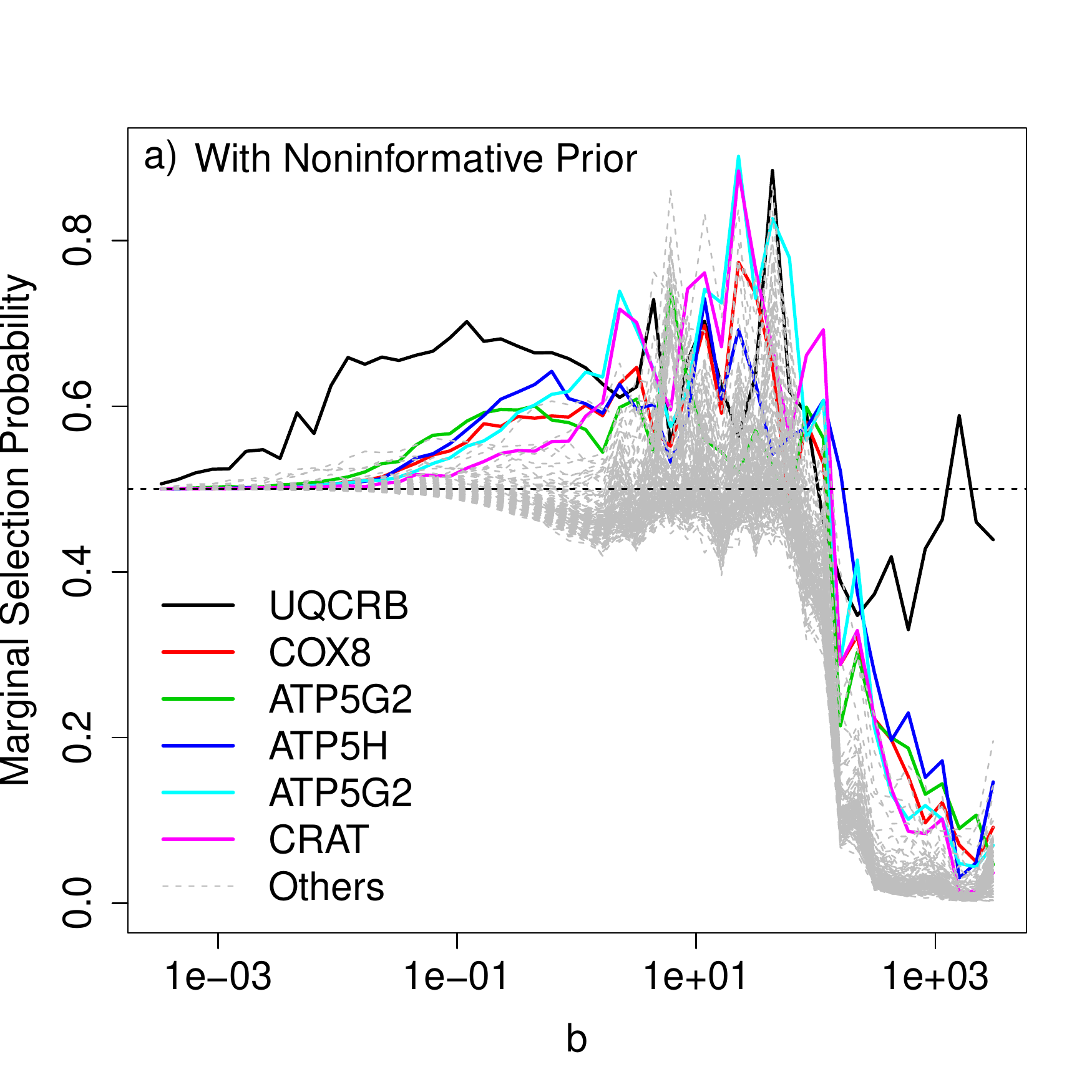}\includegraphics[height=80mm]{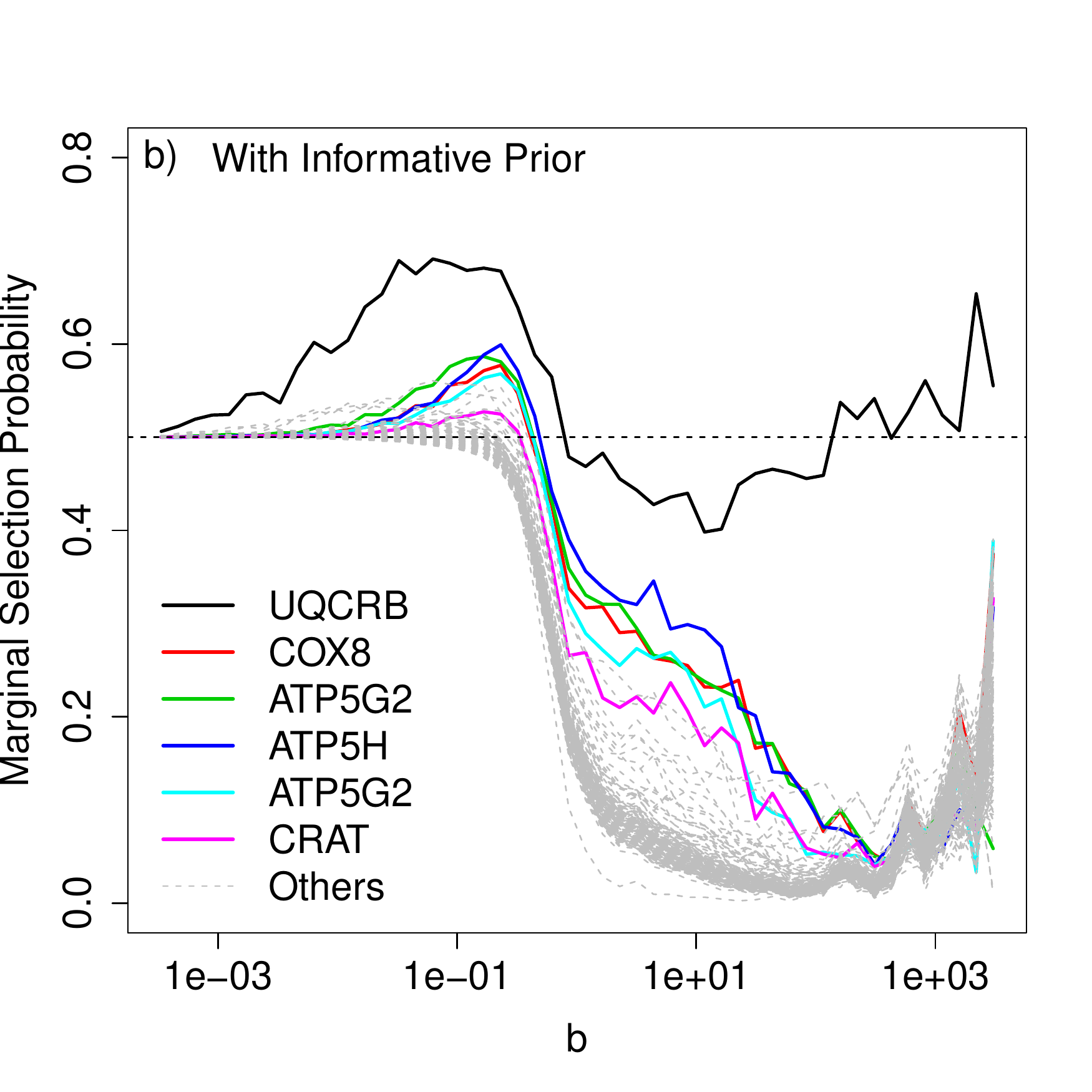}
\caption{Profile curves of the selection probability of genetic pathway data with noninformative prior for $\bga$ (a), and with informative prior as (\ref{b.e54}) (b).}\label{b.pic15}
\end{center}
\end{figure}

\begin{figure}[H]
\begin{center}
\includegraphics[height=160mm]{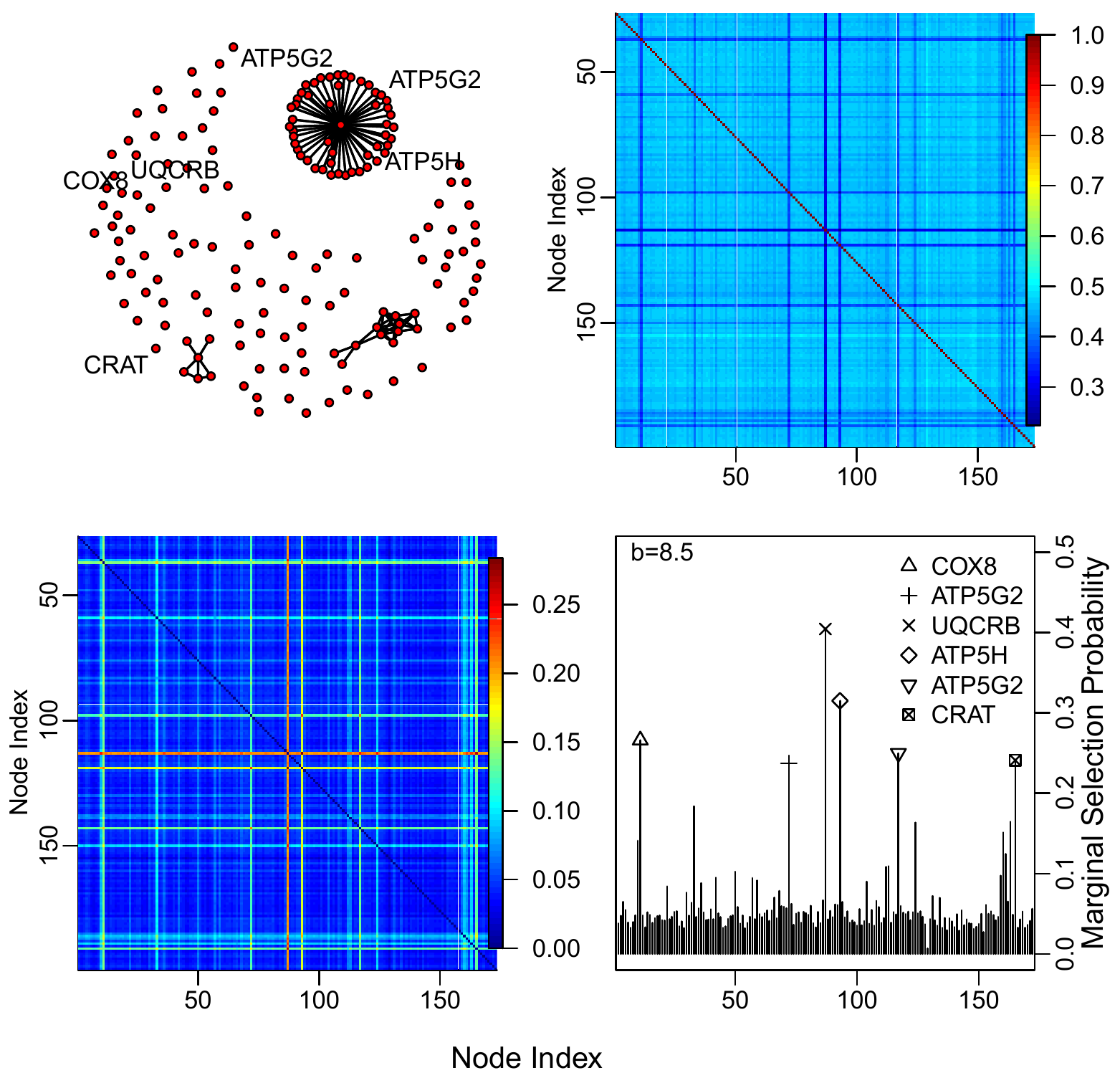}
\caption{Summary of the results for the genetic pathway data. Top left: genetic network structure of the data. Top right: the frequency matrix of two nodes aligned in the cluster over total iterations. Bottom left: the frequency matrix of two nodes anti-aligned in the cluster over total iterations. Bottom right: Selection probability with cluster algorithm at $b=8.5$ with informative prior (\ref{b.e54}).}\label{b.pic16}
\end{center}
\end{figure}

The top left plot of Figure \ref{b.pic16} shows the network of our merged gene set. Note, the prior required for our graph model is undirected graph with positive interaction only. We can see, most of the nodes are independent in this data set, and there are only three genetic clusters. Because of this, if we apply the cluster algorithm and use the adjacency matrix $\Lambda$ based this network information into expression (\ref{b.e43}), we will end up with a few nodes in the same genetic cluster that can form the clusters for the algorithm. Therefore, we consider following interaction matrix $W=\{w_{ij}\lambda_{ij}\}$ for the prior of $\bga$ with adjacency matrix $\Lambda=\{\lambda_{ij}\}$ as

\be\begin{split}\label{b.e54}
\lambda_{ij}&=1,\; i,j=1,...,p\\
w_{ij}&=\left\{\begin{array}{l l}
w,                &i\not\in S\hbox{ or }j\not\in S \\
w+\Delta w,             &i\in S \hbox{ and }j\in S ,\end{array} \right.\\
\end{split}\ee
where $S$ represents one of the three genetically networked gene clusters in the pathway network, $w$ and $\Delta w$ are small positive numbers stand for the strength of the interaction in the prior and the difference of two types of interaction. If $\Delta w=0$, we can consider (\ref{b.e54}) as a baseline graph prior for $\bga$, which is a complete graph with positive fixed interaction. Since we also vary $b$ to have an overall view about the selection probability, it is necessary to have $w\rightarrow0$ when $b\rightarrow0$ since with large shrinkage $J_{ij}\rightarrow0$ and we don't want $w_{ij}$ dominates the graph interaction. One convenient way is to express $w$ as $w=w_0\Phi[\log(b)]$ which approaches 0 as $b\rightarrow0$ and reaches the maximum $w_0$ for large $b$, where $\Phi(\cdot)$ is the CDF of standard normal. Note that the choice of $w$ is involved in the consideration of so called phase transition \citep{c15}. If $w$ is too large all the nodes will always be connected which leads to either all nodes are selected or none are selected. Now we consider $\Delta w\ne0$, say $\Delta w=5w$, so we incorporate the genetic network information into the graph prior. $\Delta w$ can not be too large, otherwise those genes in the genetic clusters will always be aligned which means in this data set they will all have small selection probability. So we choose the $w_0\approx0.01$ as small as possible to avoid the phase transition phenomena, but it is must be large enough to reduce so called region II of $b$ caused by small sample size. With this selection, we have the prior interaction $w_{ij}\approx0.01$ for two nodes not in the genetic cluster together for large $b$, and $w_{ij}\approx0.06$ for two nodes in the cluster for large $b$.

As shown in Figure \ref{b.pic15}, the effect of incorporating prior information for the graph model is obvious. We run the cluster algorithm for total $N=40000$ iterations, and discarded the first $10000$ as burn-in. So the selection probability is calculated by taking the mean of $\bga$ over $30000$ iterations. In Figure \ref{b.pic15} (a), with noninformative prior for $\bga$, we can see even though we are still able to identify several genes behaving differently from the rest (highlighted by solid lines in the plot), for the moderate value of $b$ all the curves are mixed. On the other hand, in Figure \ref{b.pic15} (b), with informative prior for $\bga$ defined as (\ref{b.e54}) the profile curves are much ``cleaner'' even for moderate $b$. Around $b=10$  we can see a bunch of curves are clearly distinguishable from the rest. We highlighted 6 nodes with highest selection probability around $b=10$ in Figure \ref{b.pic15} (b).

To examine more details of the results, in Figure \ref{b.pic16} we fixed $b=8.5$ and run the cluster algorithm for $N=60000$ iterations with first $20000$ discarded. With this shrinkage parameter, the prior interaction parameter $w_{ij}\approx0.06$ for $i$ and $j$ in the genetic cluster, and $w_{ij}\approx0.01$ otherwise. The selection probability for all nodes are shown in bottom left of Figure \ref{b.pic16} where we take a cut-off probability as $0.2$ and identify 6 nodes that have relative high selection probabilities. Among those nodes, UQCRB has the largest selection probability for all range of $b$, so it is easy to identify UQCRB as the most significant gene. We also select other five genes, COX8, ATP5G2 (two probe sets), ATP5H and CRAT at $b=8.5$. All the genes selected except CRAT are from ``Oxidative phosphorylation'' pathway which is related to ATP synthesis. It is well known ATP plays a importance role in Type II diabetes disease. CRAT is from ``Alanine-and-aspartate metabolism'' pathway. Both  ``Oxidative phosphorylation'' and ``Alanine-and-aspartate metabolism'' pathway are two top significant pathways identified using random forrest tree approach \citep{c23}.

Since our cluster algorithm forms the Wolff cluster at each iteration, a byproduct of the MCMC sampler is the frequency of two nodes being aligned or anti-aligned when they are in the cluster. The top right plot in Figure \ref{b.pic16} is the heatmap matrix of the frequency of two nodes being aligned in the cluster out of $40000$ iteration, and bottom left plot is the frequency of two nodes being anti-aligned in the cluster. The color bar of two plots shows the scale of the frequency, the darker the color the lower the frequency. In the top right plot, the dark colored lines are those genes have lower chance to be aligned to the others when they form the cluster, and in the bottom left plot, the bright colored lines are the same genes but with high chance to be anti-aligned to others if they form the cluster. Note those lines are consistent to the genes with high selection probability in the bottom right plot. This is because most of the genes have low selection probability around 0.05 then those genes with higher selection probabilities should have lower (higher) chance to be (anti-)aligned with them. For individual node, we define it is always self-aligned, so the diagonal in top right plot is 1, meanwhile an individual node is never anti-aligned to itself, so diagonal in bottom left plot has value 0. The distinguishable color of those genes in two heatmaps show that we can also use the cluster information to distinguish genes.

So far, we identify 6 genes (probe sets) with cut-off probability $0.2$, we may decrease the cut-off to select more genes. However, the selection probabilities are low for most of the genes except for UQCRB at fixed $b$. This is because of the problem of modest alterations for single gene selection, or it simply means the signals are weak. Here we selected those genes not only depending on the selection probability at fixed $b$, in stead we select them by examining their overall profile as shown in Figure \ref{b.pic15}. We  also demonstrated that the graph model variable selection can easily adopt the prior graph information, thus we can consider similar approach as \citet{c31} to select networked pathways, which may result in higher selection probability for pathways at the optimal shrinkage parameter $b$.

\section{Discussion}\label{b.sec10}
The goal of this paper is to present BVGM from two major aspects. The first is how to sample the ``in'' or ``out'' binary random variable. We pointed out that Bayesian variable selection can be considered as the binary random process on a complete graph given noninformative prior for $\gamma_j$'s, and we compared the single site and generalized Wolff cluster updating algorithm.  Another one is how to construct the interaction matrix of the complete graph, which is implemented by sampling the linear model coefficient $\beta_j$'s through the scale mixtures of normal priors. We also discussed the marginal selection probability profile under different shrinkage parameter and compared three prior settings for $\beta_j$ which represent three typical situations of shrinkage proportion. Our BVGM method possesses the advantages of simple form, easy implementation and straightforward to extension. For example, the BVGM is very easy to extend to Bayesian sparse model by representing the nonparametric function components $f_j$ as linear combination of the basis matrix $f_j=Z_j\bet_j$, then we can employ the group selection of vector $\bet_j$. Another example is to incorporate network information for $\bga$. Although this paper does not focus on how to construct the prior network structure information, the simulation and real data analysis show that it is easy to incorporating the prior graph information and improve the performance of BVGM. This paper also systematically studies the behaviors of the marginal selection probability against the shrinkage. Both theoretical and simulated results show that to have the largest gap between the signals and the noise it is critical for the scale mixture normal prior to maintain substantial proportion on small shrinkage.

However, this paper only starts a different view angle about Bayesian variable selection, further research includes but are not limited to following questions.
\begin{enumerate}
\item As shown in Theorem 3 we have $E(\beta_j|\mby,\gamma_j=1)={d\over{da_j}}\log\pi^b_j$ for orthogonal design. This equation reveals the relationship between the selection probability and the signal magnitude: the larger the signal, the further the selection probability profile being separated from the baseline. However, this relationship does not provides a cut-off rule to separate the signals from the noise. As we can see in the paper, at different shrinkage parameter $b$, the selection probabilities are different. Thus at different shrinkage parameter, the cut-off line should be different too, and we simply choose $b$ where there is a largest gap between two set of signals. However, $p(\gamma_j|\mby)>0.5$ does not mean the corresponding predictor should be selected, such as for $b\rightarrow0$ many noise predictors have selection probability no less than 0.5; and $p(\gamma_j|\mby)\rightarrow0$ does not guarantee the corresponding predictor should be removed, since all predictors have zero selection probability for very large $b$ with orthogonal design. Further research should show the consistency of selecting the predictors based on the ``largest gap'' rule, and provides more straightforward method to choose $b$.
\item Limitation of fixing $b$. The global shrinkage or temperature parameter $b$ is fixed, which limits the performance of our method since it limits the range of local shrinkage parameter. Assigning prior for $b$ is not appropriate too due to the high dimensionality such that the posterior distribution of $b$ is forced to be very small. To automatically have $b$ large and small at the same time, we can adopt a remedy similar to exchange Monte Carlo by running parallel MCMC's at two or more $b$'s with some $b$ small and the other large, and exchanging their configuration according to certain probability satisfying the detailed balance. This remedy may improve the performance of results.
\item Bayesian sparse additive model with interaction. Our BSAM does not include the interaction terms, but it should be easy to extend to include them. The only problem is figure out how to represent the interaction function components. This can be done under the spline ANOVA models similar to \citet{c28}.
\item Prior graph information. With a known networked graph prior, we have better performance in some cases because the prior reduces the searching space for $\bga$. However, there is no way to have exact knowledge about the network prior information for the predictors, and it is difficult to construct a meaningful network as the prior. So this keeps a open question as discussed by \citet{c15} and \citet{c20}.
\end{enumerate}

\label{lastpage}
\Appendix
\numberwithin{equation}{subsection}
\section{}
\subsection{Proof of Theorem 1}\label{app1}
In general, in the Markov chain of MH algorithm, the move from current state $\bga_c^0$ to the proposed state $\bga_c^*$ in the cluster has the transition probability, $P(\bga_c^0\rightarrow\bga_c^*)$, which satisfies the detailed balance condition
\be\label{b.e55}
{{P(\bga_c^0\rightarrow\bga_c^*)}\over{P(\bga_c^*\rightarrow\bga_c^0)}}={{p(\bga_c^*|\mby,\bet,\phi)}\over{p(\bga_c^0|\mby,\bet,\phi)}}=\exp\left\{-\left[U(\bga_c^*)-U(\bga_c^0)\right]\right\}.
\ee
The transition probability can be broken down into two parts:
\[\label{b.e56}
P(\bga_c^0\rightarrow\bga_c^*)=g(\bga_c^0\rightarrow\bga_c^*)A(\bga_c^0\rightarrow\bga_c^*),
\]
where $g(\cdot)$ is the selection probability, which is the probability given $\bga_c^0$ that the new target state generated, and $A(\cdot)$ is the acceptance ratio. Thus
\be\label{b.e57}
{{g(\bga_c^0\rightarrow\bga_c^*)A(\bga_c^0\rightarrow\bga_c^*)}\over{g(\bga_c^*\rightarrow\bga_c^0)A(\bga_c^*\rightarrow\bga_c^0)}}=\exp\left\{-\left[U(\bga_c^*)-U(\bga_c^0)\right]\right\}.
\ee
Now we consider the move $\bga_c^0\rightarrow\bga_c^*$, starting with a particular cluster $c$ and then adding the others to it in a particular order. Consider also the reverse move, which takes us back to $\bga_c^0$ from $\bga_c^*$, starting with exactly the same cluster (except the state in the cluster is flipped), and adding the others to it in exactly the same way as in the forward move. The probability of choosing the cluster (if the cluster is the seed node) is exactly the same in the two directions, as is the probability of adding each node to the cluster. The only difference between the two directions is the probability of ``breaking'' bonds around the edge of the cluster. Because the cluster couples with all $j\in\bar c$, for both directions, there are $|\bar c|$ bonds which have to be broken in order to flip the cluster. These broken bonds represent the affinity between the cluster and the spins which were not added to the cluster by the algorithm. We represent the probability of not adding such a node in forward move as $1-p^0_{a,j}, j\in\bar c$ and in backward move as $1-p^*_{a,j}, j\in\bar c$. Thus the probability of not adding all of them, which is proportional to the selection probability $g(\bga_c^0\rightarrow\bga_c^*)$ for the forward move, is $\prod_{j\in\bar c}(1-p^0_{a,j})$. In the reverse move then the probability of doing it is $\prod_{j\in\bar c}(1-p^*_{a,j})$. The condition of detailed balance, Equation (\ref{b.e55}), along with Equation (\ref{b.e57}), then tells us that
\be\label{b.e58}
{{g(\bga_c^0\rightarrow\bga_c^*)A(\bga_c^0\rightarrow\bga_c^*)}\over{g(\bga_c^*\rightarrow\bga_c^0)A(\bga_c^*\rightarrow\bga_c^0)}}=\prod_{j\in\bar c}\left({{1-p^0_{a,j}}\over{1-p^*_{a,j}}}\right)\cdot{{A(\bga_c^0\rightarrow\bga_c^*)}\over{A(\bga_c^*\rightarrow\bga_c^0)}}=\exp\left\{-\left[U(\bga_c^*)-U(\bga_c^0)\right]\right\}.
\ee
Note that the energy change $U(\bga_c^*)-U(\bga_c^0)$ is only determined by the bonds (the coupling between $c$ and $\bar c$) and coupling of $c$ with the external field $\mathbf h^*$, i.e.,
\be\label{b.e59}
U(\bga_c^*)-U(\bga_c^0)=\sum_{j\in\bar c}(-1)^{\gamma_j}\left(\sum_{k\in c_0}J_{jk}-\sum_{l\in c_1}J_{jl}\right)+\sum_{j\in c_1}h^*_j-\sum_{j\in c_0}h^*_j.
\ee
The first part of right hand side of Equation (\ref{b.e59}) can be decomposed as
\[\label{b.e60}
\lambda\sum_{j\in\bar c}(-1)^{\gamma_j}\left(\sum_{k\in c_0}J_{jk}-\sum_{l\in c_1}J_{jl}\right)+(1-\lambda)\sum_{j\in\bar c}(-1)^{\gamma_j}\left(\sum_{k\in c_0}J_{jk}-\sum_{l\in c_1}J_{jl}\right).
\]
With the probability of adding a node $j\in\bar c$ to the cluster, $p_{a,j}$, defined as (\ref{b.e17}),
\[\label{b.e61}
{{1-p^0_{a,j}}\over{1-p^*_{a,j}}}=\exp\left\{-\lambda(-1)^{\gamma_j}\left(\sum_{k\in c_0}J_{jk}-\sum_{l\in c_1}J_{jl}\right)\right\}
\]
Substituting above equation into Expression (\ref{b.e58}) and rearranging, we derive the acceptance ratio for the moves in the two directions as
\[\label{b.e62}
{{A(\bga_c^0\rightarrow\bga_c^*)}\over{A(\bga_c^*\rightarrow\bga_c^0)}}=\exp\left[(1-\lambda)\sum_{j\in\bar c}(-1)^{\gamma_j}\left(\sum_{k\in c_1}J_{jk}-\sum_{l\in c_0}J_{jl}\right)+\sum_{j\in c_0}h^*_j-\sum_{j\in c_1}h^*_j\right],
\]
and the acceptance probability for move from $\bga^0_c$ to $\bga^*_c$ is
\[\label{b.e63}
\alpha(\bga_c^0\rightarrow\bga_c^*)=\min\left\{{{A(\bga_c^0\rightarrow\bga_c^*)}\over{A(\bga_c^*\rightarrow\bga_c^0)}},1\right\}.
\]
As well as satisfying the detailed balance, the algorithm also guarantees the ergodicity by the fact that there is always a finite chance that any spin will be chosen as the sole member of cluster of one, which is then flipped. The appropriate succession of such moves will get us from any state to any other in a finite time as ergodicity requires.
\subsection{Proof of Theorem 2}\label{app2}
The proof of the first part in Theorem 2 is simply algebra calculation. First define $\mby^*=\mby-\sum_{k\ne j}\gamma_k\mbx_k\beta_k$, and integrate out $\beta_j$ and $\bet_{\bar j}$ separately in following expression
\be\begin{split}\label{b.e64}
p(\bga|\mby,\bta,b)&\propto\int p(\mby|\bga,\bet)p(\bet|\bta,b)d\bet\\
&\propto\int\exp\left[-{1\over2}\left(\mby^*-\gamma_j\mbx_j\beta_j\right)^2\right]p(\beta_j|\tau_j,b)d\beta_j\prod_{k\ne j}p(\beta_k|\tau_k,b)d\beta_k\\
&\propto\exp\left[{\gamma_j\over2}a^2_j\left(\gamma_j+{\tau_j\over b^2}\right)^{-1}\right]\left({{\tau_j/b^2}\over{\gamma_j+\tau_j/b^2}}\right)^{1/2}\cdot\xi(\gamma_j,\kappa_j,\bga_{\bar j},\bta_{\bar j}),
\end{split}
\ee
where $\xi(\gamma_j,\kappa_j,\bga_{\bar j},\bta_{\bar j})$ is calculated by integrating out $\bet_{\bar j}$:
\be\begin{split}\label{b.e65}
\xi(\gamma_j,\kappa_j,\bga_{\bar j},\bta_{\bar j})\propto&\int\exp\left[{\gamma_j\over2}\left(\gamma_j+{\tau_j\over b^2}\right)^{-1}\left(\sum_{k\ne j;l\ne j}\beta_k\gamma_k\mbx^T_k\mbx_j\mbx_j^T\mbx_l\beta_l\gamma_l-2\sum_{k\ne j}\mbx^T_k\mbx_j\mbx_j^T\mby\beta_k\gamma_k\right)\right]\\
&\times\exp\left(-{{{\mby^*}^T\mby^*}\over2}\right)\prod_{k\ne j}p(\beta_k|\tau_k,b)d\beta_k\\
\propto&\int\exp\left[{\gamma_j\over2}\left(\gamma_j+{\tau_j\over b^2}\right)^{-1}\left(\bet^T_{\bar j}(\mathbf c_{\bga_{\bar j}}\mathbf c^T_{\bga_{\bar j}})\bet_{\bar j}-2a_j\mathbf c^T_{\bga_{\bar j}}\bet_{\bar j}\right)\right]\\
&\times\exp\left[-{1\over2}\left(\mby-X_{\bga_{\bar j}}\bet_{\bar j}\right)^2\right]p(\bet_{\bar j}|\bta_{\bar j},b)d\bet_{\bar j}\\
\propto&\exp\left[{1\over2}\left(\mathbf a_{\bar j}-(1-\kappa_j^{\gamma_j})a_j\mathbf c_{\bga_{\bar j}}\right)^T\Omega_j^{-1}\left(\mathbf a_{\bar j}-(1-\kappa_j^{\gamma_j})a_j\mathbf c_{\bga_{\bar j}}\right)\right]|\Omega_j|^{1/2}|D_{\bar j}|^{1/2},
\end{split}\ee
where $\Omega_j=[D_{\bar j}+X^T_{\bga_{\bar j}}X_{\bga_{\bar j}}-(1-\kappa_j)^{\gamma_j}(\mathbf c_{\bga_{\bar j}}\mathbf c^T_{\bga_{\bar j}})]^{-1}$ and is easy to show it is positive definite. We also used the identity $\gamma_j/\left(\gamma_j+{\tau_j\over b^2}\right)=(1-\kappa_j^{\gamma_j})$. Note if $\gamma_j=0$, $\xi(\cdots)$ does not depend on  $\kappa_j$ or $\tau_j$.

Then by definition and above expressions,
\be\begin{split}\label{b.e66}
\pi^b_j&={{\int\sum_{\bga_{\bar j}}p(\gamma_j=1,\bga_{\bar j}|\mby,\bta,b)p(\bta)d\bta}\over{\int\sum_{\bga_{\bar j}}p(\gamma_j=0,\bga_{\bar j}||\mby,\bta,b)p(\bta)d\bta}}\\
&=\int\pi_j\cdot{{\int\sum_{\bga_{\bar j}}\xi(\gamma_j=1,\kappa_j,\bga_{\bar j},\bta_{\bar j})p(\bta_{\bar j})d\bta_{\bar j}}\over{\int\sum_{\bga_{\bar j}}\xi(\gamma_j=0,\kappa_j,\bga_{\bar j},\bta_{\bar j})p(\bta_{\bar j})d\bta_{\bar j}}}\cdot p(\tau_j)d\tau_j\\
&=\int\pi_j\xi_jp(\kappa_j)d\kappa_j
\end{split}\ee
with $x_j$ defined as (\ref{b.e25}). It is easy to show that $\xi_j=1$ for orthogonal design since $\mathbf c_{\bga_{\bar j}}=\mathbf 0$ for all $j$.

The proof of the second part for $\pi_j$ is trivial. Obviously, $\pi_j\rightarrow1$ as $\kappa_j\rightarrow1$ and $\pi_j\rightarrow0$ as $\kappa_j\rightarrow0$. For $\pi^b_j$, it is more convenient using $\pi^b_j=\int\pi_j\xi_jp(\tau_j)d\tau_j$, where $\pi_j$ and $\xi_j$ are measurable functions indexed by $b$. Both $\pi_j$ and $\xi_j$ are bounded by some positive number for all $b$. When $b\rightarrow0$, $\lim_{b\rightarrow0}\pi_j\rightarrow1$ and $\lim_{b\rightarrow0}\xi_j\rightarrow1$, thus according to Lebesgue's Dominated Convergence Theorem (DCT), $\lim_{b\rightarrow0}\pi^b_j=\lim_{b\rightarrow0}\int\pi_j\xi_jp(\tau_j)d\tau_j=\int\lim_{b\rightarrow0}(\pi_j\xi_j)p(\tau_j)d\tau_j=1$. When $b\rightarrow\infty$, $\lim_{b\rightarrow\infty}\pi_j\rightarrow0$ and $\lim_{b\rightarrow\infty}\xi_j$ equal to some finite number. Again the limit and integral commute by DCT, thus we have $\lim_{b\rightarrow\infty}\pi^b_j=\int\lim_{b\rightarrow\infty}(\pi_j\xi_j)p(\tau_j)d\tau_j=0$.

\subsection{Proof of Theorem 3}\label{app3}
The existence of $m_j$ indicates the marginal prior $p(\beta_j)=\int p(\beta_j|\tau_j)p(\tau_j)d\tau_j$ is bounded for $\beta_j\in\mathbb{R}$, which is true for Cauchy and Laplace prior. Using identity
\[\label{b.e67}
(\beta_j-a_j)p(\mby|\beta_j,\gamma_j=1)=\mbx_j^T{\partial\over{\partial\mby}} p(\mby|\beta_j,\gamma_j=1),
\]
so that
\be\begin{split}\label{b.e68}
m_j[E(\beta_j|\mby,\gamma_j=1)-a_j]&=\int(\beta_j-a_j)p(\mby|\beta_j,\gamma_j=1)p(\beta_j)d\beta_j\\
&=\int\mbx^T_j{\partial\over{\partial\mby}} p(\mby|\beta_j,\gamma_j=1)p(\beta_j)d\beta_j\\
&=\mbx^T_j{\partial\over{\partial\mby}}\log m_j.
\end{split}\ee
Following the lemma given in \citet{c24}, the interchange of the derivative and the integral is justified. The second result of Theorem 3 is straightforward by observing $m_j\propto\exp\left(-{1\over2}\mby^T\mby\right)\pi^b_j$, thus
\[\label{b.e69}
\mbx^T_j{\partial\over{\partial\mby}}\log m_j=\mbx^T_j\left(-\mby+{{d\log\pi^b_j}\over{d a_j}}{{d a_j}\over{d\mby}}\right)=-a_j+{{d\log\pi^b_j}\over{d a_j}}.
\]

For horseshoe prior, $p(\beta_j)$ is not bounded. However, using the technique introduced in \citet{c3} by defining $m^*_j=\int p(\mby|\beta_j,\gamma_j=1)p(\beta_j|\tau_j)p(\tau_j)\tau_j^{-1}d\beta_jd\tau_j$, it can be shown
\[\label{b.e70}
E(\beta_j|\mby,\gamma_j=1)=-{m^*_j\over{m_j}}\mbx^T_j{\partial\over{\partial\mby}}\log m^*_j,
\]
and similar arguments then follow. For horseshoe prior, it also can be shown
\[\label{b.e71}
{\pi^b_j\over{\pi^b_j}^*}{d\over{da_j}}\log\pi^b_j={1\over{\pi^b_j}^*}{d\over{d a_j}}\left(\int\pi_j(\tau_j^{-1}+1)p(\tau_j)d\tau_j\right)-E(\kappa_j|\mby,\gamma_j=1)a_j\]
where ${\pi^b_j}^*=\int\pi_j\tau_j^{-1}p(\tau_j)d\tau_j$.

\subsection{Proof of Theorem 4}\label{app4}
It is more convenient to use following equivalent  representation of $\pi^b_j$
\[\label{b.e72}
\pi^b_j=\int\exp\left[{a_j^2\over2}(1+\tau_j)^{-1}\right]\left[\tau_j/(1+\tau_j)\right]^{1\over2}p(\tau_j|b)d\tau_j,
\]
where $p(\tau_j|b)$ is corresponding prior of $\tau_j$ given $b$ such that $\pi^b_j=\int p(\beta_j|b)p(\tau_j)d\tau_j=\int p(\beta_j)p(\tau_j|b)d\tau_j$. Then the similar condition for $p(\tau_j|b)$ can be derived from the condition for $p(\tau_j)$ in Theorem 4, i.e.,
\be\label{b.e73}
p(\sigma^2_j|b)\sim {(\sigma^2_j)}^{\alpha-1}\exp\left(-{\lambda\over b^2}\sigma^2_j\right)L^b(\sigma^2_j)d\sigma^2_j,\hbox{ as }\sigma^2\rightarrow\infty,
\ee
where $L^b$ is the slowly varying function conditioning on parameter $b$.

Then the marginal odds $\pi^b_j$ can be expressed as
\[\label{b.e74}
\pi^b_j\propto\exp\left(a^2_j\over2\right)m^b(a_j)=\exp\left(a^2_j\over2\right)\int\omega^{-1}_j\exp\left(-{a_j^2\over\omega^2_j}\right)p(\omega^2_j|b)d\omega^2_j,
\]
where $\omega^2_j=1+\sigma^2_j$, and the integral, $m^b(a_j)$, is a scale mixture of normals. Now the proof is similar to \citet{c25}. If prior $p(\sigma^2_j|b)$ satisfies the conditions defined in (\ref{b.e73}), so does $p(\omega^2_j|b)$ satisfy similar conditions, i.e.,
\[\label{b.e75}
p(\omega^2_j|b)\sim{(\omega^2_j)}^{\alpha-1}\exp\left(-{\lambda\over b^2}\omega^2_j\right)L^b\left({\omega^2_j}\right)\hbox{ as }\omega^2_j\rightarrow\infty
\]
Then following Theorem 6.1 of \citet{c0}, as $a_j\rightarrow\infty$, $m^b(a_j)$ can be approximated as
\be\label{b.e76}
m^b(a_j)\sim\left\{\begin{array}{l  l}
  |a_j|^{2\alpha-1}L^b\left(a_j^2\right)  &if\;\; \lambda=0  \\
  |a_j|^{\alpha-1}\exp\left(-\sqrt{{2\lambda}\over b^2}|a_j|\right)L^b\left(|a_j|\right)                                               &if\;\; \lambda>0, \end{array} \right.
\ee
as $a_j\rightarrow\infty$.  The results in Theorem 4 follow by taking derivative respect to $|a_j|$ and $b$ respectively.
\subsection{The Calculation of $\pi^b_j$}\label{app5}
For orthogonal designs, $\pi^b_j$ with Laplace prior can be integrated out directly from (\ref{b.e27})
\be\begin{split}\label{b.e77}
\pi^b_j&=\int_0^1\kappa_j^{1\over2}\exp\left[{a^2_j\over2}(1-\kappa_j)\right]\cdot{1\over{2b^2}}\kappa_j^{-2}\exp\left(-{{1-\kappa_j}\over{2b^2\kappa_j}}\right)d\kappa_j\\
&={1\over{2b^2}}\left({{2\pi}\over\lambda}\right)^{1/2}\exp\left({1\over{2b^2}}+{a^2_j\over2}-{\sqrt{a_j^2}\over b}\right)\int_0^1\left({\lambda\over{2\pi}}\right)^{1/2}\kappa_j^{-3/2}\exp\left[-{{\lambda(\kappa_j-\mu)^2}\over{2\mu^2\kappa_j}}\right]d\kappa_j,
\end{split}\ee
where $\lambda=1/b^2$ and $\mu=\sqrt{1/(b^2a_j^2)}$, and the expression in the integral is the CDF of inverse Gaussian distribution. Borrowing the expression of the CDF of the inverse Gaussian, we then integrated out the integral to get expression (\ref{b.e34}).

$\pi^b_j$ with horseshoe prior for orthogonal design can also be derived directly from (\ref{b.e27}):
\be\begin{split}\label{b.e78}
\pi^b_j&=\int_0^1\kappa_j^{1\over2}\exp\left[{a^2_j\over2}(1-\kappa_j)\right]\cdot{b\over\pi}\kappa_j^{-1/2}(1-\kappa_j)^{-1/2}(1-\kappa_j+b^2\kappa_j)^{-1}d\kappa_j\\
&={1\over{\pi b}}\exp\left(a_j^2\over2\right)\int^1_0\kappa_j^{1-1}(1-\kappa_j)^{1/2-1}\left[(1-b^{-2})\kappa_j+b^{-2}\right]^{-1}\exp\left(-{a_j^2\over2}\kappa_j\right)d\kappa_j,
\end{split}\ee
where the expression in the integral is the transformation of the hypergeometric inverted-beta distribution which was shown to be represented by degenerate hypergeometric functions \citep{c6,c27}, thus we can follow \citet{c27} to express it as (\ref{b.e35}).

\subsection{Lancaster and \v{S}alkauskas Basis for Natural Cubic Spline}\label{app6}

In this paper, we follow \citet{c4} to employ the cubic spline LS basis described by \citet{c14}. Consider the $j$th function $f_j(x)$, and let $\boldsymbol\nu_j=(\nu_{1j},...,\nu_{K_jj})$ be the set of $100\times{{k-1}\over{K_j-1}}\%, k=1,...,K_j$ quantile of $x_{ij}, i=1,...,n$. Thus $\nu_{1j}=\min_i(x_{ij})$ and $\nu_{K_jj}=\max_i(x_{ij})$. $K_j$ denotes the number of knots for the spline functions. Then the cubic spline expansion of $f_j(x)$ is expressed as
\be\begin{split}\label{b.e79}
f_j(x_{ij})&=\sum_{k=1}^{K_j}[\Phi_{kj}(x_{ij})g_{kj}+\Psi_{kj}(x_{ij})s_{kj}]\\
&=\Phi_j(x_{ij})^T\mathbf g_j+\Psi_j(x_{ij})^T\mathbf s_j,
\end{split}\ee
where $\mathbf g_j=(g_{1j},...,g_{K_jj})^T$ and $\mathbf s_j=(s_{1j},...,s_{K_jj})^T$ are the coefficients of this expression, $\Phi_j(x_{ij})=[\Phi_{1j}(x_{ij}),...\Phi_{K_jj}(x_{ij})]^T$ and $\Psi_j(x_{ij})=[\Psi_{1j}(x_{ij}),...\Psi_{K_jj}(x_{ij})]^T$ are two basis vectors, and the basis functions $\{\Phi_{kj}(x)\}_{k=1}^{K_j}$ and $\{\Psi_{kj}(x)\}_{k=1}^{K_j}$ are defined as
\be\begin{split}\label{b.e80}
\Phi_{kj}(x)&\propto\left\{\begin{array}{l l}
 0,                                                        & x<\nu_{k-1,j}           \\
 -(2/h_{kj}^3)(x-\nu_{k-1,j})^2(x-\nu_{kj}-0.5h_{kj}),     & \nu_{k-1,j} \leq x<\nu_{kj}           \\
(2/h_{k+1,j}^3)(x-\nu_{k+1,j})^2(x-\nu_{kj}+0.5h_{k+1,j}), & \nu_{kj} \leq x<\nu_{k+1,j}           \\
0                                                          & x\geq\nu_{k+1,j},\end{array} \right.\\
\Psi_{kj}(x)&\propto\left\{\begin{array}{l l}
 0,                                                        & x<\nu_{k-1,j}           \\
 (1/h_{kj}^2)(x-\nu_{k-1,j})^2(x-\nu_{kj}),                & \nu_{k-1,j} \leq x<\nu_{kj}           \\
(1/h_{k+1,j}^2)(x-\nu_{k+1,j})^2(x-\nu_{kj}),              & \nu_{kj} \leq x<\nu_{k+1,j}           \\
0                                                          & x\geq\nu_{k+1,j},\end{array} \right.\\
\end{split}\ee
where $h_{kj}=\nu_{kj}-\nu_{k-1,j}$. Note that $\Phi_{1j},\Psi_{1j}$ and $\Phi_{K_jj},\Psi_{K_jj}$ are defined by last two lines and first two lines of above expressions respectively. $\mathbf g_j$ and $\mathbf s_j$ are interpreted the ordinate and slope of $f_j(x)$. Since $f_j(x)$ is a natural cubic splines with the second derivative equal to zero at two end points, and continuous derivative at knot points, both $\mathbf g_j$ and $\mathbf s_j$ are constrained \citep{c14} by $\mathbf s_j=A_j^{-1}C_j\mathbf g_j$, where
\[\label{b.e81}
A_j= \left( \begin{array}{ccccccccc}
2           & 1           & 0        & 0        & 0      & \cdots & 0                  & 0      & 0\\
\omega_{2j} & 2           & \mu_{2j} & 0        & 0      & \cdots & 0                  & 0      & 0\\
0           & \omega_{2j} & 2        & \mu_{2j} & 0      & \cdots & 0                  & 0      & 0\\
\vdots      & \cdots      & \ddots   & \ddots   & \ddots & \cdots & \vdots             & \vdots & \vdots\\
0           & 0           & 0        & 0        & 0      & \cdots & \omega_{K_j-1,j}   & 2      & \mu_{K_j-1,j}\\
0           & 0           & 0        & 0        & 0      & \cdots & 0                  & 1      & 2\end{array} \right),\]
and
\[\label{b.e82}
C_j= \left( \begin{array}{ccccccccc}
-{1\over{h_{2j}}}  & {1\over{h_{2j}}} & 0  & 0   & \cdots & 0   & 0   & 0\\
-{{\omega_{2j}}\over{h_{2j}}} & {{\omega_{2j}}\over{h_{2j}}}-{{\mu_{2j}}\over{h_{3j}}}    & {{\mu_{2j}}\over{h_{3j}}} & 0        & \cdots & 0                  & 0      & 0\\
0  & -{{\omega_{3j}}\over{h_{3j}}} & {{\omega_{3j}}\over{h_{3j}}}-{{\mu_{3j}}\over{h_{4j}}}        & {{\mu_{3j}}\over{h_{4j}}} & \cdots & 0                  & 0      & 0\\
\vdots      & \cdots      & \ddots   & \ddots   & \cdots & \vdots             & \vdots & \vdots\\
0  & 0  & 0  & 0  & \cdots & -{{\omega_{K_j-1,j}}\over{h_{K_j-1}}} & {{\omega_{K_j-1,j}}\over{h_{K_j-1}}}-{{\mu_{K_j-1,j}}\over{h_{K_j}}}& {{\mu_{K_j-1,j}}\over{h_{K_j}}}\\
0  & 0  & 0  & 0  & \cdots & 0  & -{1\over{h_{K_jj}}}   & {1\over{h_{K_ij}}}\end{array} \right),\]
where $\omega_{kj}=h_{kj}/(h_{kj}+h_{k+1,j})$ and $\mu_{kj}=1-\omega_{kj}$ for $k=2,...,K_j$.
With this constraints, $\mathbf s_j$ can be replaced from the function expression (\ref{b.e79}),
\be\begin{split}\label{b.e83}
f_j(x_{ij})&=[\Phi_j(x_{ij})^T+\Psi_j(x_{ij})^TA_j^{-1}C_j]\mathbf g_j\\
&=\mathbf t_j^T(x_{ij})\mathbf g_j,
\end{split}\ee
where $\mathbf t_j^T(x_{ij})=(t_{1j}(x_{ij}),...,t_{K_jj}(x_{ij}))=\Phi_j(x_{ij})^T+\Psi_j(x_{ij})^TA_j^{-1}C_j$. Furthermore, consider the identifying constraints, $\sum_kg_{kj}=0$, we can express $g_{1j}=-(g_{2j}+\cdots+g_{K_jj})$, thus
\[\label{b.e84}
f_j(x_{ij})=\mathbf t_j^T(x_{ij})\mathbf g_j=[t_{2j}(x_{ij})-t_{1j}(x_{ij})]g_{2j}+\cdots+[t_{K_jj}(x_{ij})-t_{1j}(x_{ij})]g_{K_jj}={\mbz_j^*}^T(x_{ij})\bet_j,
\]
where $\bet_j=(g_{2j},...,g_{K_jj})^T$ and we define matrix
\[\label{b.e85}
Z^*_j=\left( \begin{array}{c}
{\mbz_j^*}^T(x_{1j})  \\
\vdots   \\
{\mbz_j^*}^T(x_{nj}) \end{array} \right).
\]
Now the $j$th nonparametric function expressed by the natural cubic spline basis is $f_j(\mbx_j)=Z^*_j\bet_j$. In order to incorporate the assumption of a priori smoothness, \citet{c4} consider a prior distribution on $\bet_j$'s as,
\be\label{b.e86}
[\bet_j|\sigma^2_{ej},\sigma^2_{dj}]\sim N\left[\mathbf 0, \Delta_j^{-1}T_j(\Delta_j^{-1})^T\right],
\ee
where $N$ is the $K_j-1$ dimensional multivariate normal distribution, and
\[\label{b.e87}
T_j=\left(\begin{array}{ccc}
\sigma^2_{ej} & 0                     & 0\\
0             &\sigma^2_{dj}I_{K_j-3} & 0\\
0             & 0                     &\sigma^2_{ej}
\end{array}\right),
\]
where two variance components $\sigma^2_{ej}$ and $\sigma^2_{dj}$ are selected here because of the different normal assumptions for the differences of the ordinates and the differences of slopes. $\Delta_j$ is given by
\[\label{b.e88}
\Delta_j= \left( \begin{array}{cccccccc}
{2\over{h_{2j}}}  & {1\over{h_{2j}}} & {1\over{h_{2j}}}  & {1\over{h_{2j}}}  & {1\over{h_{2j}}} & \cdots & {1\over{h_{2j}}}   &{1\over{h_{2j}}}\\
{1\over{h_{2j}}}  & -\left({1\over{h_{2j}}}+{1\over{h_{3j}}}\right) & {1\over{h_{3j}}} & 0      &0  & \cdots & 0                      & 0\\
0  & {1\over{h_{3j}}}  & -\left({1\over{h_{3j}}}+{1\over{h_{4j}}}\right)        & {1\over{h_{4j}}} &0  & \cdots & 0                   & 0\\
\vdots      & \vdots      & \vdots   & \vdots   & \vdots & \cdots             & \vdots & \vdots\\
0  & 0  & 0  & 0 &0 & \cdots &  -{1\over{h_{K_jj}}}   & {1\over{h_{M_jj}}}\end{array} \right).
\]

So far the construction of function $f_j(\mbx_j)$ is exactly the same as \citet{c4}. Note that $f_j(\mbx_j)=Z^*_j\bet_j$ with the prior of $\bet_j$ given by (\ref{b.e86}) is equivalent to have $f_j(\mbx_j)=Z^*_j\Delta^{-1}_j\bet_j$ with $[\bet_j]\sim N(\mathbf0, T_j)$. Henceforth, we define the final $n\times M_j$ basis matrix $Z_j=Z_j^*\Delta^{-1}_j$ such that $f_j(\mbx_j)=Z_j\bet_j$, where $M_j=K_j-1$. Define $\tau_{ej}=\sigma^{-2}_{ej}$ and $\tau_{bj}=\sigma^{-2}_{bj}$, and modify the one variance component prior algorithm in Section \ref{b.sec7}, then we can easily employ the LS basis into BSAM.

\pagebreak

\end{document}